\documentclass[12pt]{elsarticle}

\usepackage{graphicx, import}
\usepackage{amsfonts,textgreek}
\usepackage{mathtools}
\usepackage{dsfont}
\usepackage{multicol}
\usepackage{bbm}
\usepackage{url}
\usepackage[dvipsnames]{xcolor}
\usepackage[colorlinks=true,allcolors=blue]{hyperref}
\usepackage[nameinlink,noabbrev,capitalize]{cleveref}
\usepackage{xfrac}
\usepackage{caption}
\usepackage{subcaption}
\usepackage{xspace}
\usepackage{stmaryrd}
\usepackage{enumerate}
\usepackage{soul}
\usepackage{lscape}

\usepackage[utf8]{inputenc}
\usepackage{wrapfig,bm}
\usepackage{lineno}
\usepackage{siunitx} 
\usepackage{caption}
\usepackage{psfrag}
\usepackage{upgreek}
\usepackage{isomath}
\usepackage{latexsym}
\usepackage{array}
\usepackage{multirow}
\usepackage{booktabs}
\usepackage{colortbl}
\usepackage{verbatim}
\usepackage{graphicx}
\usepackage{epstopdf,overpic}
\usepackage{float}
\usepackage{soul}
\usepackage{color}
\usepackage{dirtytalk}
\usepackage{textcomp}
\usepackage{algorithm, algpseudocode, lmodern, mathtools}
\usepackage{tikz}
\usetikzlibrary{shapes.geometric, arrows}
\usepackage{amssymb}
\usepackage{amsmath}
\usepackage{mathrsfs}  
\usepackage{graphicx}
\usepackage{titlesec}
\usepackage{indentfirst}
\usepackage{bigints}
\usepackage{breqn}
\usepackage[export]{adjustbox}
\usepackage{pgfplots}
\usetikzlibrary{plotmarks,spy,calc}
\newlength{\figureheight}
\newlength{\figurewidth}

\usepackage[margin=2.5cm]{geometry}

\usepackage{multirow}

\usepgfplotslibrary{groupplots}

\def\centerarc[#1](#2)(#3:#4:#5);%
    {
    \draw[#1]([shift=(#3:#5)]#2) arc (#3:#4:#5);
    }

\biboptions{sort&compress}

\pgfplotsset{compat=1.15}

\tikzstyle{decision} = [diamond, draw, fill=blue!20, 
    text width=4.5em, text badly centered, node distance=3cm, inner sep=0pt]
\tikzstyle{block1} = [rectangle, draw, fill=blue!20, 
    text width=12em, text centered, rounded corners, minimum height = 3em]
\tikzstyle{block2} = [rectangle, draw, fill=blue!20, 
    text width=20em, text centered, rounded corners, minimum height = 3em]
\tikzstyle{block3} = [rectangle, draw, fill=blue!20, 
    text width=16em, text centered, rounded corners, minimum height = 3em]
\tikzstyle{block4} = [rectangle, draw, fill=blue!20, 
    text width=14em, text centered, rounded corners, minimum height = 3em]
\tikzstyle{block5} = [rectangle, draw, fill=blue!20, 
    text width=20em, text centered, rounded corners, minimum height = 3em]
\tikzstyle{block6} = [rectangle, draw, fill=blue!20, 
    text width=6em, text centered, rounded corners, minimum height = 3em]    

\tikzstyle{redblock1} = [rectangle, draw, fill=red!20, 
    text width=20em, text centered, rounded corners, minimum height = 3em]
\tikzstyle{redblock2} = [rectangle, draw, fill=red!20, 
    text width=20em, text centered, rounded corners, minimum height = 3em]
\tikzstyle{redblock3} = [rectangle, draw, fill=red!20, 
    text width=28em, text centered, rounded corners, minimum height = 4em]
\tikzstyle{redblock4} = [rectangle, draw, fill=red!20, 
    text width=24em, text centered, rounded corners, minimum height = 3em]
\tikzstyle{redblock5} = [rectangle, draw, fill=red!20, 
    text width=20em, text centered, rounded corners, minimum height = 3em]
\tikzstyle{redblock6} = [rectangle, draw, fill=red!20, 
    text width=6em, text centered, rounded corners, minimum height = 3em] 
    
\tikzstyle{line} = [draw, -latex', line width=1.0pt]
\tikzstyle{line2} = [draw, line width=1.0pt]
\tikzstyle{cloud} = [draw, ellipse,fill=red!20,text width=12em, text centered,
    minimum height=2em]
\tikzstyle{cloud2} = [draw, chamfered rectangle,fill=red!20,text width=14em, text centered,
    minimum height=2em]
\tikzstyle{cloud3} = [draw, chamfered rectangle,fill=red!20,text width=15em, text centered,
    minimum height=2em] 

\begin{document}
\begin{frontmatter}
 
\title{Adaptive shape optimization with NURBS designs and PHT-splines for solution approximation in time-harmonic acoustics}

\author[]{Javier Videla$^a$}
\author[]{Ahmed Mostafa Shaaban$^b$}
\author[]{Elena Atroshchenko$^{a,}$ \footnote{Corresponding author, eatroshch@gmail.com, e.atroshchenko@unsw.edu.au}}

\address[add1]{School of Civil and Environmental Engineering, University of New South Wales, Sydney, Australia}
\address[add2]{Institute of Structural Mechanics, Bauhaus-Universität Weimar, Weimar, Germany}


\begin{abstract}
Geometry Independent Field approximaTion (GIFT) was proposed as a generalization of Isogeometric analysis (IGA), where different types of splines are used for the parameterization of the computational domain and approximation of the unknown solution. GIFT with Non-Uniform Rational B-Splines (NUBRS) for the geometry and PHT-splines for the solution approximation were successfully applied to problems of time-harmonic acoustics, where it was shown that in some cases, adaptive PHT-spline mesh yields highly accurate solutions at lower computational cost than methods with uniform refinement. Therefore, it is of interest to investigate performance of GIFT for shape optimization problems, where NURBS are used to model the boundary with their control points being the design variables and PHT-splines are used to approximate the solution adaptively to the boundary changes during the optimization process. 

In this work we demonstrate the application of GIFT for 2D acoustic shape optimization problems and, using three benchmark examples, we show that the method yields accurate solutions with significant computational savings in terms of the number of degrees of freedom and computational time.

\end{abstract}


\end{frontmatter}


\section{Introduction}

Due to rapid civil and transport development, noise pollution has become an important public health concern, requiring efficient noise control solutions, such as design of structures (e.g. noise barriers) with optimal acoustic performance. Numerical design is based on the solution of the wave propagation problem, modeled by the Helmholtz equation. 

Numerical methods for the Helmholtz equation encounter two major challenges: the so-called ``pollution error" and treatment of unbounded domains. The pollution error results from the numerical dispersion error, which is related to the discrepancy between the exact and numerical wave number $(k)$. It has been proven theoretically that for 2D and 3D problems, the pollution error cannot be fully eliminated \cite{doi:10.1137/S0036142994269186}. It is controlled by adapting the mesh size to the wave length, i.e. the element size is held a few times smaller than the wave length. Hence, the number of elements grows proportionally to the wave number, making computations for large $k$ unfeasible. However, pollution error is inversely proportional to the order of shape functions and hence can be significantly reduced by the use of higher-order approximations. This makes isogeometric analysis (IGA), with its natural feature of polynomial degree elevation or $p-$refinement, an attractive alternative for solution of the Helmholtz equation, as demonstrated in \cite{COOX2016441,WU201563}.

Another source of error in modeling wave propagation in unbounded domains results from introducing a boundary truncation domain (usually a circle of radius $R$ in 2D and a sphere of radius $R$ in 3D), in which the Sommerfeld radiation condition is modeled by the Absorbing Boundary Condition (ABC). The accuracy of modeling the ABC increases with $R$, nonetheless, larger computational domain leads to a larger system and higher computational cost. Another approach to reduce domain truncation error is by using higher order ABCs. This requires higher order derivatives of the shape functions, which present a problem in linear FEM, but can be easily overcome by the use of higher order NURBS in IGA. A detailed study of various ABCs in the context of Isogeometric Collocation was performed in \cite{Atr-IGA-C}. 

Isogeometric analysis (IGA) and its subsequent variations have been a topic of active research since IGA was originally introduced in 2005 by Hughes et al. \cite{hughes2005isogeometric}. The main objective of IGA is to connect the Computer-Aided Design (CAD) model directly with numerical analysis. In most cases, Non-Uniform Rational B-Spline (NURBS) are able to preserve the original geometry replacing Lagrange polynomials in the FEM discretization. It has been successfully proven that, IGA could be implemented in several engineering applications, such as structural mechanics \cite{zhang2017nurbs}, fluid-structure interaction \cite{bazilevs2008isogeometric}, fracture mechanics \cite{de2011x}, electromagnetics \cite{BUFFA20101143}, Helmholtz equation \cite{COOX2016441,WU201563}, among others. Ref. \cite{nguyen2015isogeometric} presents a complete review for the application of IGA in different engineering aspects. 

The necessity to truncate the boundary presents a serious disadvantage of domain-type methods in comparison with the boundary-type methods, such as boundary element method (BEM) and its isogeometric variation, namely IGABEM. In BEM, the Sommerfeld radiation condition at infinity is automatically satisfied by the fundamental solutions and the mesh burden is reduced to the boundary discretization \cite{SommerfeldPartial,KELLER1989172}. BEM and IGABEM have been widely applied to acoustic problems in exterior domains \cite{SIMPSON2014265}. Moreover, IGABEM has been efficiently paired with optimization methods for acoustic shape optimization in different research works \cite{CHEN2019926,SHAABAN2021113950,JIANG2021124,SHAABAN2022108410}.

One of the fundamental problems of the numerical formulas which utilize either NURBS or B-splines as basis functions (such as IGA) is that, in 2D and 3D applications, basis functions are generated as a tensor-product of 1D structures without the ability of local refinement. This makes them expensive in terms of computational resources. Different basis functions have been proposed to overcome this disadvantage, such as T-splines \cite{sederberg2004t}, truncated hierarchical B-splines (THB-splines) \cite{GIANNELLI2012485} and Polynomial splines over Hierarchical T-meshes (PHT-Splines) \cite{deng2006dimensions, deng2008polynomial}. All of the aforementioned basis functions have been efficiently applied to the problems of statics, dynamics, acoustics. See \cite{BAZILEVS2010229,GIANNELLI2012485,videla2019h}.


Geometry Independent Field approximaTion (GIFT) is a numerical scheme proposed in \cite{atroshchenko2018weakening} as an alternative to iso-parametric methods that decouples the geometry and field spaces. It consists in employing certain basis functions (for example, NURBS) to model the computational domain, whereas employing different basis functions to approximate the field. GIFT with NURBS to model the geometry and PHT-splines to model the unknown solution were applied to problems of linear elasticity and Laplace equation \cite{atroshchenko2018weakening}, bending of cracked Kirchhoff-Love plates \cite{videlaapplication}, and time-harmonic acoustics \cite{videla2019h}. It was demonstrated that, the adaptive local refinement of the solution can obtain optimal convergence rates in the cases of solutions of reduced continuity. Additionally, Jansari et al. \cite{JANSARI2022106728} developed the research work of \cite{videla2019h} by enriching the PHT-splines solution with the partition of unity property using plane waves, showing in some cases that, the enrichment provides an enhancement of several orders of magnitude in the overall solution error. 

In the field of optimization, IGA formulations have been applied to many engineering applications, such as: structural shape optimization \cite{SUN201826,LOPEZ20211004}, composite structural optimization \cite{GAO2022115263}, acoustic shape optimization \cite{SHAABAN2020115598}, piezoelectric energy harvesters \cite{PERALTA2020115521,CALDERONHURTADO2022116503}, thermal meta-materials \cite{JANSARI2022123201}, heat conduction problems \cite{KOSTAS2018600} and fluids \cite{Nortoft2013} (for a comprehensive review, the reader is referred to \cite{IGA_shape_optimization_review}). In addition to the better accuracy of IGA over FEM per degree of freedom because of the NURBS higher order and higher continuity, IGA addresses an important issue in FEM shape optimization associated with the boundary representation and the evolution of mesh following the changing boundary. In IGA, the set of control variables that parameterize the boundary is naturally chosen as design variables, providing a tight link between the design, analysis and optimization models. The same approach to shape optimization is applicable in GIFT.

The optimization methods are classified in two families: gradient-free and gradient-based methods. The family of gradient-free methods includes, for example, Particle Swarm Optimization (PSO) \cite{Kennedy,Eberhart} implemented in shape optimization problems in \cite{SUN201826,SHAABAN2021113950}, Genetic Algorithm \cite{RENNER2003709} and its optimization applications in \cite{BARBIERI2013356}, among others. The main advantage of gradient-free methods is their ability to find global minimum without any sensitivity analysis, however, at a much higher computational cost than gradient-based methods. Gradient-based methods \cite{Sokolowski,ALLAIRE20211} have much higher convergence rate, however, depending on the initial guess, they may converge to a local minimum. The performance of gradient-based optimization methods can be further improved by providing exact gradients, obtained from the shape derivatives of the objective function, the weak form of the problem and constraints to perform the sensitivity analysis \cite{CHEN2019926,LOPEZ20211004}. However, in many applications shape derivatives are difficult to obtain. In such case, gradients are calculated using finite-difference approximations. This can be done efficiently for small and medium-size problems. In this work, we use Sequential Quadratic Programming (SQP) algorithm, which is considered as one of the most effective methods for solving nonlinear constraint optimization problems \cite{BOGGS2000123}. In SPQ, the sequence of quadratic sub-problems is solved at each iteration to obtain the search direction. The algorithm is efficiently implemented within Matlab \verb $fmincon$ function.

Combining shape optimization with adaptive refinement can be traced back to the 80's. For example, the works of Kikuchi et al. \cite{kikuchi1986adaptive} and Canales et al. \cite{CANALES1993131}, are devoted to study the shape optimization problems for linear elasticity using FEM and adaptive refinements. Both works show that it is possible to achieve a process of automatic mesh generation and shape optimization, while the main drawback is that the mesh is prone to distortion, which is a usual problem with FE formulations. In a more recent work, Mohite and Upadhyay \cite{MOHITE201519} proposed an adaptive shape optimization framework for laminated composite plates, showing the advantage of adaptive refinement over uniform refinement.

As pointed out by the review study from Upadhyay et al. \cite{UPADHYAY2021102992}, adaptive mesh refinement is not fully integrated in practical applications, since the mesh refinement requires accesses to the exact geometry, which means, an automatic communication with CAD model is needed. The later is a drawback that can be alleviated with our GIFT formulation.

In the literature related to adaptive optimization with splines, Chen et al. \cite{chen2020adaptive} proposed an adaptive shape optimization scheme using T-splines and finite cell method for structural shape optimization problems. Topology optimization using morphable components and Hierarchical B-splines (HB-splines) \cite{XIE2020112696}, as well as Truncated Hierarchical B-splines (THB-Splines) has been introduced by Xie et al.\cite{XIE2021131} showing superior numerical performance over its uniform counterpart.

It is of particular interest the work of Gupta et al. \cite{GUPTA2022114993} on adaptive topology optimization using GIFT. In their work, they used NURBS to model the geometry, and PHT-splines to approximate the unknown displacement field and the density function in the framework of Solid Isotropic Material with Penalization (SIMP) method. It was shown that adaptive GIFT can achieve up to 90\% CPU time reduction in comparison with uniform meshes.

In this paper, an adaptive shape optimization scheme for Helmholtz problems is proposed. GIFT scheme is employed, using PHT-splines to discretize the sound field, while NURBS are utilized to model the boundary and the interior of the computational domain. The shape optimization is performed over the control points that define the NURBS boundary, while adaptive refinement is done over the PHT-spline basis functions. This allows us to reduce both the number of the degrees of freedom, as well as the computational time of the full optimization process. The adaptive optimization in this work is based on the recovery-based error estimator, previously proposed in \cite{videla2019h} for the Helmholtz equation and PHT-splines. 

The remainder of the paper is organized as follows. Section \ref{prelim_Section} is assigned for the preliminaries, where the boundary value problem for the Helmholtz equation, NURBS, PHT-splines, GIFT formulation, and a generic shape optimization problem are introduced. Section \ref{adaptive_ref} in particular is devoted to the adaptive optimization. In Section \ref{Num-Res-shape}, numerical results for three benchmark examples are discussed. Conclusions are drawn in Section \ref{conclusions}.

\section{Preliminaries}\label{prelim_Section}
\subsection{Boundary Value Problem (BVP) for the Helmholtz equation}\label{section:BVP}

\begin{figure}[!ht]
\centering
\begin{tikzpicture}[scale = 1, classical/.style={thick, ->,>=stealth}]

\draw [black, thick, line width=1.5pt] plot [smooth cycle] coordinates {(1,0) (1.125, 0.5) (1.125, 1) (0.75,1) (0,1.15) (-0.5,0.75) (-1,0.75) (-1.125,0.375) (-0.9,0) (-0.75, -0.375) (-0.8, -0.625) (-0.375, -0.8) (0,-0.75) (0.5, -0.5) (1,-0.6)};

 \draw [dashed,line width=1.5pt ] (0,0) circle(4.4);;
 
\node[] at (-2.5, -2.9) {$\Omega$};

\node[] at (-1.125, 1.05) {$\Gamma_D$};

\node[] at (1.0, 1.4) {$\Gamma_N$};

\node[] at (0.75, -0.85) {$\Gamma_R$};

\node[] at (0.1, 0.18) {scatterer};

 \node[]   at (0.25,0.5) { $\mathbf{n}$};

 \draw [black] [->] [thick,line width=0.75pt ] plot coordinates {  (0,1.15) (0.2,0.65) };

\node[] at (-3.3,3.5) {$\Sigma$};

\draw [thick,->](0,0) -- (-4.35, -0.7);

\node[] at (-2.5,-0.11) {$R$};
 
\end{tikzpicture}
\caption{Helmholtz acoustic problem.}\label{fig:BVP}
\end{figure}
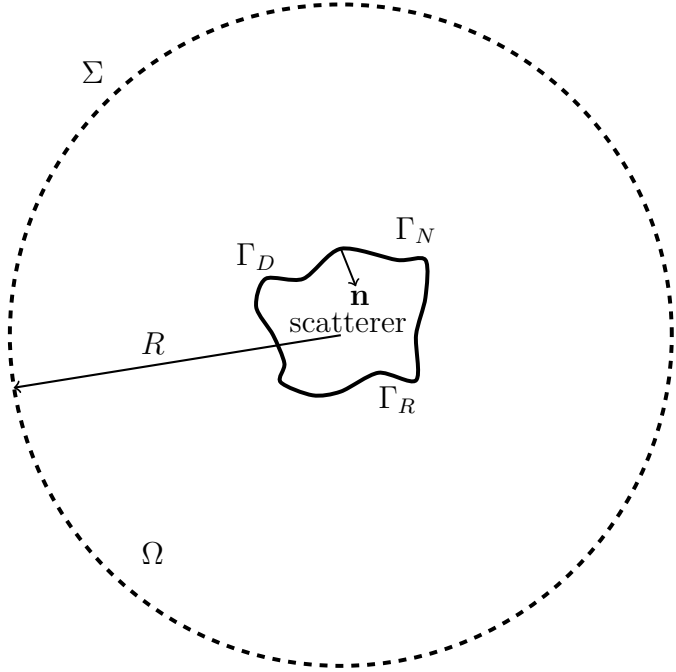

The boundary value problem for the Helmholtz equation, as explained in Figure \ref{fig:BVP}, in domain $\Omega\in\mathbb{R}^{2}$ with boundary $\Gamma$ consists in finding the spacial component of the acoustic pressure, $u$, such that:
\begin{subequations}
\begin{alignat}{2}
\Delta u + k^2 u &= 0 \quad && \text{in} \quad \Omega \\
u &= g \quad &&\text{on} \quad \Gamma_{D} \\
\dfrac{\partial u}{\partial \boldsymbol{n}} &= ikh \quad &&\text{on} \quad \Gamma_{N} \\
\dfrac{\partial u}{\partial \boldsymbol{n}} + \alpha u  &= f  \quad &&\text{on} \quad \Gamma_{R}
\end{alignat}
\label{eqn:Boundary_conditions_DNR}
\end{subequations}
where $\Delta$ is the scalar Laplace operator, $\boldsymbol{n}$ is a unit normal vector on $\Gamma$, outward to $\Omega$, $k$ is the wave number, and $i$ is the imaginary unit. Dirichlet, Neumann and Robin boundary conditions are prescribed on parts of the boundary, $\Gamma_{D} ,\Gamma_{N}$ and $\Gamma_{R}$ respectively ($\Gamma = \Gamma_{D}\cup\Gamma_{N}\cup\Gamma_{R}$ and $\Gamma_D\cap\Gamma_N = \Gamma_D\cap\Gamma_R = \Gamma_R\cap\Gamma_N = \emptyset$), in which $g$, $h$, $\alpha$ and $f$ are prescribed functions. For exterior problems, $u$ refers to a scattered wave and the boundary conditions on the scatterer boundary are written in terms of the incoming wave $u^{\text{inc}}$ and its normal derivative, \cite{videla2019h}, i.e.:
\begin{equation}
\begin{split}
    u &= -u^{\text{inc}} \quad\text{on} \quad \Gamma_{D} \\
    \dfrac{\partial u}{\partial\boldsymbol{n}} &= -\dfrac{\partial u^{\text{inc}}}{\partial\boldsymbol{n}} \quad \text{on} \quad \Gamma_{N}
\end{split}
\end{equation}

Additionally, the solution satisfies the Sommerfeld radiation condition prescribed at infinity to truncate all possible reflections of spurious acoustic waves from the far-field. This condition is formulated for 2D problems as follows \cite{SommerfeldPartial}:
\begin{equation}
\lim\limits_{r\rightarrow\infty}  \sqrt{r} \left(\dfrac{\partial u}{\partial r} - i k u\right) = 0. \label{eq:Sommerfeld_condition}
\end{equation} where $r$ is the distance from the origin. In domain-type methods, the Sommerfeld radiation condition is replaced by the Absorbing Boundary Condition (ABC) on the truncation boundary $\Sigma$ (usually given by a circle of radius $R$ in 2D) \cite{Bayliss1980, Atr-IGA-C}. In this work we use Bayliss–Gunzburger–Turkel condition of order 1, referred as BGT1, formulated in polar coordinates $(r, \theta)$ as: 
\begin{equation}
    \dfrac{\partial u}{\partial r} + \mathcal{B}u = 0,
\end{equation}
where
\begin{equation}
\mathcal{B}u = \left(\dfrac{1}{2 R} - i k\right)u \label{b_1} 
\end{equation}

Note, that due to higher continuity of splines, it is also possible to accommodate ABCs of higher order in the framework of IGA and GIFT. 

The weak form of the boundary value problem is written as follows: Find $u^h \in \mathscr{U}^h \subseteq \mathscr{U} = \left\lbrace u \in \mathbb{H}^1 ({\Omega}), u=g \hspace{0.15cm} \textrm{on} \ {\Gamma}_D \right\rbrace$ such that $\forall v^h \in \mathscr{V}^h_0 \subseteq \mathscr{V}_0 = \left\lbrace v \in \mathbb{H}^1 ({\Omega}), v=0 \hspace{0.15cm} \textrm{on} \ {\Gamma}_D \right\rbrace$,
\begin{equation}
a(u^h,v^h) = l(v^h)
\label{eq:weak_form}
\end{equation}
\noindent with
\begin{equation}
a(u^h,v^h) = \int _{\Omega} \triangledown  u^h \triangledown  v^h d\Omega -k^{2} \int _{\Omega}   u^h  v^h d\Omega + \alpha \int _{\Gamma_{R}} u^h v^h d\Gamma + \int _{\Sigma} \mathcal{B}u^h v^h d\Gamma
\label{eq:weak_form_a}
\end{equation}
\begin{equation}
l(v^h) =  i k \int _{\Gamma_{N}}h v^h d\Gamma + \int _{\Gamma_{R}} f v^h d\Gamma
\label{eq:weak_form_l}
\end{equation} 

\subsection{Non-Uniform Rational B-Splines (NURBS)}
A set of $n$ B-spline basis functions of degree $p$ is defined in the parametric space $\xi \in [0,1]$ on a knot vector $\Xi = \{\xi_1,\cdots, \xi_i, \cdots ,\xi_{n+p+1}\}$ (a non-decreasing sequence of knots $\xi_i$ with $\xi_1 = 0$ and $\xi_{n+p+1} = 1$). Then, the $i$-th B-spline basis function of degree $p$, $N_{i, p}(\xi)$ is defined recursively as \cite{piegl1996nurbs}:

\begin{equation}  \label{eq:7}
    N_{i,0}(\xi) = \Bigg\{ 
    \begin{array}{lcl}
     1, \;\;\;\;\;\;\;\;\;\; \xi_i\leq\xi<\xi_{i+1} \\
     0, \;\;\;\;\;\;\;\;\;\; \mbox{ otherwise }
    \end{array}	 
\end{equation} and for $p\geq1$  
    
\begin{equation}  \label{eq:8}	
 N_{i,p}(\xi)=\frac{\xi-\xi_i}{\xi_{i+p}-\xi_i} N_{i,p-1}(\xi)+\frac{\xi_{i+p+1}-\xi}{\xi_{i+p+1}-\xi_{i+1}} N_{i+1,p-1}(\xi)
\end{equation} \\
    
The NURBS basis function is written as:
\begin{equation}  \label{eq:9}
R_{i,p}(\xi) = \frac{N_{i,p}(\xi)w_i}{\sum\limits_{\hat{i}=1}^{n} N_{\hat{i},p}(\xi)w_{\hat{i}}} 
\end{equation}  where $w_i$ is the $i$-th weight.

A NURBS curve is parameterized by a set of NURBS basis functions and control points $B_i$ as follows:
\begin{equation}  \label{eq:1D-NURBS-parameterization}	
 \begin{bmatrix}
x(\xi) \\
y(\xi) 
\end{bmatrix} =\sum\limits_{i=1}^{n} R{_{i,p}} (\xi)B{_i
 } \end{equation} 
 
Two-dimensional NURBS basis functions are obtained as a tensor-product of one-dimensional NURBS. For each direction, a knot vector is defined: $\Xi=\{\xi_1,\cdots, \xi_i, \cdots ,\xi_{n+p+1}\}$ and $\Phi =\{\eta_1,\cdots, \eta_i, \cdots ,\eta_{m+q+1}\}$ (where $m$ is the number of basis functions in direction $\eta$ ($\eta\in[0, 1]$) and $q$ is their degree). Then, the corresponding NURBS basis functions are written as
 
\begin{equation}  \label{eq:NURBS_2D}	
 R_{ij}(\xi,\eta)=\frac{N_{i,p}(\xi) N_{j,q}(\eta) w_{ij}} {\sum\limits_{\hat{i}=1}^{n} \sum\limits_{\hat{j}=1}^{m} N_{\hat{i},p}(\xi) N_{\hat{j},q}(\eta) w_{\hat{i} \hat{j}}} 
  \end{equation}  where $w_{ij}$ is the weight associated with the control point $P_{ij}$. The NURBS surface is parameterized as
  
\begin{equation}  \label{eq:2D-NURBS-parameterization}	
\begin{bmatrix}
x(\xi, \eta) \\
y(\xi, \eta) 
\end{bmatrix} = \sum\limits_{i=1}^{n} \sum\limits_{j=1}^{m} R_{ij} (\xi,\eta)P_{ij} \end{equation}   

  

\subsection{PHT-splines}

Polynomial Splines over hierarchical T-meshes (PHT-Splines) were proposed by Deng et al. \cite{deng2006dimensions, deng2008polynomial} as a generalization of B-splines over hierarchical T-meshes in order to provide both local refinement and adaptability using a polynomial basis, which is able to parameterize the geometry. The following section briefly covers the formulation of PHT-splines.

Consider $\mathscr{T}$ as a hierarchical T-mesh, $b_{i}(\xi,\eta)$, $i=1,2,...,n$ as a set of B\'ezier splines (B-splines), $\xi$ and $\eta$ as two parametric coordinates defined on the space $\left[0,1 \right] \times \left[ 0,1\right]$, and $P_{i}$ as the control points. Then, the polynomial spline surface over $\mathscr{T}$ at level 0 is defined by:

\begin{equation}
    S(\xi,\eta) = \sum_{i=1}^n b_{i}(\xi,\eta)P_{i} 
\end{equation}

A spline basis function is represented locally as a linear combination of Bernstein polynomials. In particular, each B-Spline $b_{i}(\xi,\eta)$ is defined as

\begin{equation}
    b_{i} (\xi,\eta) = \sum_{j=1}^{p+1}\sum_{k=1}^{p+1}C^{i}_{jk}\hat{B}_{j,k}(\xi,\eta)
\end{equation}

\noindent where $\hat{B}_{j,k}(\xi,\eta) = B_{j}(\xi)B_{k}(\eta)$ is a tensor product of Bernstein polynomials defined on the reference interval $\left[-1,1\right]$ : 

\begin{equation}
    B_{j}(\xi) = \frac{1}{2^{p}}\binom{p}{j-1}\left(1-\xi\right)^{p-j+1}\left(1+\xi\right)^{j-1}  
\end{equation}

The procedure to compute B\'ezier coefficients $C^{i}_{jk}$ based on a particular NURBS curve is detailed in \cite{borden2011isogeometric}. Then, based on a B-spline surface written over an initial regular T-mesh (level-$k$), we briefly explain how to compute the new coefficients given a refinement on a level-$k+1$: 








\begin{enumerate}
    \item Construct a B\'ezier representation of the basis functions. This is done using Bernstein polynomials and B\'ezier coefficients (other authors refer them as B\'ezier ordinates).
    \item Once a level refinement is applied, the shape functions whose support was lying on refined element must be modified. Given the B\'ezier representation of the basis function, the process consists in truncating the basis function by zeroing out B\'ezier coefficients. This process is also explained in the introductory paper to PHT-splines \cite{deng2008polynomial} and in the work of Garau and V\'asquez \cite{garau2018algorithms}. 
    \item After the truncation of coefficients, the insertion of new basis functions is performed. For each new basis function, the process includes the computation of the corresponding B\'ezier ordinates and the assignment of a new global basis index. The new basis functions are standard (non-truncated) B-splines computed based on the local knot vector information. In the case of a removed T-junction, the local knot vectors are to be computed based on the neighbour connectivity of the elements stored in the quad-tree structure. Finally, the new basis indices corresponding to the new basis functions are determined. 
\end{enumerate}

It is worth noting that this work is based on the open-source package IGAPACK (\url{https://github.com/canitesc/IGAPack}) which contains all the aforementioned PHT-splines routines. (See also \cite{anitescu2018recovery})

\subsection{Geometry Independent Field approximaTion (GIFT)}

Geometry Independent Field approximaTion (GIFT) \cite{atroshchenko2018weakening} can be seen as an extension of the concept of isoparametric elements in standard IGA to sub- and super-parametric elements. In other words, computational domain can be kept in its original, coarse CAD parameterization, given by NURBS, e.g. eq. (\ref{eq:2D-NURBS-parameterization}), while other type of splines can be chosen to approximate the unknown solution. In this work we use PHT-splines, as it was implemented before in \cite{anitescu2018recovery,videla2019h,videlaapplication,JANSARI2022106728}. The representations for the solution and the geometric quantities are substituted into the weak form (\ref{eq:weak_form})-(\ref{eq:weak_form_a})-(\ref{eq:weak_form_l}) and after the integration is carried out, the weak form is transformed into a system of linear algebraic equations for unknown control points.  




\subsection{Shape optimization problem.}

The aim of shape optimization problems in acoustic is to improve the acoustic performance of the system by modifying the boundary of the structure. Improving acoustic performance is achieved by minimizing certain quantity of interest $J(u)$ (e.g. level of sound in the protected zone) subjected to some design constrains. Notation $J(u)$ implies that $J$ is function of $u$, i.e. the solution of the Helmholtz boundary value problem (\ref{eqn:Boundary_conditions_DNR}a)-(\ref{eqn:Boundary_conditions_DNR}d). One of the main features of spline-based boundary parameterization used in this work is that the boundary is described by a set of control points $\boldsymbol{x}$, which are used as design variables. Hence, the problem is written as follows:

\begin{equation}
\begin{split}
    \text{Objective function:   }& \min\limits_{\boldsymbol{x}} J(u), \text{   subject to eq.(\ref{eqn:Boundary_conditions_DNR}a)-(\ref{eqn:Boundary_conditions_DNR}d)}\\
    \text{Such that:   }& a_i \leq x_i \leq b_i, \\
    \text{             }& m(\boldsymbol{x}) = 0, \\
    \text{             }& n(\boldsymbol{x}) \leq 0, \\
\end{split}
\end{equation}
where each design variable $x_i$ is set to take values within interval $[a_i, b_i]$,  and functions $m(\boldsymbol{x})$ and $n(\boldsymbol{x})$ represent the non-linear constraints, such as design limit on the area or volume of the structure. 

\subsection{Recovery-based adaptive refinement and optimization}\label{adaptive_ref}

This section illustrates the adaptive GIFT shape optimization framework proposed in this study. 

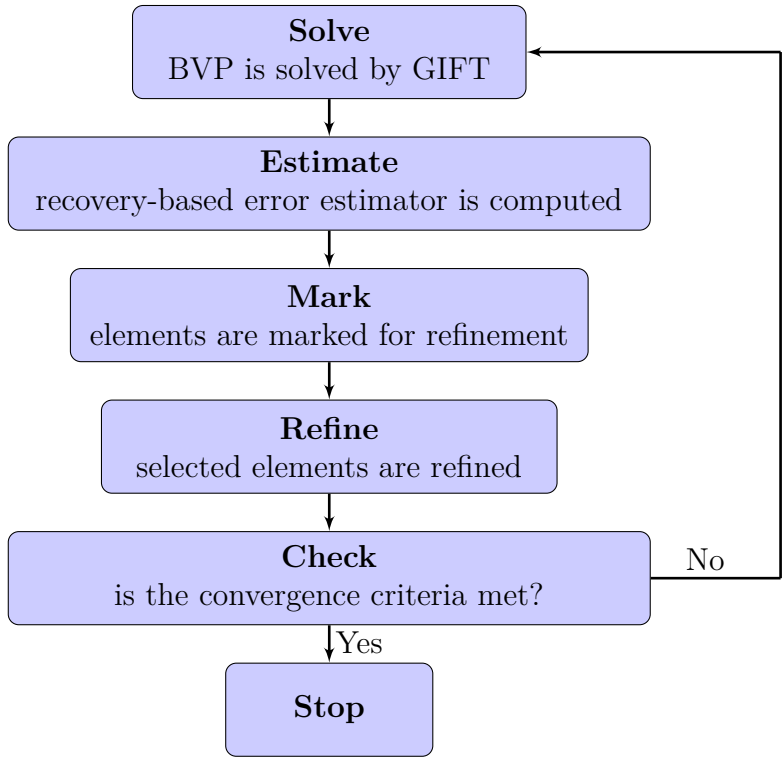
\begin{figure}[ht!]
      \centering
      \begin{tikzpicture}[node distance = 2cm, auto]

    \node [block1]  at (0, 0) (block1) {\textbf{Solve} \\ BVP is solved by GIFT};
    \node [block2]  at (0, -1.75) (block2) {\textbf{Estimate} \\ recovery-based error estimator is computed};  
    \node [block3]  at (0, -3.5) (block3) {\textbf{Mark} \\ elements are marked for refinement}; 
    \node [block4]  at (0, -5.25) (block4) {\textbf{Refine} \\ selected elements are refined}; 
    \node [block5]  at (0, -7) (block5) {\textbf{Check} \\ is the convergence criteria met?};
    \node [block6]  at (0, -8.75) (block6) {\textbf{Stop}};
    
    \node []  at (0.4, -7.85) {Yes};
    \node []  at (5, -6.75) {No};
    
    \path [line] (block1) -- (block2);
    \path [line] (block2) -- (block3);
    \path [line] (block3) -- (block4);
    \path [line] (block4) -- (block5);
    \path [line] (block5) -- (block6);
    
    \path [line2] (block5) -- (6, -7);
    \path [line2] (6, -7) -- (6, 0);
    \path [line] (6, 0) -- (block1);
    
    
     
     
\end{tikzpicture}
      \caption{Classical scheme for adaptive refinement.}
      \label{fig:scheme_refinement}
  \end{figure}

In a classical adaptive refinement scheme (Figure \ref{fig:scheme_refinement}), there are five main steps: Solve, Estimate, Mark, Refine and Check. As the name suggests, in the first step, the BVP is solved by the selected numerical method. Then, the error metrics are computed (e.g. recovery-based error estimator used in this work). The ``Mark" step consists in using the marking algorithm to select elements which need to be refined. Subsequently, in the ``Refine" step, those selected elements are refined. After that, depending on the convergence criteria, one can either repeat the process or stop it.
 
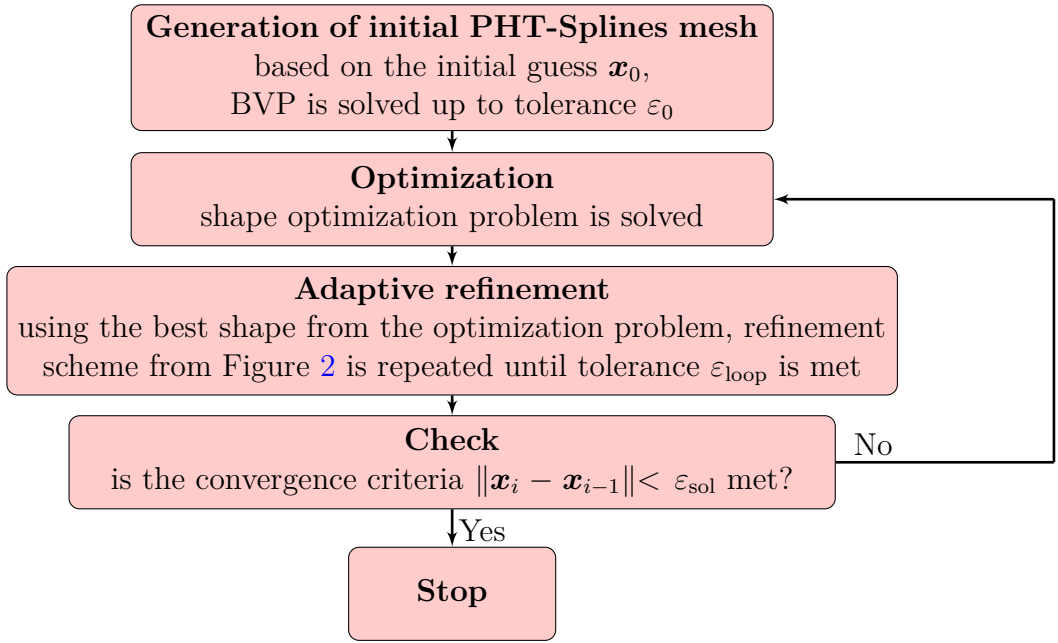
\begin{figure}[ht!]
      \centering
      \begin{tikzpicture}[node distance = 2cm, auto]

    \node [redblock1]  at (0, 0) (block1) {\textbf{Generation of initial PHT-Splines mesh} \\ based on the initial guess $\boldsymbol{x}_0$, BVP is solved up to tolerance $\varepsilon_0$};
    \node [redblock2]  at (0, -1.75) (block2) {\textbf{Optimization} \\ shape optimization problem is solved};  
    \node [redblock3]  at (0, -3.5) (block3) {\textbf{Adaptive refinement} \\ using the best shape from the optimization problem, refinement scheme from Figure \ref{fig:scheme_refinement} is repeated until tolerance $\varepsilon_{\text{loop}}$ is met}; 
    \node [redblock4]  at (0, -5.25) (block4) {\textbf{Check} \\ is the convergence criteria $\|\boldsymbol{x}_i - \boldsymbol{x}_{i - 1}\| < \varepsilon_{\text{sol}}$ met?}; 
    \node [redblock6]  at (0, -7.0) (block5) {\textbf{Stop}};
    
    \node []  at (0.4, -6.15) {Yes};
    \node []  at (5.6, -5) {No};
    
    \path [line] (block1) -- (block2);
    \path [line] (block2) -- (block3);
    \path [line] (block3) -- (block4);
    \path [line] (block4) -- (block5);
    
    \path [line2] (block4) -- (8, -5.25);
    \path [line2] (8, -5.25) -- (8, -1.75);
    \path [line] (8, -1.75) -- (block2);
    
    
     
     
\end{tikzpicture}
      \caption{Scheme of algorithm 1.}
      \label{fig:scheme_optimization}
  \end{figure}

In this work, as stated in the previous section, the BVP is solved using a GIFT formulation with PHT-splines and NURBS. The shape optimization is performed on the physical space, which is parameterized by NURBS basis functions, and therefore, does not affect the PHT-splines basis functions. This means that the optimization process employs a fixed PHT-splines mesh. Consequently, one can sequentially apply optimization and adaptive refinement in a step-by-step way, which means that the optimization is performed over the NURBS control points, without affecting the PHT-splines, and then, after the optimization process has finished, the adaptive refinement is done over the PHT-splines mesh, keeping the NURBS structure unchanged. This scheme is further explained in Algorithm \ref{alg:adaptive_optimisation}. The algorithm is controlled by four parameters: $\varepsilon_0$ - initial error tolerance, $\varepsilon_{\text{loop}}$ - error tolerance in each iteration, $\varepsilon_{\text{sol}}$ - error tolerance for the optimization solution, $N_{\text{Max steps}}$ - maximum number of steps. The process is also explained in Figure \ref{fig:scheme_optimization}.

\begin{algorithm}[ht!]
\caption{Adaptive shape optimization pseudo-algorithm }
\label{alg:adaptive_optimisation}
\begin{algorithmic}
\State \textbf{Input parameters:} $\boldsymbol{x}_{0}$, $N_{\text{Max steps}}$, $\varepsilon_{0}$, $\varepsilon_{\text{loop}}$, $\varepsilon_{\text{sol}}$
\State \textbf{Step 1:} create initial NURBS structure based on $\boldsymbol{x}_{0}$ and PHT-splines meshe
\State \textbf{Step 2:} perform adaptive refinement over the PHT-splines basis, driven by the recovery-based error indicator, controlled by $\varepsilon_{0}$ 
\State $i = 0$
\While{true}
    \State $i = i + 1$
    \State perform shape optimization over the NURBS structure, obtaining $\boldsymbol{x}_{i}$ - best solution at $i-$th step
    \State generate new NURBS structure using $\boldsymbol{x}_{i}$ 
    \State perform recovery-based adaptive refinement based on the optimal solution $\boldsymbol{x}_{i}$ until error estimator reaches the error tolerance $\varepsilon_{\text{loop}}$
    \If{$i > N_{\text{Max steps}}$ }
        \State end while-loop
    \EndIf
    \If{ $\left \| \boldsymbol{x}_{i} - \boldsymbol{x}_{i-1} \right \| < \varepsilon_{\text{sol}}$}
        \State end while-loop
    \EndIf
    \State update solution and update PHT-splines mesh
\EndWhile
\end{algorithmic}
\end{algorithm}

The recovery-based error estimator proposed in \cite{videla2019h} for the Helmholtz equation is employed. 
The ``Dörfler strategy’’ is used to mark the elements that need to be refined \cite{Dorfler1996}. The strategy consists in sorting the elements according to their error contribution to the total error and then refining the elements, whose error contribution is bigger than a certain percentage of the error estimator. In the subsequent numerical experiments, 50\% is utilized for the marking algorithm.
\cleardoublepage
\section{Numerical results}\label{Num-Res-shape}
We present three numerical examples to demonstrate the accuracy of the adaptive GIFT shape optimization. We devise one benchmark problem  with a known solution and known local minimum. Next, we analyze the shape optimization problem of a horn. And lastly, we analyze the shape optimization problem of a sound barrier. Both problems have been extensively studied in the literature, mostly in the context of isogeometric boundary element methods (IGABEM) \cite{MOSTAFASHAABAN2020156,CHEN2018507}. As pointed out previously, NURBS are employed to model the geometry, and PHT-splines of degree $p=3$ are employed to model the sound pressure.
\subsection{Plane wave scattering problem by a cylinder in a truncated domain}
\begin{figure}[!ht]
\centering
\begin{tikzpicture}[scale = 1, classical/.style={thick, ->,>=stealth}]
\draw[classical] (-3,0) -- (-2,0)  node[midway,above] {$u^{\text{inc}}$};
\draw[thick] (-2.75,-0.1)--(-2.75,0.1) ;
\draw[thick] (-2.7,-0.1)--(-2.7,0.1) ;
\draw (1.5,0) circle [radius=2.];
\node[] at (0.5,-0.75) {$\Gamma$};
\node[] at (-0.5,-2) {$\Sigma$};
\draw[classical] (-3,0) -- (-2,0)  node[midway,above] {$u^{\text{inc}}$};
\draw[thick] (-2.75,-0.1)--(-2.75,0.1) ;
\draw[thick] (-2.7,-0.1)--(-2.7,0.1) ;
\draw[black, thick] (1.5,0) circle [radius=1.0];
\node[] at (1.5 + 1.5,0.25) {$\Omega$};
\node[] at (0.5,-0.75) {$\Gamma$};
\draw[classical] (1.5,0)--(1.5+1.4,-1.4); 
\node[] at (1.5 + 1.2, -0.8) {$R$};
\draw[classical] (1.5,0)--(1.5 + 0.95,-0.3); 
\node[] at (2.05, 0.0) {$a$};
\end{tikzpicture}
\caption{Truncated domain for the cylinder problem.}
\label{fig:cylinder_geo}
\end{figure}
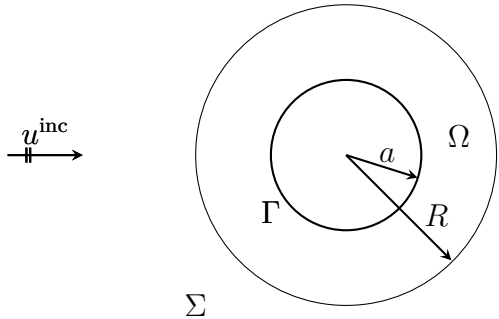

In the first numerical example, the problem of a plane wave scattered by a cylinder of radius $a$ is considered. For numerical simulations, infinite domain is truncated by an artificial boundary: circle of radius $R$. This truncation boundary is denoted by $\Sigma$ and the domain enclosed by $\Gamma$ and $\Sigma$ is denoted by $\Omega$, as shown in Figure \ref{fig:cylinder_geo}. The boundary value problem with BGT1 boundary condition consists in finding $u$ such that:

\begin{equation}
\begin{split}
    \Delta u + k^2 u &= 0 \text{   in   }\Omega\\
    \dfrac{\partial u}{\partial \boldsymbol{n}} &= - \dfrac{\partial u^{\text{inc}}}{\partial \boldsymbol{n}}  \text{   on   }\Gamma\\
     \dfrac{\partial u}{\partial r} + \left(\dfrac{1}{2 R} - i k\right)u &= 0\text{   on   }\Sigma
     \end{split}\label{BVP-cylinder-BGT1}
\end{equation}

The incoming wave $u^{\text{inc}}$ travels in the horizontal direction, i.e:
\begin{equation}
u^{\text{inc}} = e^{ikx}\label{u_inc_full}
\end{equation}
In order to derive the analytical solution, the incoming wave is expanded in polar coordinates into a series:
\begin{equation}
e^{ikr\cos\theta} = \sum\limits_{n = 0}^{\infty}\kappa_n i^n J_{n}(kr)\cos n\theta,\label{exp_series}
\end{equation}
where 
\begin{equation}
\kappa_0 = 1, \quad \kappa_n = 2\text{    for } n \geq 1
\end{equation}

\noindent and $J_n(k r)$ is Bessel function of the first kind of order $n$. 
The detailed derivation of the exact solution $u^{\text{exact}}$ of boundary value problem (\ref{BVP-cylinder-BGT1}) is given in \cite{Atr-IGA-C}. Here only the main results are recalled. 

The solution is sought in the form:
\begin{equation}
u^{\text{exact}} = \sum\limits_{n = 0}^{\infty}(a_n H_n^{(1)}(k r) + b_n H_n^{(2)}(k r))\cos n\theta,\label{u_general_form}
\end{equation}
where $H_n^{(1)}(k r)$ and $H_n^{(2)}(k r)$ are the Hankel functions of the first and second types, respectively. The coefficients $a_n$ and $b_n$ are obtained by substituting eq.(\ref{u_general_form}) to boundary conditions of (\ref{BVP-cylinder-BGT1}) on $r = a$ and $r = R$, and matching the coefficients of Hankel functions. This procedure leads to the following system of equations for unknown $a_n, b_n$:
\begin{equation}
\left(
\begin{matrix}
    A_{11}       & A_{12} \\
    A_{21}       & A_{22} \\
\end{matrix}\right)
\left(
\begin{matrix}
    a_{n} \\
    b_{n} \\
\end{matrix}\right)
=
\left(
\begin{matrix}
    b \\
    0 \\
\end{matrix}\right)\label{sys-BGT1}
\end{equation}
where 
\begin{equation}
A_{11} = \dfrac{d}{d r}H^{(1)}_{n}(k r)|_{r = a}, \quad A_{12} = \dfrac{d}{d r}H^{(2)}_{n}(k r)|_{r = a}, \,\,\, b = -\kappa_n i^n \dfrac{d}{d r} J_{n}(k r)|_{r = a}\label{BGT_BC_Neumann}
\end{equation}
and
\begin{equation}
\begin{split}
A_{21} &= k H^{(1)}_{n - 1}(k R) - \left(i k + \dfrac{(2 n - 1)}{2 R}\right) H^{(1)}_{n}(k R),\\
A_{22} &= k H^{(2)}_{n - 1}(k R) - \left(i k + \dfrac{(2 n - 1)}{2 R}\right) H^{(2)}_{n}(k R)
\end{split}
\end{equation}

The objective function is considered as follows:
\begin{equation}
    J(u) = -\int\limits_{\Omega}\left | u \right |^2 dx \label{obj_circ}
\end{equation}

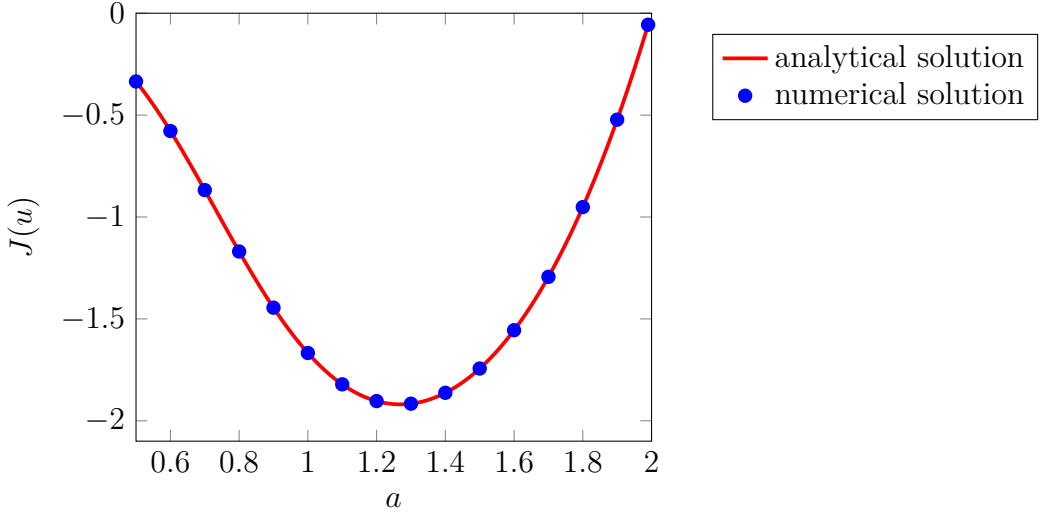
\begin{figure}[!ht]
	\centering
	\begin{tikzpicture}[scale=1]
\begin{axis}[cycle list name=exotic,
        xlabel = $a$,
        ylabel = $J(u)$,
	xmin = 0.5,
	xmax = 2,
	ymin = -2.1,
	ymax = 0.0,
	legend style={at={(1.75,0.75)},anchor=south east}
    ]
   \addplot [color = red,
             solid,
             smooth,
             line width = 0.5mm]
    table [x expr = \thisrowno{0}, y expr = \thisrowno{1}] {Sections/Shape-optimization-examples/Benchmark-cylinder/Benchmark-cylinder-Figures/benchmark-obj-analytical-data-BGT1.txt};
   
   \addlegendentry{analytical solution}
   
   \addplot [color = blue,
             mark=*,
             only marks,
             line width = 0.5mm]
    table [x expr = \thisrowno{0}, y expr = \thisrowno{1}] {Sections/Shape-optimization-examples/Benchmark-cylinder/Benchmark-cylinder-Figures/benchmark-obj-numerical-data-BGT1.txt};
    
    \addlegendentry{numerical solution}
    
\end{axis}
\end{tikzpicture}
\caption{Cylinder example: objective function $J(u)$ as a function of radius $a$.} \label{fig:obj_benchmark_ABC}
\end{figure}

In the numerical results, $R = 2$ and $k = 0.25\pi$ are used. Shape optimization results are presented below for three different cases: 

{\bf [case 1]:} In the first case, we restrict the design space to circular geometries. Hence, the shape of the inner circle can be described with one design variable: its radius. The objective function as a function of radius $a$ is plotted in Figure \ref{fig:obj_benchmark_ABC}. The minimum of the objective function $\min J(u) = -1.9193$ is reached at $a = 1.2689$. In Figure \ref{fig:obj_benchmark_ABC} both the analytical and numerical solutions are presented to validate the implementation. 

\begin{figure}[!ht]
	\centering
	\begin{tikzpicture}[scale = 1, classical/.style={thick, ->,>=stealth}]
\begin{axis}[cycle list name=exotic,
	xmin = 0,
	xmax = 1.5,
	ymin = 0,
	ymax = 1.5,
    axis lines = left, xtick=\empty, ytick=\empty,
    clip = false,
    ]
\node[] at (1.2, 0.1) {$P_1(x_1, 0)$};
\node[] at (1.5, 0.1) {$x$};
\node[] at (-0.1, 1.5) {$y$};
\draw[thick, dashed] (0, 0) -- (0.2*4.75528, 0.2*1.54508);
\node[] at (1.45, 0.4) {$P_2(x_1, x_1\tan(\pi/10))$};
\draw[thick, dashed] (0, 0) -- (0.24*4.04508, 0.24*2.93893);
\node[] at (1.55, 0.8) {$P_3(x_2 \cos(\pi/5), x_2\sin(\pi/5))$};
\draw[thick, dashed] (0, 0) -- (0.3*2.93893, 0.3*4.04508);
\node[] at (1.55, 1.2) {$P_4(x_3 \cos(3\pi/10), x_3\sin(3\pi/10))$};
\draw[thick, dashed] (0, 0) -- (0.2*1.54508, 0.2*4.75528);
\node[fill=white] at (0.25, 1.2) {$P_5(x_1\tan(\pi/10), x_1)$};
\node[] at (-0.2, 1) {$P_6(0, x_1)$};

\addplot [color = blue,
             smooth,
             solid,
             line width = 0.25mm]
    table [x expr = \thisrowno{0}, y expr = \thisrowno{1}] {Sections/Shape-optimization-examples/Benchmark-cylinder/Benchmark-cylinder-Figures/cylinder_optim_inner_boundary.txt};

\addplot [color = red,
             solid,
             mark = *,
             line width = 0.25mm]
    table [x expr = \thisrowno{0}, y expr = \thisrowno{1}] {Sections/Shape-optimization-examples/Benchmark-cylinder/Benchmark-cylinder-Figures/cylinder_cpts_test.txt};
    
\end{axis}   
\end{tikzpicture}
\caption{Cylinder example: control points on the inner boundary given by three design variables $(x_1, x_2, x_3)$. In case 2, all weights are equal to 1. In case 3, weights of points $P_2$, $P_3$, $P_4$, $P_5$ are additional design variables $x_4, x_5, x_6$, $x_7$, respectively.} \label{fig:cylinder_shapes_case_2_3}
\end{figure}
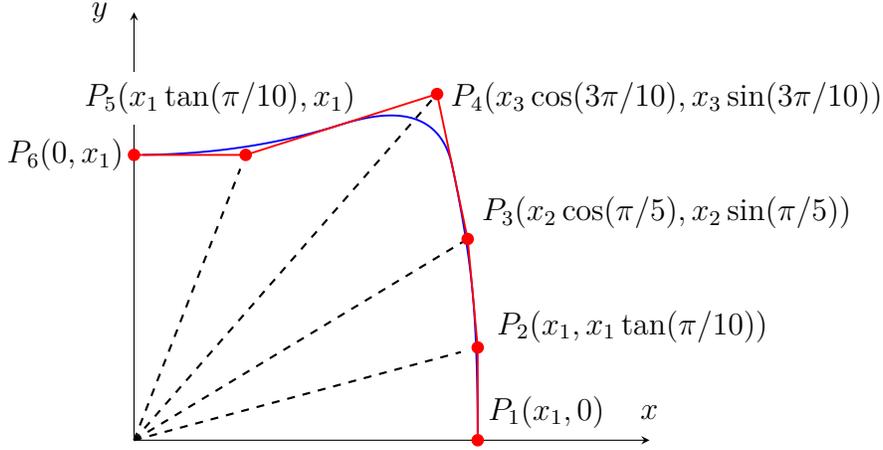

{\bf [case 2]:} In the second case, to describe the shape of the inner boundary, we use six control points, distributed uniformly in the circumferential direction in the first quadrant, parameterized with three design variables ($x_{1}, x_{2}, x_{3}$), as shown in Figure \ref{fig:cylinder_shapes_case_2_3}. Such set up implies that control points are allowed to move only in the radial direction. The shape is created in the first quadrant and then rotated about the center to create shapes in the second, third and forth quadrants. The weights of all control points on the design boundary are set to 1. The design variables vary in the interval: $0.5 \leq x_1, x_2, x_3 \leq 1.8$.  Note that such parameterization cannot represent a circle exactly. 

{\bf [case 3]:} In this study case, we use the same set up as in case 2, but adding weights of the internal points ($P_2$, $P_3$, $P_4$, $P_5$) as design variables ($x_4, x_5, x_6, x_7$, respectively). We assume that $0.5 \leq x_1, x_2, x_3 \leq 1.8$ and $ 0.75 \leq x_4, x_5, x_6, x_7 \leq 2$.  

\begin{figure}[ht!]
    \centering
    \begin{subfigure}[b]{0.3\textwidth}
        \includegraphics[width=\textwidth]{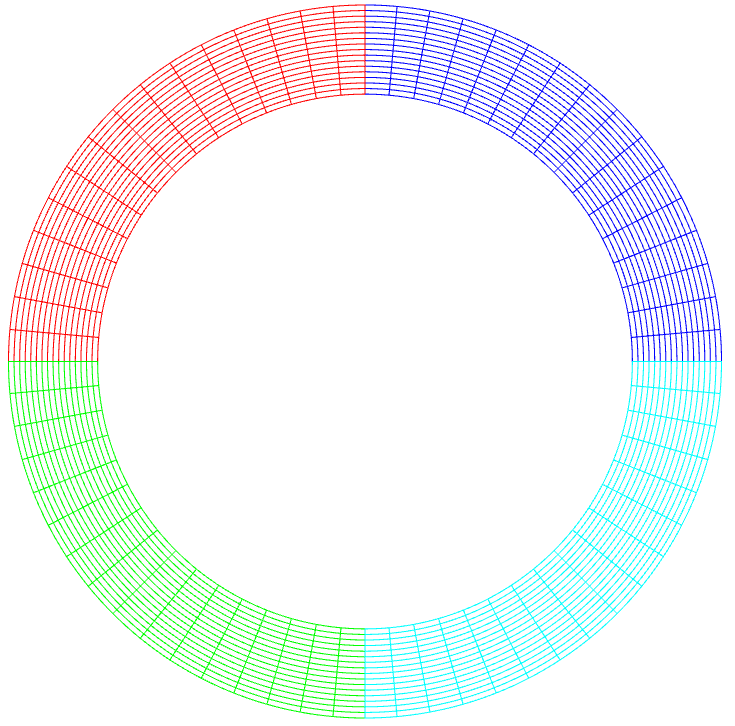}
        \caption{Initial geometry, a circle of radius $a = 1.5$}
    \end{subfigure}
    \quad
    \begin{subfigure}[b]{0.3\textwidth}
        \includegraphics[width=\textwidth]{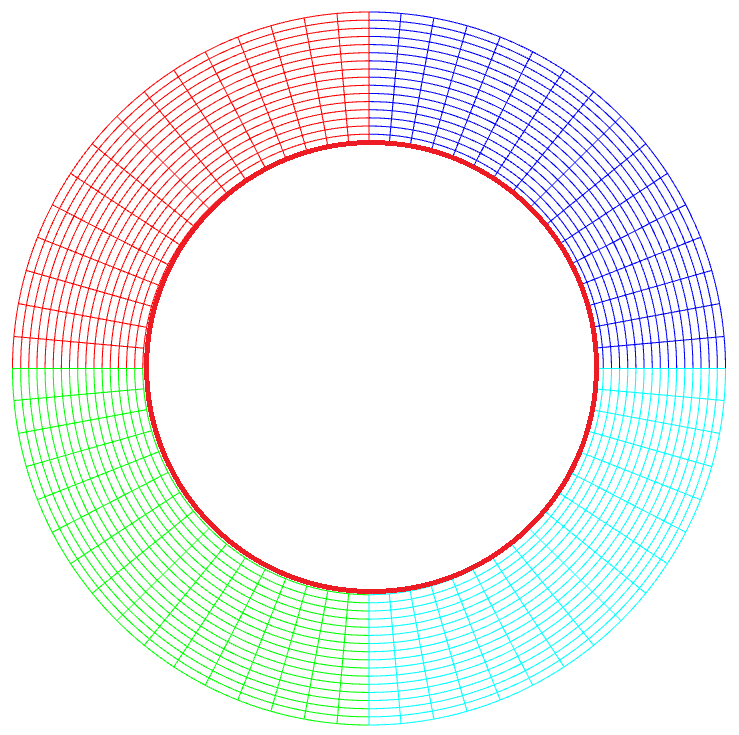}
        \caption{Final geometry, a circle of radius $a = 1.2689$ - uniform mesh}
    \end{subfigure}
    \quad
    \begin{subfigure}[b]{0.3\textwidth}
        \includegraphics[width=\textwidth]{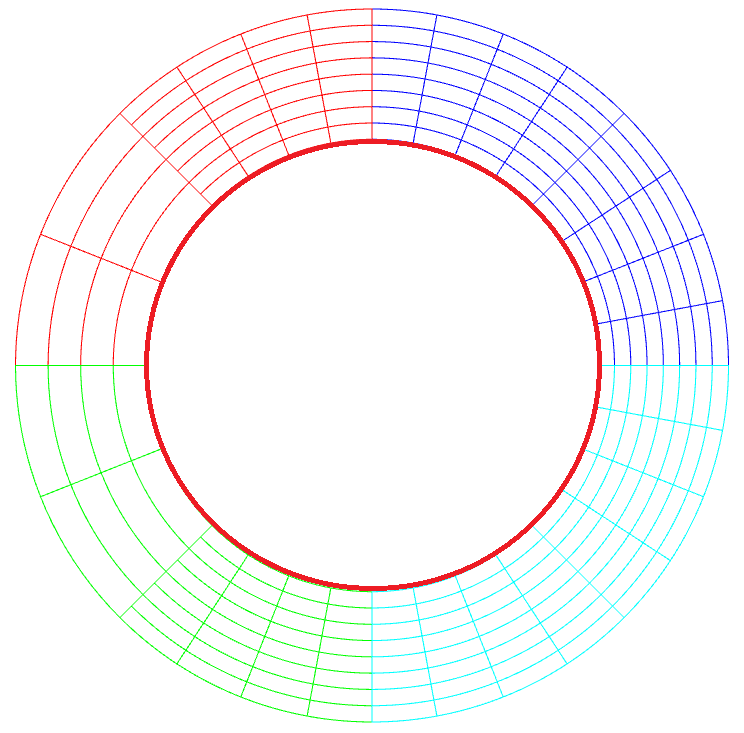}
       \caption{Final geometry, a circle of radius $a = 1.2689$ - adaptive mesh}
    \end{subfigure}
    ~ 
    \caption{Cylinder example, [case 1]: initial and final shapes.}
    \label{fig:case-1-shapes-uniform-vs-adaptive}
\end{figure}

\begin{figure}[ht!]
    \centering
    \begin{subfigure}[b]{0.3\textwidth}
        \includegraphics[width=\textwidth]{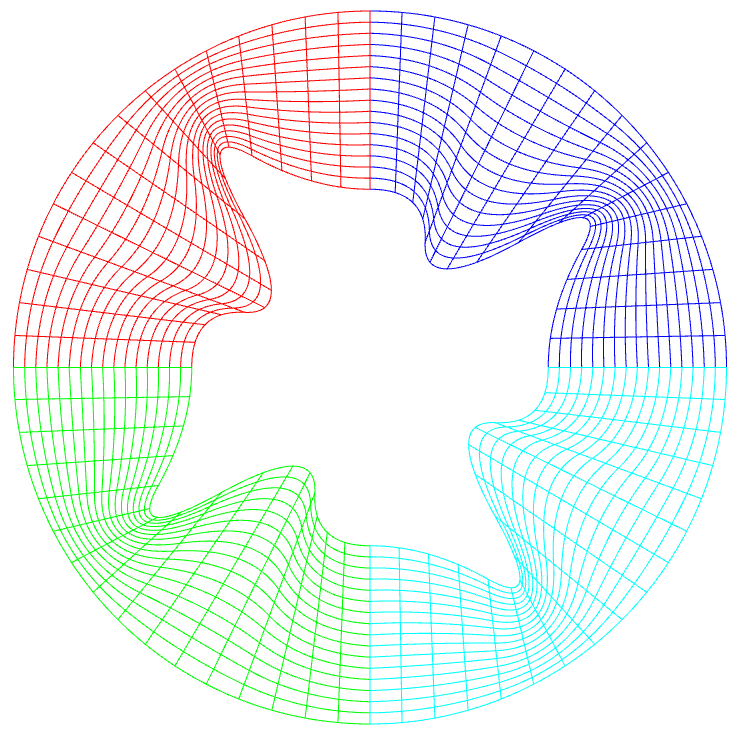}
        \caption{Initial geometry, given by $x_1 = 1$, $x_2 = 1.7$, $x_3 = 0.5$ \vspace{0.35cm}}
    \end{subfigure}
    \quad
    \begin{subfigure}[b]{0.3\textwidth}
        \includegraphics[width=\textwidth]{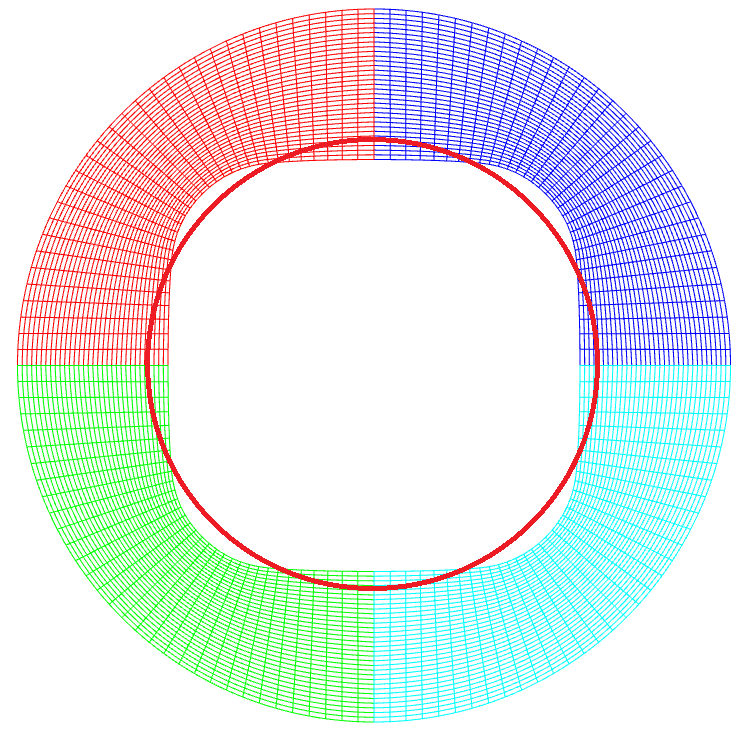}
        \caption{Final geometry, given by  $x_1 = 1.1554$, $x_2 = 1.3788$, $x_3 = 1.3788$. - uniform mesh}
    \end{subfigure}
    \quad
    \begin{subfigure}[b]{0.3\textwidth}
        \includegraphics[width=\textwidth]{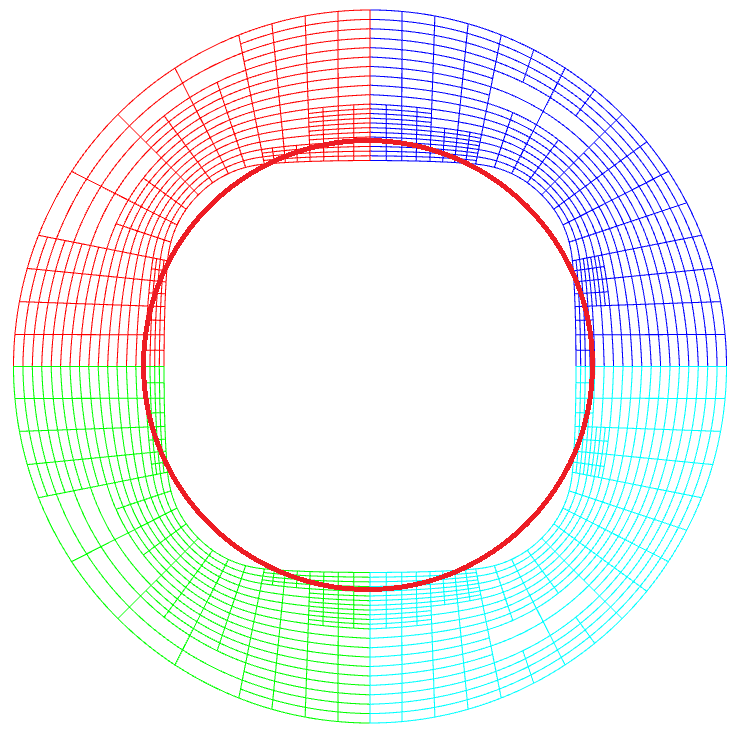}
       \caption{Final geometry, given by  $x_1 = 1.1554$, $x_2 = 1.3788$, $x_3 = 1.3788$. - adaptive mesh}
    \end{subfigure}
    ~ 
    \caption{Cylinder example, [case 2]: initial and final shapes.}
    \label{fig:case-2-shapes-uniform-vs-adaptive}
\end{figure}

\begin{figure}[ht!]
    \centering
    \begin{subfigure}[b]{0.3\textwidth}
        \includegraphics[width=\textwidth]{Sections/Shape-optimization-examples/Benchmark-cylinder/Benchmark-cylinder-Figures/initial_uniform_mesh_3_ctp_no_line_x.png}
        \caption{Initial geometry, given by $x_1 = 1$, $x_2 = 1.7$, $x_3 = 0.5$, $x_4 = 1$, $x_5 = 1$, $x_6 = 1$, $x_7 = 1$ \vspace{0.35cm}}
    \end{subfigure}
    \quad
    \begin{subfigure}[b]{0.3\textwidth}
        \includegraphics[width=\textwidth]{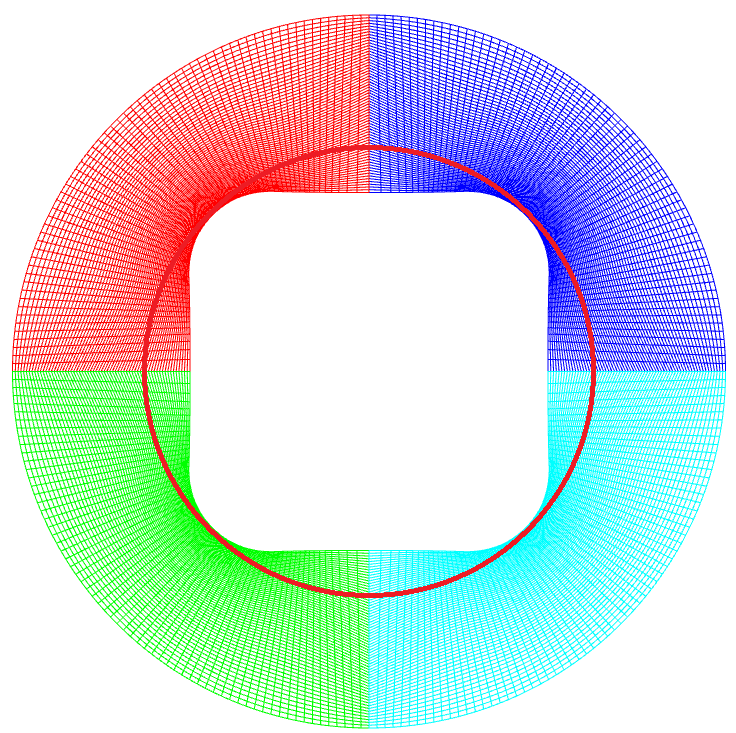}
        \caption{ Final geometry, given by  $x_1 = 1.0042$, $x_2 = 1.2570$, $x_3 = 1.2570$, $x_4 = 1.8534$, $x_5 = 1.8248$, $x_6 = 1.8248$, $x_7 = 1.8534$ - uniform mesh}
    \end{subfigure}
    \quad
    \begin{subfigure}[b]{0.3\textwidth}
        \includegraphics[width=\textwidth]{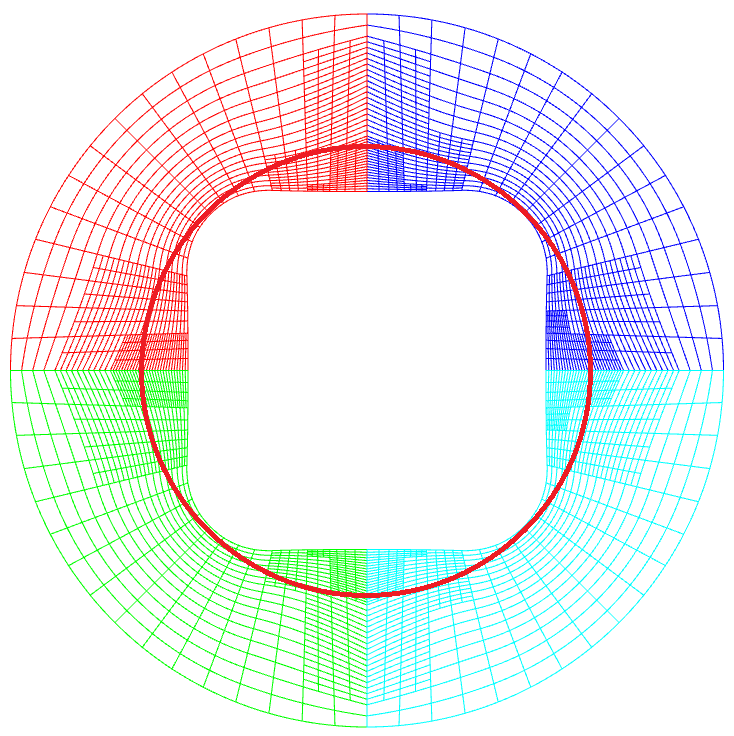}
       \caption{ Final geometry, given by $x_1 = 1.0042$, $x_2 = 1.2570$, $x_3 = 1.2570$, $x_4 = 1.8534$, $x_5 = 1.8248$, $x_6 = 1.8248$, $x_7 = 1.8534$ - adaptive mesh}
    \end{subfigure}
    ~ 
    \caption{Cylinder example, [case 3]: initial and final shapes.}
    \label{fig:case-3-shapes-uniform-vs-adaptive}
\end{figure}

Numerical results for all three cases are shown next. For case 1, in Figure \ref{fig:case-1-shapes-uniform-vs-adaptive} an initial geometry (a circle with radius $a = 1.5$) and a final geometry (a circle with radius $a = 1.2689$) can be seen. The optimization converges to the exact solution. For cases 2 and 3, an initial geometry is set up as a highly distorted shape, shown in Figure \ref{fig:case-2-shapes-uniform-vs-adaptive}a. In both design spaces, optimization procedure, starting from this initial state, converges to two local minima, different from the one in the design space of case 1, shown in Figure \ref{fig:case-2-shapes-uniform-vs-adaptive} and Figure \ref{fig:case-3-shapes-uniform-vs-adaptive}.   

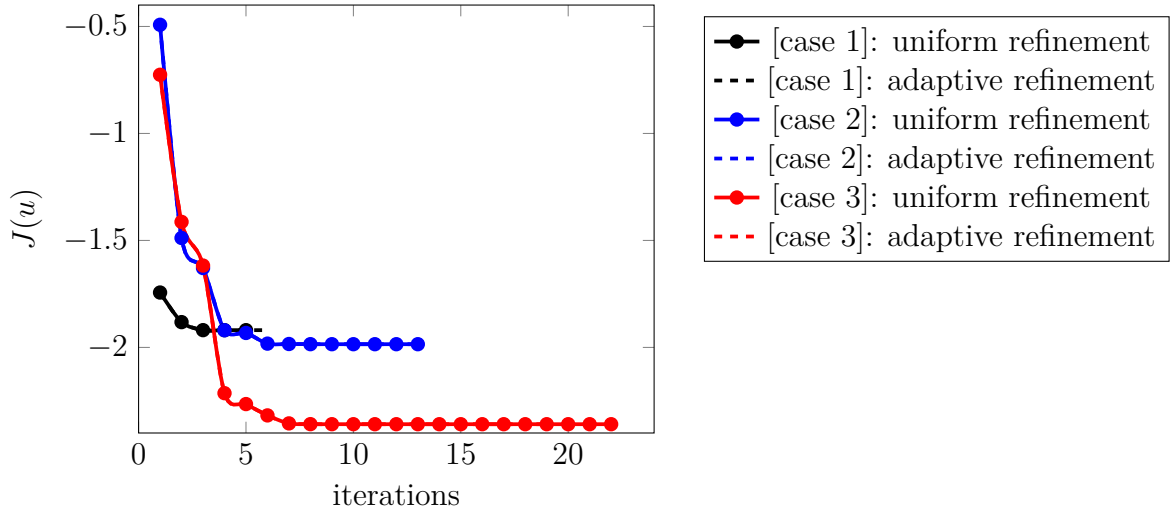
\begin{figure}[!ht]
	\centering
	\begin{tikzpicture}[scale=1]
\begin{axis}[cycle list name=exotic,
        xlabel = iterations,
        ylabel = $J(u)$,
       	xmin = 0,
	    xmax = 24,
     	ymin = -2.4,
	    ymax = -0.4,
	    legend style={at={(2.0,0.4)},anchor=south east}
    ]
    
    
   
    
    
    
   
     \addplot [color = black,
             solid,
             mark=*,
             smooth,
             line width = 0.5mm]
    table [x expr = \thisrowno{0}, y expr = \thisrowno{2}] {Sections/Shape-optimization-examples/Benchmark-cylinder/Benchmark-cylinder-Figures/benchmark-obj-convergence-1-var-uniform.txt};
    
   \addlegendentry{[case 1]: uniform refinement};
    
   \addplot  [color = black,
              dashed,
              smooth,
              line width = 0.5mm]
   table [x expr = \thisrowno{0}, y expr = \thisrowno{2}] {Sections/Shape-optimization-examples/Benchmark-cylinder/Benchmark-cylinder-Figures/benchmark-obj-convergence-1-var-adaptive.txt};
    
   \addlegendentry{[case 1]: adaptive refinement};   
   
      \addplot [color = blue,
             solid,
             mark=*,
             smooth,
             line width = 0.5mm]
    table [x expr = \thisrowno{0}, y expr = \thisrowno{2}] {Sections/Shape-optimization-examples/Benchmark-cylinder/Benchmark-cylinder-Figures/benchmark-obj-convergence-3-var-uniform.txt};
    
   \addlegendentry{[case 2]: uniform refinement};
    
   \addplot  [color = blue,
              dashed,
              smooth,
              line width = 0.5mm]
   table [x expr = \thisrowno{0}, y expr = \thisrowno{2}] {Sections/Shape-optimization-examples/Benchmark-cylinder/Benchmark-cylinder-Figures/benchmark-obj-convergence-3-var-adaptive.txt};
    
   \addlegendentry{[case 2]: adaptive refinement};   
      \addplot [color = red,
             solid,
             mark=*,
             smooth,
             line width = 0.5mm]
    table [x expr = \thisrowno{0}, y expr = \thisrowno{2}] {Sections/Shape-optimization-examples/Benchmark-cylinder/Benchmark-cylinder-Figures/benchmark-obj-convergence-6-var-uniform.txt};
    
   \addlegendentry{[case 3]: uniform refinement};
    
   \addplot  [color = red,
              dashed,
              smooth,
              line width = 0.5mm]
   table [x expr = \thisrowno{0}, y expr = \thisrowno{2}] {Sections/Shape-optimization-examples/Benchmark-cylinder/Benchmark-cylinder-Figures/benchmark-obj-convergence-6-var-adaptive.txt};
    
   \addlegendentry{[case 3]: adaptive refinement};

\end{axis}
\end{tikzpicture}
\caption{Cylinder example: convergence of the objective function $J(u)$ during the optimization process.} \label{fig:benchmark-convergence-all}
\end{figure}

The convergence process is shown in Figure \ref{fig:benchmark-convergence-all}. Optimization algorithm converges to the only minimum in the design space of case 1 with $\min J(u) = -1.9193$. In a richer space of case 2, the algorithm finds a local minimum with $\min J(u) = -1.985$, and in the design space of case 3, the local minimum with even smaller value of the objective function, $\min J(u) = -2.4461$, is found. Note that the results in terms of the objective function values for uniform and adaptive refinements are quasi-identical. It is an expected result, since at each step, the solution is refined (either adaptively or uniformly) until the error tolerance is lower than $\varepsilon_{\text{loop}} = 10^{-3}$.        

    ~ 

\begin{figure}[!ht]
	\centering
	\begin{tikzpicture}
\begin{axis}[cycle list name=exotic,
        xlabel = DOFs,
        ylabel = $H_{1}-$error estimator,
        ymode=log,
       	xmin = 100,
	    xmax = 1e5,
     	ymin = 1e-5,
	    ymax = 1.0,
	    xmode=log,
	    legend style={at={(2.0,0.4)},anchor=south east}
    ]
    
         \addplot [color = black,
             mark = *,
             solid,
             line width=1.0pt,
             mark size=2.0pt]
    table [x expr = \thisrowno{0}, y expr = \thisrowno{1}] {Sections/Shape-optimization-examples/Benchmark-cylinder/Benchmark-cylinder-Figures/benchmark-estimator-conv-1-var-uniform.txt};
    
   \addlegendentry{[case 1]: uniform refinement};
    
    \addplot [color = black,
             dashed,
             line width=1.0pt,
             mark size=1.5pt]
    table [x expr = \thisrowno{0}, y expr = \thisrowno{1}] {Sections/Shape-optimization-examples/Benchmark-cylinder/Benchmark-cylinder-Figures/benchmark-estimator-conv-1-var-adapt.txt};
    
   \addlegendentry{[case 1]: adaptive refinement};
   
   \addplot [color = blue,
             mark = *,
             solid,
             line width=1.0pt,
             mark size=2.0pt]
    table [x expr = \thisrowno{0}, y expr = \thisrowno{1}] {Sections/Shape-optimization-examples/Benchmark-cylinder/Benchmark-cylinder-Figures/benchmark-estimator-conv-3-vars-uniform.txt};
    
   \addlegendentry{[case 2, 3]: uniform refinement};
    
   \addplot  [color = blue,
             dashed,
             line width=1.0pt,
             mark size=1.5pt]
   table [x expr = \thisrowno{0}, y expr = \thisrowno{1}] {Sections/Shape-optimization-examples/Benchmark-cylinder/Benchmark-cylinder-Figures/benchmark-estimator-conv-3-vars-adapt.txt};
    
   \addlegendentry{[case 2, 3]: adaptive refinement};
   
    
    
    
 \addplot [color=black,solid,forget plot]
  table[row sep=crcr]{%
700	0.000564049\\
4000	2.88893E-05\\
};
  \addplot [color=black,solid,forget plot]
  table[row sep=crcr]{%
900	0.000137804 \\
2000	3.5319E-05 \\
900  3.5319E-05 \\
900	0.000137804 \\
};
\node[below, align=center, inner sep=0mm, text=black]
at (axis cs:1200,3e-05,0) {$1$};
\node[left, align=right, inner sep=0mm, text=black]
at (axis cs:800,7e-05,0) {$1.7$};
   
    \addplot [color=blue,solid,forget plot]
  table[row sep=crcr]{%
5000	3.22E-03 \\
35000	1.63E-04 \\
};

    \addplot [color=blue,solid,forget plot]
  table[row sep=crcr]{%
7500	3.46E-03\\
25000	5.46E-04\\
25000 3.46E-03\\
7500	3.46E-03\\
};

\node[below, align=center, inner sep=0mm, text=black]
at (axis cs:15000,7e-03,0) {$1$};
\node[left, align=right, inner sep=0mm, text=black]
at (axis cs:50000,1e-03,0) {$1.5$};
\end{axis}
\end{tikzpicture}
\caption{Cylinder example: convergence of the $H_{1}-$error estimator for the initial shape in terms of DOFs.} \label{fig:benchmark-convergence-all-estimator}
\end{figure}
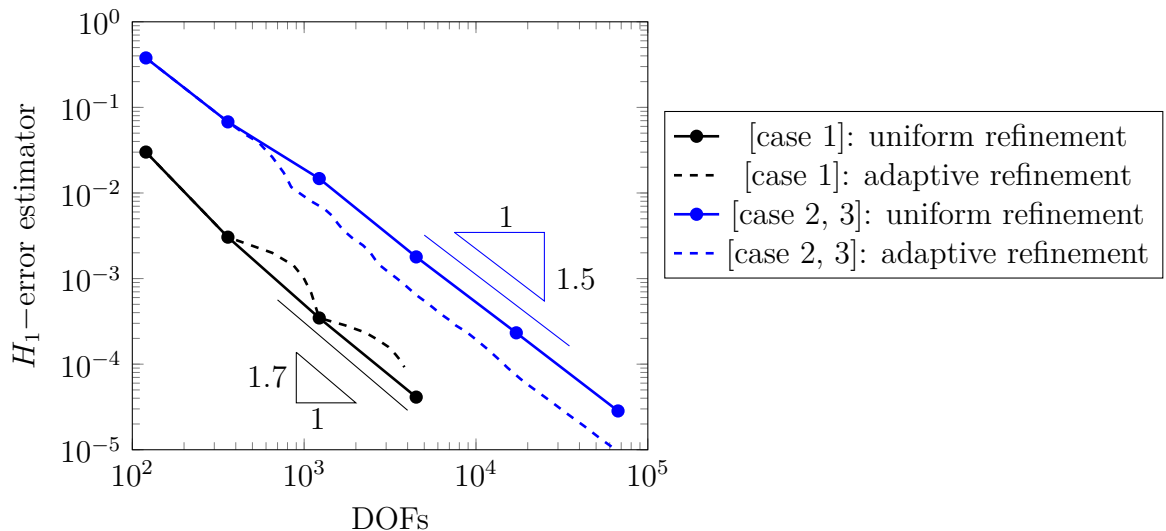

Next, the adaptive shape optimization algorithm \ref{alg:adaptive_optimisation} is studied. First, we compare the convergence plots of the recovery-based error estimator for the uniform and adaptive refinement for the initial shape in three cases in Figure \ref{fig:benchmark-convergence-all-estimator}. As it can be seen from the Figure, the solution for the initial shape in case 2(3) benefits slightly from the adaptive refinement due to the fact that, the shape is geometrically more complex than a circle in case 1. However, since this shape does not have any corners, the convergence rate for both, uniform and adaptive refinement, is optimal $(\sim h^{p/2})$.  

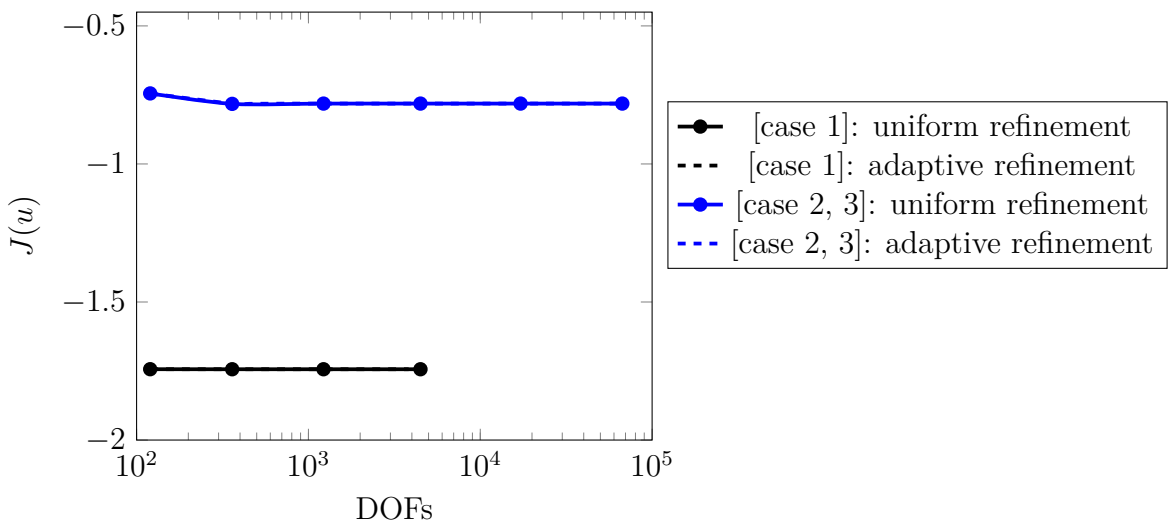
\begin{figure}[!ht]
 	\centering
 	\begin{tikzpicture}
\begin{axis}[cycle list name=exotic,
        xlabel = DOFs,
        ylabel = $J(u)$,
       	xmin = 100,
	    xmax = 1e5,
     	ymin = -2,
	    ymax = -0.45,
	    xmode = log,
	    legend style={at={(2.0,0.4)},anchor=south east}
    ]
    
         \addplot [color = black,
             solid,
             mark=*,
             smooth,
             line width = 0.5mm]
    table [x expr = \thisrowno{0}, y expr = \thisrowno{2}] {Sections/Shape-optimization-examples/Benchmark-cylinder/Benchmark-cylinder-Figures/benchmark-estimator-conv-1-var-uniform.txt};
    
   \addlegendentry{[case 1]: uniform refinement};
    
    \addplot [color = black,
              dashed,
              smooth,
              line width = 0.5mm]
    table [x expr = \thisrowno{0}, y expr = \thisrowno{2}] {Sections/Shape-optimization-examples/Benchmark-cylinder/Benchmark-cylinder-Figures/benchmark-estimator-conv-1-var-adapt.txt};
    
   \addlegendentry{[case 1]: adaptive refinement};
   
   \addplot [color = blue,
             solid,
             mark=*,
             smooth,
             line width = 0.5mm]
    table [x expr = \thisrowno{0}, y expr = \thisrowno{2}] {Sections/Shape-optimization-examples/Benchmark-cylinder/Benchmark-cylinder-Figures/benchmark-estimator-conv-3-vars-uniform.txt};
    
   \addlegendentry{[case 2, 3]: uniform refinement};
    
   \addplot  [color = blue,
              dashed,
              smooth,
              line width = 0.5mm]
   table [x expr = \thisrowno{0}, y expr = \thisrowno{2}] {Sections/Shape-optimization-examples/Benchmark-cylinder/Benchmark-cylinder-Figures/benchmark-estimator-conv-3-vars-adapt.txt};
    
   \addlegendentry{[case 2, 3]: adaptive refinement};
   
    
    
    

\end{axis}
\end{tikzpicture}
 \caption{Cylinder example: convergence of the objective function $J(u)$ in terms of DOFs.} \label{fig:benchmark-convergence-cost-function-dof-all}
\end{figure}

In Figure \ref{fig:benchmark-convergence-cost-function-dof-all}, comparison in terms of the convergence of the objective function is shown for the initial shapes in case 1 and case 2(3). It can be seen that the results for uniform and adaptive refinements are quasi-identical for any number of DOFs. This can be explained by the fact that, in this particular example with simple regular geometries, objective function (in the form of an integral) converges very fast.

Then, uniform and adaptive shape optimization results are compared. Tables \ref{tab:cylinder_table_1_var_comparison}, \ref{tab:cylinder_table_3_vars_comparison} and \ref{tab:cylinder_table_6_vars_comparison} show the results of those comparisons in terms of DOFs, computational time and the optimal objective function obtained. In the three cases, the initial tolerance employed for the error estimator is $\varepsilon_0 = 10^{-2}$, while the loop tolerances for the error estimator are taken $\varepsilon_{\text{loop}} = 10^{-3}$ and  $\varepsilon_{\text{loop}} = 10^{-4}$. The last row shows the relative difference between the corresponding quantities for the uniform and adaptive refinements, calculated as shown for the time difference below:
\begin{equation}
    \text{Difference} = \dfrac{\text{time}_{\text{uniform}} - \text{time}_{\text{adaptive}}}{\text{time}_{\text{uniform}}}\times{100}\%
\end{equation}
Results for cases 2 and 3 show that for both values of $\varepsilon_{\text{loop}}$, adaptive optimization overcomes uniform optimization in terms of DOFs and computational time. The results for case 1, on the other hand, show that adaptive refinement can deliver the optimal result with less DOFs, but with no conclusive difference in terms of computational time. This result can be explained by the fact that the geometry in case 1 is always smooth and simple, therefore there is no significant advantage in using adaptive refinement in this case. 

\begin{table}[ht!]
\centering
\begin{tabular}{lrrrr}
\multicolumn{1}{c}{} & \multicolumn{4}{c}{$\varepsilon_{\text{loop}} = 10^{-3}$} \\ \cline{2-5} 
                     & \multicolumn{1}{l}{DOFs} & \multicolumn{1}{l}{ Time (s)} & \multicolumn{1}{l}{ Objective function} & \multicolumn{1}{l}{Best solution}\\ \hline
Uniform              & 1224 & 9.568 & -1.91936 & 1.26889 \\  
Adaptive             & 992 & 7.673 & -1.91936 & 1.26889 \\ \hline
Difference (\%)      & 18.95 & 19.81 & 0  & 0 \\       
                     & \multicolumn{1}{l}{}     & \multicolumn{1}{l}{}                       & \multicolumn{1}{l}{}                      \\
\multicolumn{1}{c}{} & \multicolumn{4}{c}{$\varepsilon_{\text{loop}} = 10^{-4}$}                                                                              \\ \cline{2-5} 
                     & \multicolumn{1}{l}{DOFs} & \multicolumn{1}{l}{ Time (s)} & \multicolumn{1}{l}{ Objective function} & \multicolumn{1}{l}{Best solution}\\ \hline
Uniform              & 4488 & 23.334 & -1.91936 & 1.26889 \\
Adaptive             & 3344 & 36.356 & -1.91936 & 1.26889 \\ \hline
Difference (\%)      & 25.49 & -55.81 & 0 & 0
\end{tabular}
\caption{Cylinder example: uniform and recovery-based adaptive refinement results for [case 1] with 1 design variable. $\varepsilon_{\text{0}} = 10^{-2}$ and $N_{\text{Max steps}}=10$.}
\label{tab:cylinder_table_1_var_comparison}
\end{table}

\begin{table}[ht!]
\centering
\begin{tabular}{lrrrr}
\multicolumn{1}{c}{} & \multicolumn{4}{c}{$\varepsilon_{\text{loop}} = 10^{-3}$}  \\ \cline{2-5} 
                     & \multicolumn{1}{l}{DOFs} & \multicolumn{1}{l}{ Time (s)} & \multicolumn{1}{l}{ Objective function} & \multicolumn{1}{l}{Best solution} \\ \hline
Uniform              & 4488 & 122.787 & -1.985  & [1.1554, 1.3788, 1.3788] \\  
Adaptive             & 1536 & 44.566 & -1.985 & [1.1554, 1.3788, 1.3788] \\ \hline
Difference (\%)      & 65.78 & 63.70 & 0 & 0  \\       
                     & \multicolumn{1}{l}{}     & \multicolumn{1}{l}{}                       & \multicolumn{1}{l}{}                      \\
\multicolumn{1}{c}{} & \multicolumn{4}{c}{$\varepsilon_{\text{loop}} = 10^{-4}$} \\ \cline{2-5} 
                     & \multicolumn{1}{l}{DOFs} & \multicolumn{1}{l}{ Time (s)} & \multicolumn{1}{l}{Objective function} & \multicolumn{1}{l}{Best solution} \\ \hline
Uniform              & 17160 & 188.196 & -1.985  & [1.1554, 1.3788, 1.3788] \\
Adaptive             & 5304 & 85.902 & -1.985 & [1.1554, 1.3788, 1.3788] \\ \hline
Difference (\%)      & 69.09 & 54.36 & 0 & 0
\end{tabular}
\caption{Cylinder example: uniform and recovery-based adaptive refinement results for [case 2] with 3 design variables. $\varepsilon_{\text{0}} = 10^{-2}$ and $N_{\text{Max steps}}=10$.}
\label{tab:cylinder_table_3_vars_comparison}
\end{table}

\begin{table}[ht!]
\centering
\resizebox{\columnwidth}{!}{%
\begin{tabular}{lrrrr}
\multicolumn{1}{c}{} & \multicolumn{4}{c}{$\varepsilon_{\text{loop}} = 10^{-3}$}                                                                              \\ \cline{2-5} 
                     & \multicolumn{1}{l}{DOFs} & \multicolumn{1}{l}{ Time (s)} & \multicolumn{1}{l}{ Objective function} & \multicolumn{1}{l}{Best solution} \\ \hline
Uniform              & 17160                    & 861.643                                   & -2.4461 & [1.0042, 1.2570, 1.2570, 1.8534, 1.8248, 1.8248, 1.8534]                                   \\  
Adaptive             & 2976                     & 234.360                                   & -2.4461  & [1.0042, 1.2570, 1.2570, 1.8534, 1.8248, 1.8248, 1.8534]                                \\ \hline 
Difference (\%)      & 82.66                        & 72.80                                 & 0 & 0  \\                                    \\
                     & \multicolumn{1}{l}{}     & \multicolumn{1}{l}{}                       & \multicolumn{1}{l}{}                      \\
\multicolumn{1}{c}{} & \multicolumn{4}{c}{$\varepsilon_{\text{loop}} = 10^{-4}$}                                                                              \\ \cline{2-5} 
                     & \multicolumn{1}{l}{DOFs} & \multicolumn{1}{l}{ Time (s)} & \multicolumn{1}{l}{Objective function} &\multicolumn{1}{l}{Best solution}  \\ \hline
Uniform              & 67080                    & 2759.996                                  & -2.4461 & [1.0042, 1.2570, 1.2570, 1.8534, 1.8248, 1.8248, 1.8534]                                    \\
Adaptive             & 17160                    & 666.817                                   & -2.4461 & [1.0042, 1.2570, 1.2570, 1.8534, 1.8248, 1.8248, 1.8534]                                    \\ \hline
Difference (\%)      & 74.42                       & 75.84                               & 0 & 0
\end{tabular}
}
\caption{Cylinder example: uniform and recovery-based adaptive refinement results for [case 3] with 7 design variables. $\varepsilon_{\text{0}} = 10^{-2}$ and $N_{\text{Max steps}}=10$.}
\label{tab:cylinder_table_6_vars_comparison}
\end{table}

\cleardoublepage
\subsection{Acoustic horn }
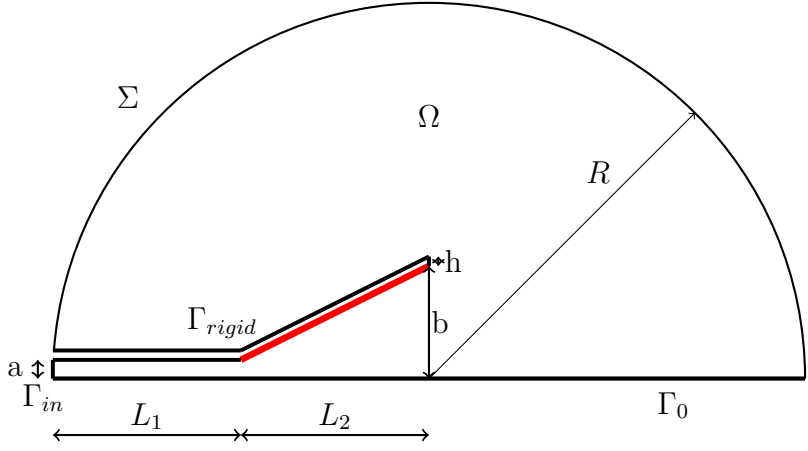
\begin{figure}[ht!]
	\centering
 	\hspace*{-50pt}
\begin{tikzpicture}[scale = 5, classical/.style={thick}]

\begin{scope}
    \clip  (-1.1,0) rectangle (1.1,1.05);
    \centerarc[thick,black](0,0)(0:175.711:1);
    \draw [thick,red, line width=2.5pt](0.0,0.3) -- (-0.5,0.05);  
    \draw [thick, line width=1.5pt](-0.5,0.05) -- (-1.0,0.05);  
     \draw [thick, line width=1.5pt](-1.0,0.05) -- (-1.0,0.0); 
     \draw [thick,line width=1.5pt](-1.0,0) -- (1.0,0);
    \draw [thick, line width=1.5pt](-1.0,0.075) -- (-0.5,0.075); 
    \draw [thick, line width=1.5pt](-0.5,0.075) -- (0.0,0.325);
    \draw [thick, line width=1.5pt](0.0,0.325) -- (0.0,0.3);
 \draw [->] (0.0,0.0) -- (0.707,0.707);
\end{scope}
     \draw [thick,line width=1.5pt](-1.0,0) -- (1.0,0);

\node[] at (-1.025,-0.05) {$\Gamma_{in}$};
\node[] at (-0.8,0.75) {$\Sigma$};
\node[] at (0.65,-0.07) {$\Gamma_0$};
\node[] at (-0.55,0.15) {$\Gamma_{rigid}$};
\node[] at (0,0.7) {$\Omega$};
\node[] at (0.45,0.55) {$R$};

\draw [thick ] [<->] (-1.04,0) -- (-1.04,0.05) ;
 \node[] at (-1.1,0.025) {a};

\draw [thick ] [<->] (0,0) -- (0,0.3) ;
 \node[] at (0.03,0.15) {b};
 
\draw [thick ] [<->] (0.025,0.3) -- (0.025,0.325) ;
 \node[] at (0.065,0.3125) {h};
 
\draw [thick ] [<->] (-1,-0.15) -- (-0.5,-0.15) ;
 \node[] at (-0.75,-0.1) {$L_1$};
 
\draw [thick ] [<->] (-0.5,-0.15) -- (-0,-0.15) ;
 \node[] at (-0.25,-0.1) {$L_2$}; 
\end{tikzpicture}
	\caption{The horn problem domain.}
	\label{fig:horn}
\end{figure}

The acoustic horn problem studied in this section aims to reduce the back reflection of the incoming wave transmission in order to provide an efficient wave distribution towards the far-field. The horn is modeled as a 2D planar symmetry problem. The horn problem domain $\Omega$ is shown in Figure \ref{fig:horn} where $\Gamma_{in}$, $\Gamma_{rigid}$ and $\Gamma_{0}$ refer to the in-going wave inlet, the horn rigid boundary and the symmetry line, respectively. In order to model infinite domain, a truncation artificial boundary $\Sigma$ is defined as a semi-circle with radius $R$. \\


 The boundary value problem for this problem can be written as follows:
\begin{equation}
    \begin{split}
\Delta u + k^2u &= 0 \qquad \qquad \, \, \text{in} \; \Omega, \\
\frac{ \partial u}{ \partial \boldsymbol{n}} + \left(ik+\frac{1}{ 2R}\right)u &= 0     \qquad  \qquad \, \text{on} \; \Sigma, \\
\frac{ \partial u}{ \partial \boldsymbol{n}} + iku &= 2ikA_m     \qquad \text{on} \; \Gamma_{in},\\
\frac{ \partial u}{ \partial \boldsymbol{n}} &= 0     \qquad \qquad \, \text{on} \; \Gamma_{rigid},  \\
\frac{ \partial u}{ \partial \boldsymbol{n}} &= 0     \qquad \qquad \, \text{on} \;   \Gamma_{0}  
    \end{split}
\end{equation}

\noindent in which $A_m$ is the incident wave amplitude transferring from $\Gamma_{in}$ into the acoustic domain.

The wave reflection coefficient $\mathcal{R}$ to be minimized in this optimization problem is defined as $\mathcal{R} = B_m/A_m$ on $\Gamma_{in}$, where $B_m$ is the reflected wave amplitude computed as $B_m=|u_{in}-A_m|$ and $u_{in}$ is the predicted acoustic pressure averaged on $\Gamma_{in}$ as $u_{in}= \frac{1}{a} \int_{\Gamma_{in}} u d\Gamma$ \cite{doi:10.1002/nme.2132,MOSTAFASHAABAN2020156}. No constraint is defined in this optimization problem. Therefore, the objective function can be written directly as follows: 
 
  \begin{equation}        \label{eq:obj_horn}
    J(u) = \mathcal{R} 
 \end{equation} 

The red line in Figure \ref{fig:horn} represents the part of the horn, which is optimized in order to achieve the minimum of $\mathcal{R}$, where the $x$ and $y$ coordinates of the control points are set as design variables. 

The initial geometry of the horn is considered as follows: the horn thickness is $h=2.5 \mathrm{cm}$ and the dimensions $a$ and $b$ are set as $5$ and $30 \mathrm{cm}$, respectively. The first dimension $a$ is constant for a length of $L_1=50 \mathrm{cm}$, after that, a consequent flaring of $L_2=50 \mathrm{cm}$ length is defined. The incident wave amplitude $A_m$ is chosen as $1$ and the radius $R$ is taken as $1 \mathrm{m}$. 

\begin{figure}[ht!]
        \includegraphics[width=\textwidth]{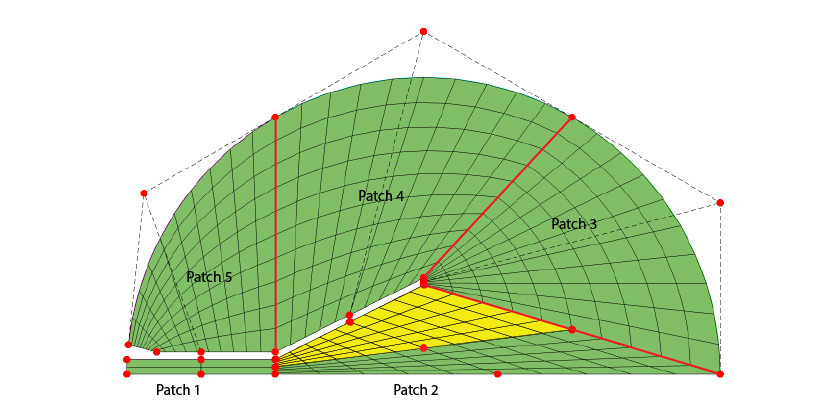}
\caption{Horn optimization problem: NURBS parameterization.}
\label{fig:fig:horn-nurbs}
\end{figure}

For the GIFT implementation, the computational domain is split into five patches, as shown in Figure \ref{fig:fig:horn-nurbs}. The optimization will be performed only on the location of the control points that define the horn wall on the patch number 2. Note that, if the geometry parameterization is not refined during the solution refinement process, then a change in the horn boundary parameterization will lead to the change in the parameterization of the entire patch 2. Hence, the stiffness matrix for patch 2 needs to be re-calculated every time when the design parameters change during the optimization process. This is an unnecessary computational burden which can be avoided by reducing the continuity of the geometry parameterization in the radial direction. Continuity of the geometry basis is reduced by inserting knot value at $0.5$ twice. Then all changes in the system matrix are localized in the yellow area of patch 2, as shown in Figure \ref{fig:fig:horn-nurbs}.

Next, we present shape optimization results and then analyze the performance of the adaptive algorithm. 

\subsubsection{Shape optimization results} 
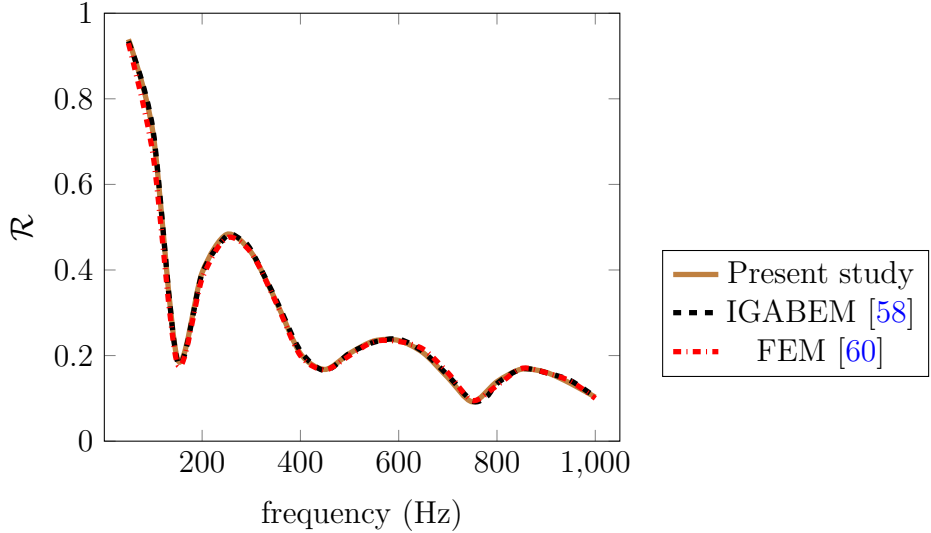
\begin{figure}[ht!]
	\centering
	\begin{tikzpicture}[scale=1]
\begin{axis}[cycle list name=exotic,
        xlabel= frequency (Hz),
        ylabel= $\mathcal{R}$,
	xmin = 1,
	xmax = 1050,
	ymin = 0.0,
	ymax = 1.0,
	legend style={at={(1.6,0.15)},anchor=south east}
    ]

    \addplot [brown, solid, smooth, line width=2.0pt]
    table [x expr = \thisrowno{0}, y=y] {Sections/Shape-optimization-examples/Horn/Figures/initial-shapes-spetrum/IGA.txt};
    
    \addlegendentry{Present study};

    \addplot [black, dashed, smooth,  line width=2.0pt]
    table [x expr = \thisrowno{0}, y=y] {Sections/Shape-optimization-examples/Horn/Figures/initial-shapes-spetrum/IGABEM.txt};
    
   \addlegendentry{IGABEM \cite{MOSTAFASHAABAN2020156}};

  \addplot [red, dash dot, smooth, line width=2.0pt]
    table [x expr = \thisrowno{0}, y=y] {Sections/Shape-optimization-examples/Horn/Figures/initial-shapes-spetrum/FEM.txt};
    
   \addlegendentry{FEM \cite{doi:10.1002/nme.2132}};

\end{axis}
\end{tikzpicture}
\caption{Reflection spectra of the initial horn.}
\label{fig:initail-spectra}
\end{figure}

The reflection spectrum of the initial shape is computed using the proposed GIFT model and compared in Figure \ref{fig:initail-spectra} against previously published spectra obtained by IGABEM \cite{MOSTAFASHAABAN2020156} and FEM \cite{doi:10.1002/nme.2132} where excellent agreement can be seen between the three spectra.

\begin{figure}[ht!]
    \centering
    \begin{subfigure}[b]{0.45\textwidth}
        \includegraphics[width=\textwidth]{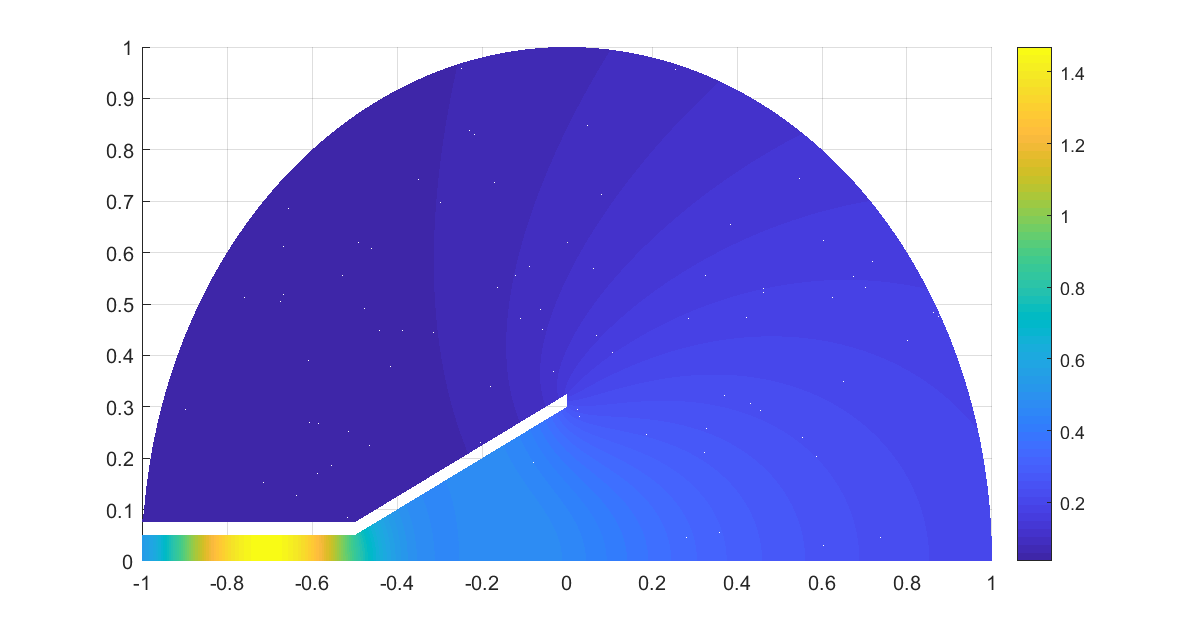}
        \caption{Initial sound pressure for $f=280$ Hz}
    \end{subfigure}
    ~ 
    \begin{subfigure}[b]{0.45\textwidth}
        \includegraphics[width=\textwidth]{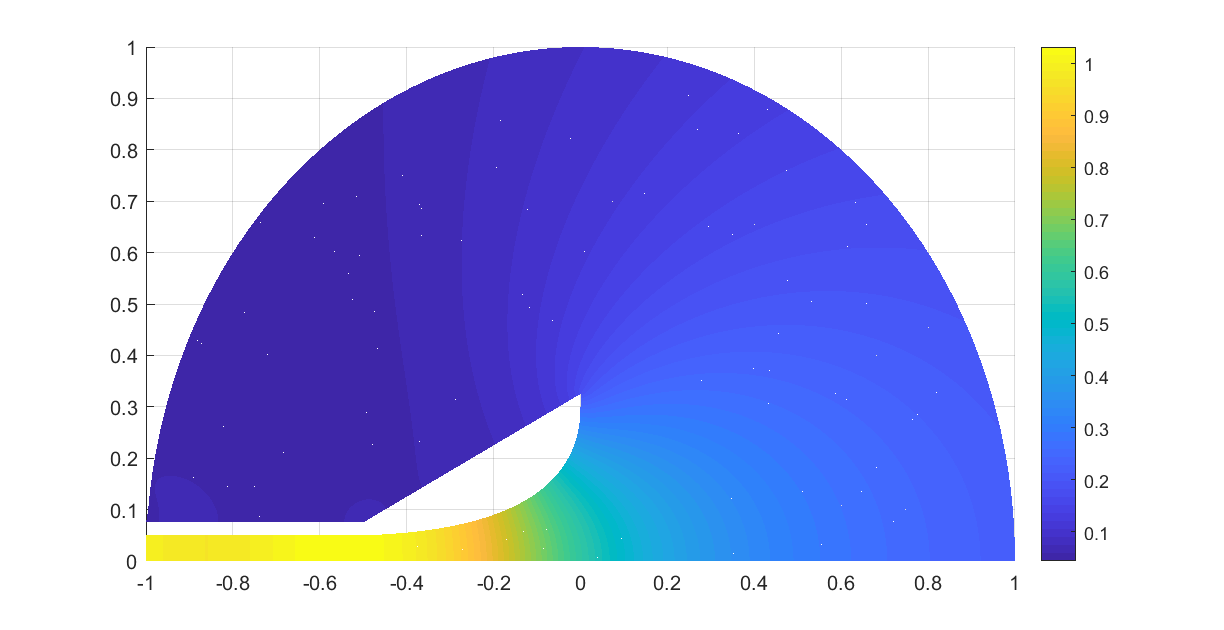}
       \caption{Optimized sound pressure for $f=280$ Hz}
    \end{subfigure}    
    ~
    \begin{subfigure}[b]{0.45\textwidth}
        \includegraphics[width=\textwidth]{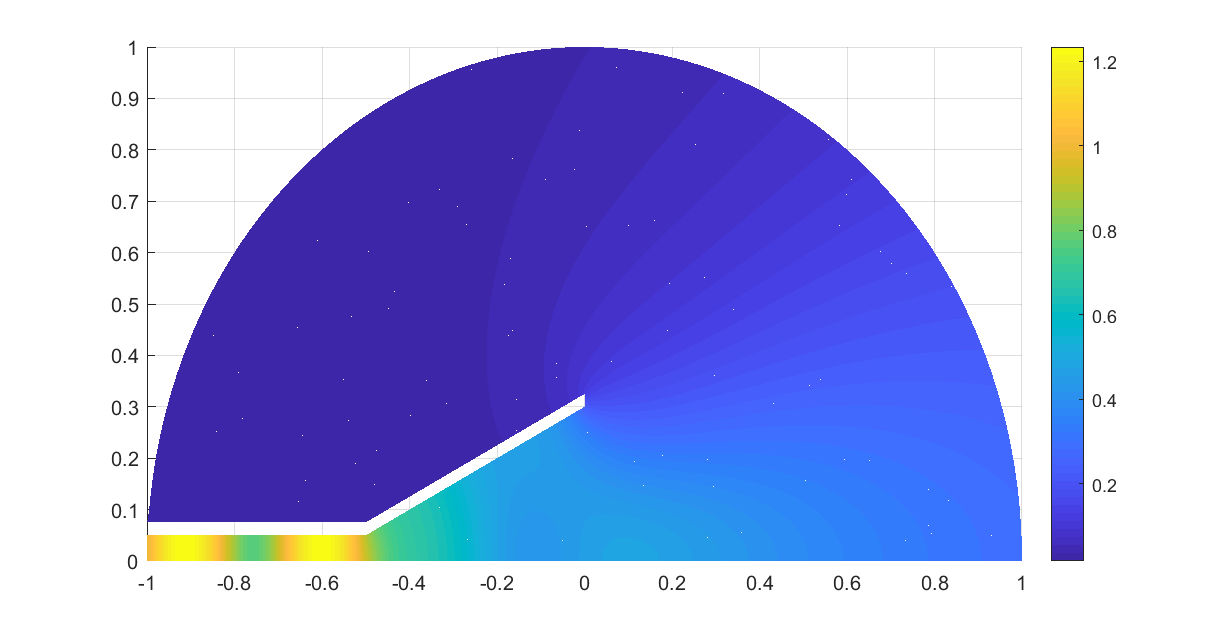}
        \caption{Initial sound pressure for $f=550$ Hz}
    \end{subfigure}
    ~ 
    \begin{subfigure}[b]{0.45\textwidth}
        \includegraphics[width=\textwidth]{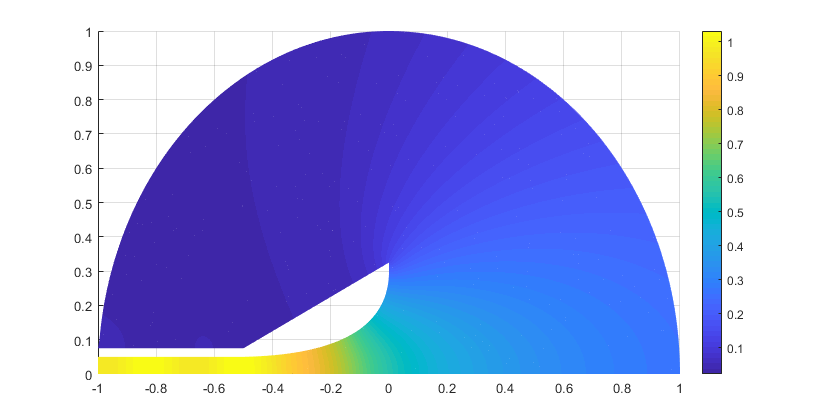}
       \caption{Optimized sound pressure for $f=550$ Hz}
    \end{subfigure} 
    ~
    \begin{subfigure}[b]{0.45\textwidth}
        \includegraphics[width=\textwidth]{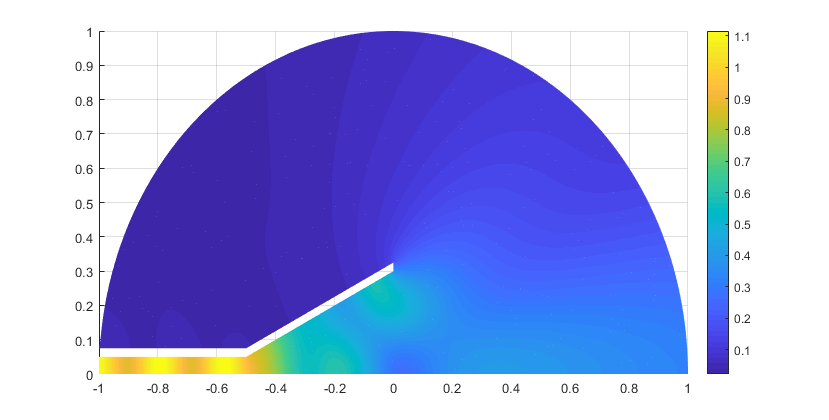}
        \caption{Initial sound pressure for $f=780$ Hz}
    \end{subfigure}
    ~ 
    \begin{subfigure}[b]{0.45\textwidth}
        \includegraphics[width=\textwidth]{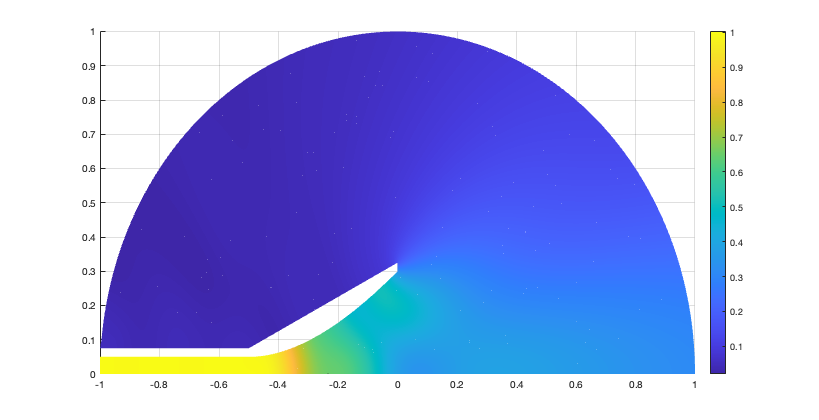}
       \caption{Optimized sound pressure for $f=780$ Hz}
    \end{subfigure}
    ~
    \begin{subfigure}[b]{0.45\textwidth}
        \includegraphics[width=\textwidth]{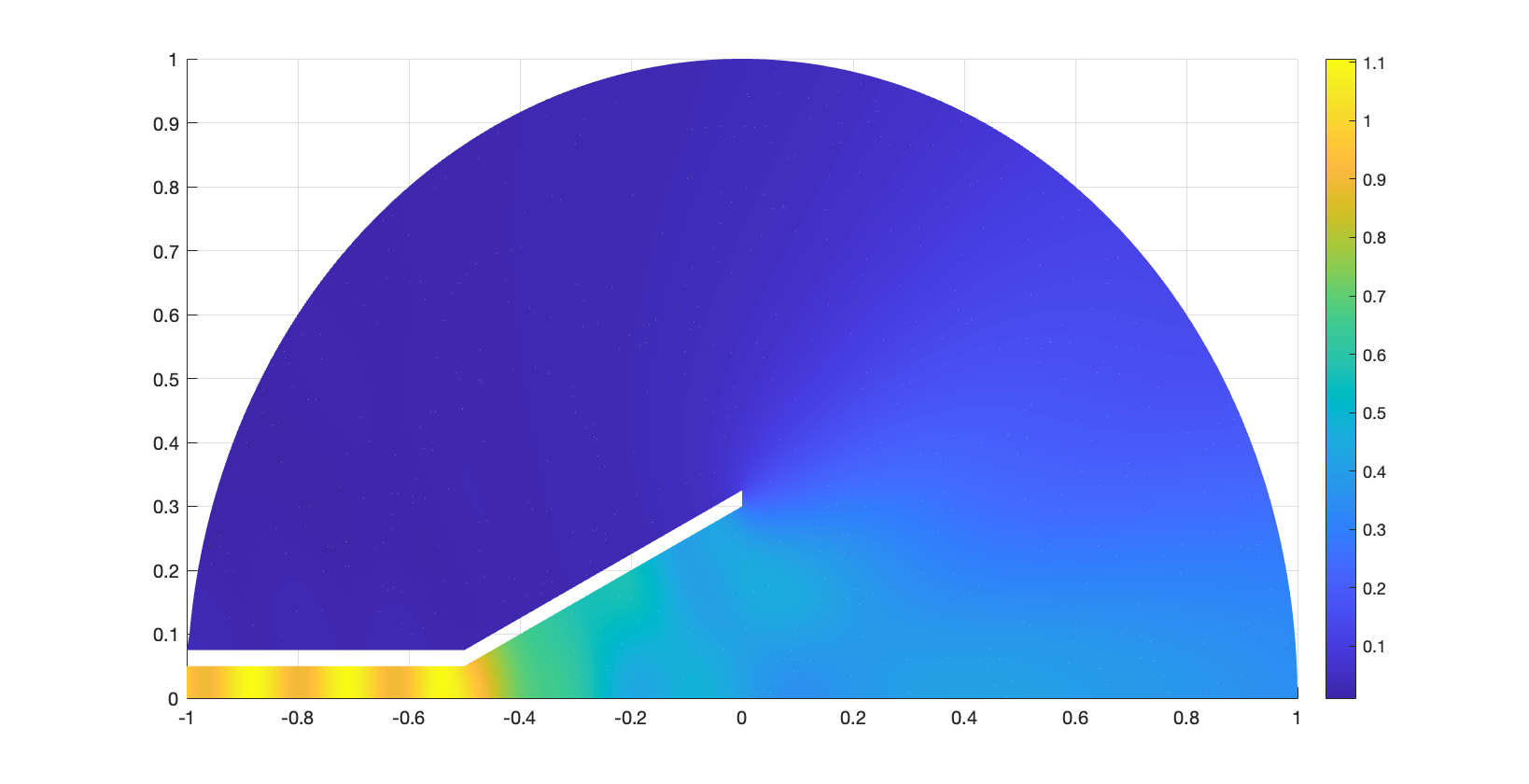}
        \caption{Initial sound pressure for $f=1000$ Hz}
    \end{subfigure}
    ~ 
    \begin{subfigure}[b]{0.45\textwidth}
        \includegraphics[width=\textwidth]{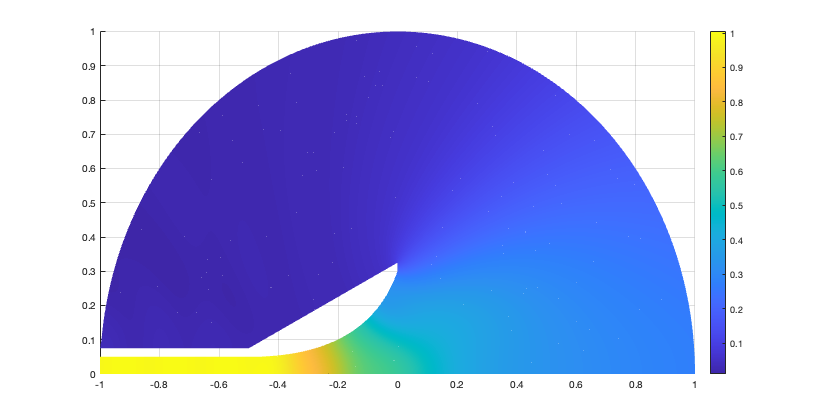}
       \caption{Optimized sound pressure for $f=1000$ Hz}
    \end{subfigure} 
 \caption{Sound pressure distribution (absolute value) for the initial and optimized horn shapes using 1 control point - 2 design variables.}   
     \label{fig:horn_initial_optimized_sound_pressure}
\end{figure}

The results obtained from the optimization problem with GIFT are compared against those obtained from both IGABEM with PSO and FEM with Genetic Algorithm reported in \cite{MOSTAFASHAABAN2020156} and \cite{BARBIERI2013356}, respectively. The optimization models of IGABEM in \cite{MOSTAFASHAABAN2020156} use two design variables (the $x$ and $y$ coordinates of one control point) getting benefit from the automatic smoothness of IGA, while FEM models in \cite{BARBIERI2013356} use only one design variable with applying additional curve smoothness using Hermite polynomials. Figure \ref{fig:2DV-all-shape} shows the optimized shapes for all comparative methods for different single frequencies $f = 280, 550, 780, 1000 \mathrm{Hz}$ with sound speed $c = 345\mathrm{m/s}$ and wave-number $k = 5.09940, 10.01667, 14.20546, 18.21213$, respectively. Figure \ref{fig:obj_fx_number_vs_iter_horn_1ctp} illustrates the variation of the reflection coefficient $\mathcal{R}$ through the optimization process, while the reflection spectra for initial and optimized horn shapes are included in Figure \ref{fig:2DV-all-spectra}. Figure \ref{fig:horn_initial_optimized_sound_pressure} shows the domain sound pressure values of the initial and optimized horns for frequencies $f = 280$, $550$, $780$ and $1000$ $\mathrm{Hz}$. 

\begin{figure}[ht!]
	\centering
	\begin{tikzpicture}[scale=1]
    \pgfplotsset{footnotesize,samples=10}
    \begin{groupplot}[group style = {group size = 2 by 2, horizontal sep = 40pt, vertical sep = 45pt},
    width = 6.0cm, height = 5.0cm]
    
    \nextgroupplot[ title = {a) $f$=280 Hz},
                     legend style = {
            column sep = 10pt, legend columns = 2},
            legend to name = grouplegendshape2,
             width=0.5 \textwidth,
             height=0.3\textwidth,
	         xmin = 0.49 - 1,
	         xmax = 1.01 - 1,
        	ymin = 0.0,
	        ymax = 0.31,
	        xtick = {-0.5, -0.25, 0},
        	ytick  = {0, 0.05, 0.1, 0.15, 0.2, 0.25, 0.3},
	]

        \addplot [color=brown,
              solid,
              smooth,
              line width=2.0pt]
    table [x expr = \thisrowno{0}, y expr = \thisrowno{1}] {Sections/Shape-optimization-examples/Horn/Figures/280-2DV/shape_280_GIFT.txt};
  \addlegendentry{\textcolor{black}{Shape - present study}};
  
         \addplot [color= brown,
               only marks,
              mark=*,
              mark options={solid}]
    table [x expr = \thisrowno{0} , y expr = \thisrowno{1}] {Sections/Shape-optimization-examples/Horn/Figures/280-2DV/280-IGA-cpt.txt};
  \addlegendentry{\textcolor{black}{Design variables - present study}};

                \addplot [dashed,
              smooth,
              line width=2.0pt]
    table [x expr = \thisrowno{0} - 1, y expr = \thisrowno{1}] {Sections/Shape-optimization-examples/Horn/Figures/280-2DV/280-IGABEM-shape.txt};

    \addlegendentry{\textcolor{black}{IGABEM shape \cite{MOSTAFASHAABAN2020156}}};
    
        \addplot [color=black,
               only marks,
              mark=*,
              mark options={solid}]
    table [x expr = \thisrowno{0} - 1, y expr = \thisrowno{1}] {Sections/Shape-optimization-examples/Horn/Figures/280-2DV/280-IGABEM-cpt.txt};
    
\addlegendentry{\textcolor{black}{IGABEM design variables \cite{MOSTAFASHAABAN2020156}}};

 \addplot [color=red,
              solid,
              smooth,
              line width=2.0pt]
    table [x expr = \thisrowno{0} - 1, y expr = \thisrowno{1}] {Sections/Shape-optimization-examples/Horn/Figures/280-2DV/280-FEM-shape.txt};
    
    \addlegendentry{\textcolor{black}{FEM shape \cite{BARBIERI2013356} }};

\nextgroupplot[ title = {b) $f$=550 Hz}, 
             width=0.5\textwidth,
             height=0.3\textwidth,
	         xmin = 0.49 - 1,
	         xmax = 1.01 - 1,
        	ymin = 0.0,
	        ymax = 0.31,
	        xtick = {-0.5, -0.25, 0.0},
        	ytick  = {0, 0.05, 0.1, 0.15, 0.2, 0.25, 0.3},
	]
	
\addplot [color=brown,
              solid,
              smooth,
              line width=2.0pt]
    table [x expr = \thisrowno{0}, y expr = \thisrowno{1}] {Sections/Shape-optimization-examples/Horn/Figures/550-2DV/shape_550_GIFT.txt};

         \addplot [color= brown,
               only marks,
              line width=0.5 0pt,
              mark=*,
              mark options={solid}]
    table [x expr = \thisrowno{0} , y expr = \thisrowno{1}] {Sections/Shape-optimization-examples/Horn/Figures/550-2DV/550-IGA-cpt.txt};

                \addplot [color=black,
              dashed,
              smooth,
              line width=2.0pt]
    table [x expr = \thisrowno{0} - 1, y expr = \thisrowno{1}] {Sections/Shape-optimization-examples/Horn/Figures/550-2DV/550-IGABEM-shape.txt};
    
         \addplot [color= black,
               only marks,
              line width=0.5 0pt,
              mark=*,
              mark options={solid}]
    table [x expr = \thisrowno{0} - 1, y expr = \thisrowno{1}] {Sections/Shape-optimization-examples/Horn/Figures/550-2DV/550-IGABEM-cpt.txt};

    \addplot [color=red,
              solid,
              smooth,
              line width=2.0pt]
    table [x expr = \thisrowno{0} - 1, y expr = \thisrowno{1}] {Sections/Shape-optimization-examples/Horn/Figures/550-2DV/550-FEM-shape.txt};

    \nextgroupplot[ title = {c) $f$=780 Hz}, 
             width=0.5 \textwidth,
             height=0.3\textwidth,
	         xmin = 0.49-1,
	         xmax = 1.01-1,
        	ymin = 0.0,
	        ymax = 0.31,
	        xtick = {-0.5, -0.25, 0},
        	ytick  = {0, 0.05, 0.1, 0.15, 0.2, 0.25, 0.3},
	]
	
	        \addplot [color=brown,
              solid,
              smooth,
              line width=2.0pt]
    table [x expr = \thisrowno{0}, y expr = \thisrowno{1}] {Sections/Shape-optimization-examples/Horn/Figures/780-2DV/shape_780_GIFT.txt};
    
         \addplot [color= brown,
               only marks,
              line width=0.5 0pt,
              mark=*,
              mark options={solid}]
    table [x expr = \thisrowno{0}, y expr = \thisrowno{1}] {Sections/Shape-optimization-examples/Horn/Figures/780-2DV/780-IGA-cpt.txt};

                \addplot [color=black,
              dashed,
              smooth,
              line width=2.0pt]
    table [x expr = \thisrowno{0}-1, y expr = \thisrowno{1}] {Sections/Shape-optimization-examples/Horn/Figures/780-2DV/780-IGABEM-shape.txt};
    
         \addplot [color= black,
               only marks,
              line width=0.5 0pt,
              mark=*,
              mark options={solid}]
    table [x expr = \thisrowno{0}-1, y expr = \thisrowno{1}] {Sections/Shape-optimization-examples/Horn/Figures/780-2DV/780-IGABEM-cpt.txt};

    \addplot [color=red,
              solid,
              smooth,
              line width=2.0pt]
    table [x expr = \thisrowno{0}-1, y expr = \thisrowno{1}] {Sections/Shape-optimization-examples/Horn/Figures/780-2DV/780-FEM-shape.txt};

    \nextgroupplot[ title = {d) $f$=1000 Hz}, 
             width=0.5 \textwidth,
             height=0.3\textwidth,
	         xmin = 0.49-1,
	         xmax = 1.01-1,
        	ymin = 0.0,
	        ymax = 0.31,
	        xtick = {-0.5, -0.25, 0},
        	ytick  = {0, 0.05, 0.1, 0.15, 0.2, 0.25, 0.3},
	]
	
	       \addplot [color=brown,
              solid,
              smooth,
              line width=2.0pt]
    table [x expr = \thisrowno{0}, y expr = \thisrowno{1}] {Sections/Shape-optimization-examples/Horn/Figures/1000-2DV/shape_1000_GIFT_unif.txt};

         \addplot [color= brown,
               only marks,
              line width=0.5 0pt,
              mark=*,
              mark options={solid}]
    table [x expr = \thisrowno{0}, y expr = \thisrowno{1}] {Sections/Shape-optimization-examples/Horn/Figures/1000-2DV/1000-IGA-cpt.txt};

                \addplot [color=black,
              dashed,
              smooth,
              line width=2.0pt]
    table [x expr = \thisrowno{0}-1, y expr = \thisrowno{1}] {Sections/Shape-optimization-examples/Horn/Figures/1000-2DV/1000-IGABEM-shape.txt};
    
         \addplot [color= black,
               only marks,
              line width=0.5 0pt,
              mark=*,
              mark options={solid}]
    table [x expr = \thisrowno{0}-1, y expr = \thisrowno{1}] {Sections/Shape-optimization-examples/Horn/Figures/1000-2DV/1000-IGABEM-cpt.txt};

     \addplot [color=red,
              solid,
              smooth,
              line width=2.0pt]
    table [x expr = \thisrowno{0}-1, y expr = \thisrowno{1}] {Sections/Shape-optimization-examples/Horn/Figures/1000-2DV/1000-FEM-shape.txt};

    \end{groupplot}
    \node at ($(7.5 cm ,-7 cm)$) {\ref{grouplegendshape2}}; 
\end{tikzpicture}
\caption{Optimized horn shapes using 1 control point - 2 design variables.}
\label{fig:2DV-all-shape}
\end{figure}
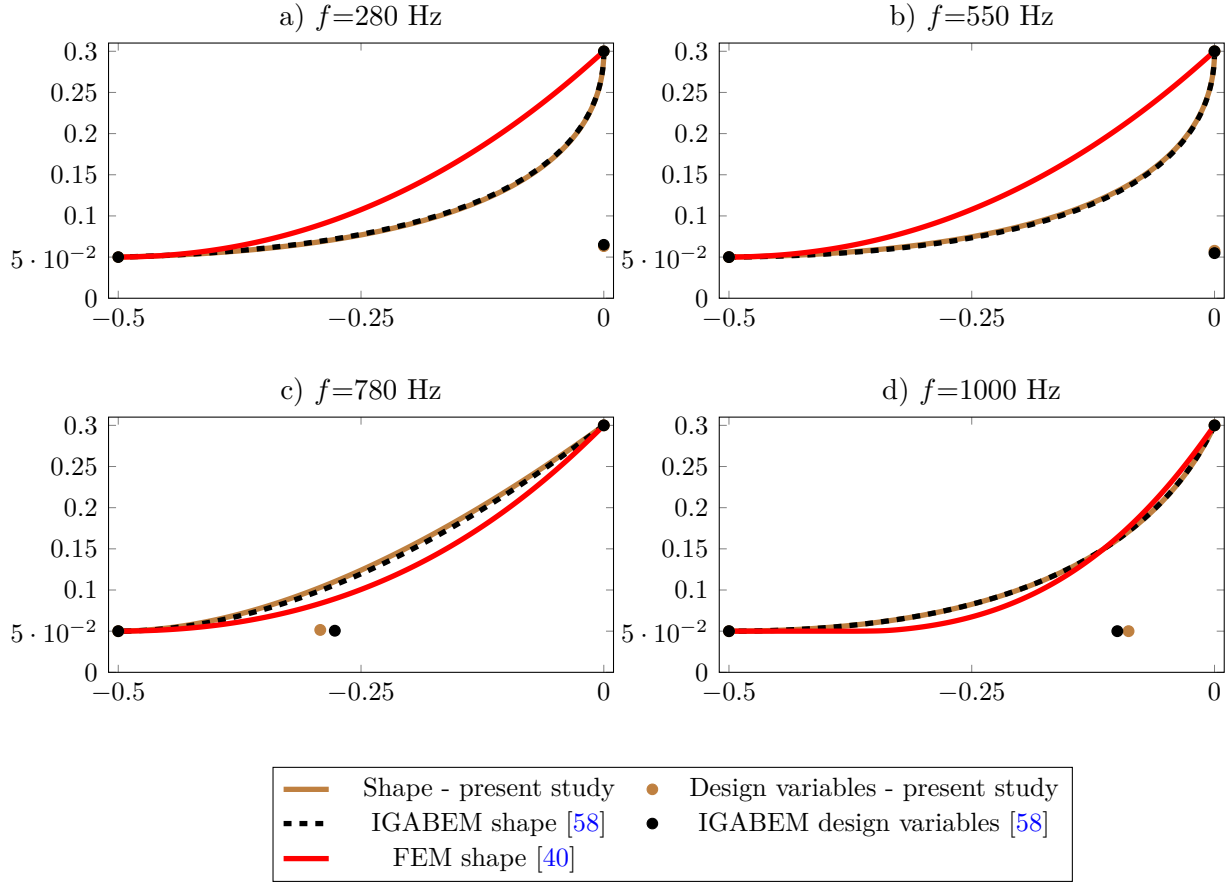

\begin{figure}[ht!]
    \centering
    \begin{tikzpicture}[scale=1]
\begin{semilogyaxis}[cycle list name=exotic,
        xlabel = iterations,
        ylabel = $J(u)$,
	xmin = 0,
	xmax = 32,
	ymin = 1e-4,
	ymax = 1.0,
	legend style={at={(1.55,0.6)},anchor=south east}
    ]

 \addplot[blue, solid, smooth, line width=1.5pt, mark=*] table [x expr = \thisrowno{0}, y expr = \thisrowno{1}]{Sections/Shape-optimization-examples/Horn/Figures/280-2DV/280-history.txt};
    
  \addlegendentry{$f$ = 280 Hz};
  
 \addplot[red, solid, smooth, line width=1.5pt, mark=*] table [x expr = \thisrowno{0}, y expr = \thisrowno{1}] {Sections/Shape-optimization-examples/Horn/Figures/550-2DV/550-history.txt};
    
  \addlegendentry{$f$ = 550 Hz};
  
\addplot[brown, solid, smooth, line width=1.5pt, mark=*] table [x expr = \thisrowno{0}, y expr = \thisrowno{1}] {Sections/Shape-optimization-examples/Horn/Figures/780-2DV/780-history.txt};
    
  \addlegendentry{$f$ = 780 Hz};
  
\addplot[cyan, solid, smooth, line width=1.5pt, mark=*] table [x expr = \thisrowno{0}, y expr = \thisrowno{1}] {Sections/Shape-optimization-examples/Horn/Figures/1000-2DV/1000-history.txt};
    
  \addlegendentry{$f$ = 1000 Hz}; 
  
\end{semilogyaxis}
\end{tikzpicture}
    \caption{Convergence of the objective function $J(u) = \mathcal{R}$ in terms of number of iterations for the horn optimization problem with 1 control point - 2 design variables.}
    \label{fig:obj_fx_number_vs_iter_horn_1ctp}
\end{figure}

\begin{figure}[!ht]
	\centering
	\begin{tikzpicture}[scale=1]
\begin{semilogyaxis}[cycle list name=exotic,
        xlabel= frequency (Hz),
        ylabel= $\mathcal{R}$,
	xmin = 1,
	xmax = 1050,
	ymin = 0.0004,
	ymax = 1.0,
	legend style={at={(2.15,0.15)},anchor=south east}
    ]

      \addplot [black, solid, line width=1.5pt]
    table [x expr = \thisrowno{0}, y=y] {Sections/Shape-optimization-examples/Horn/Figures/initial-shapes-spetrum/IGA.txt};
    
   \addlegendentry{Initial shape - present study};

        \addplot [blue, solid, line width=1.5pt]
        table [x expr = \thisrowno{0}, y expr =  \thisrowno{1}] {Sections/Shape-optimization-examples/Horn/Figures/280-2DV/280-spectra-GIFT.txt};
    
   \addlegendentry{$f$=280 Hz - present study};
    
    \addplot [red, solid, line width=1.5pt]
    table [x expr = \thisrowno{0}, y expr =  \thisrowno{1}] {Sections/Shape-optimization-examples/Horn/Figures/550-2DV/550-spectra-GIFT.txt};
    
    \addlegendentry{$f$=550 Hz - present study};
    
    \addplot [brown, solid, line width=1.5pt]
    table [x expr = \thisrowno{0}, y=y] {Sections/Shape-optimization-examples/Horn/Figures/780-2DV/780-spectra-GIFT.txt};
    
    \addlegendentry{$f$=780 Hz - present study};
    
    \addplot [cyan, solid,line width=1.5pt]
    table [x expr = \thisrowno{0}, y=y] {Sections/Shape-optimization-examples/Horn/Figures/1000-2DV/1000-spectra-GIFT.txt};
    
    \addlegendentry{$f$=1000 Hz - present study};

    \addplot [blue, dashed, line width=1.5pt]
    table [x expr = \thisrowno{0}, y=y] {Sections/Shape-optimization-examples/Horn/Figures/280-2DV/280-spectra-IGABEM.txt};
    
   \addlegendentry{$f$=280 Hz - IGABEM \cite{MOSTAFASHAABAN2020156}};
    
    \addplot [red, dashed, line width=1.5pt]
    table [x expr = \thisrowno{0}, y=y] {Sections/Shape-optimization-examples/Horn/Figures/550-2DV/550-spectra-IGABEM.txt};
    
    \addlegendentry{$f$=550 Hz - IGABEM \cite{MOSTAFASHAABAN2020156}};
    
    \addplot [brown, dashed, line width=1.5pt]
    table [x expr = \thisrowno{0}, y=y] {Sections/Shape-optimization-examples/Horn/Figures/780-2DV/780-spectra-IGABEM.txt};
    
    \addlegendentry{$f$=780 Hz - IGABEM \cite{MOSTAFASHAABAN2020156}};
    
    \addplot [cyan, smooth, dashed, line width=1.5pt]
    table [x expr = \thisrowno{0}, y=y] {Sections/Shape-optimization-examples/Horn/Figures/1000-2DV/1000-spectra-IGABEM.txt};
    
    \addlegendentry{$f$=1000 Hz - IGABEM \cite{MOSTAFASHAABAN2020156}}; 
\end{semilogyaxis}
\end{tikzpicture}
\caption{Reflection spectra for the optimized horn shapes using 1 control point - 2 design variables.}
\label{fig:2DV-all-spectra}
\end{figure}


In the next study, 2 control points (4 design variables) are studied in the optimization process. The sound speed is again $c = 345$ $\mathrm{m/s}$ and the horn is optimized for four different single frequencies $f = 280$, $400$, $550$ and $780$ $\mathrm{Hz}$. Figure \ref{fig:4DV-all-shape} shows the optimized shapes for all the comparative methods. Figure \ref{fig:obj_fx_number_vs_iter_horn_2ctp} shows the convergence plot for each optimized shape, while Figure \ref{fig:4DV-all-spectra} presents the reflection spectra plot for each shape. Overall, the results are in agreement with the references for the first three frequencies, on the other hand, for 
$f = 780 \mathrm{Hz}$, the obtained results are quite different than the references. Still, Figure \ref{fig:4DV-all-spectra} shows improvement in the obtained result versus the initial shape. 

\begin{figure}[!ht]
	\centering
	\begin{tikzpicture}[scale=1]
    \pgfplotsset{footnotesize,samples=10}
    \begin{groupplot}[group style = {group size = 2 by 2, horizontal sep = 40pt, vertical sep = 45pt},
    width = 6.0cm, height = 5.0cm]
    
    \nextgroupplot[ title = {a) $f$=280 Hz},
                     legend style = {
            column sep = 10pt, legend columns = 2},
            legend to name = grouplegendshape4,
             width=0.5 \textwidth,
             height=0.3\textwidth,
	         xmin = 0.49 - 1,
	         xmax = 1.01 - 1,
        	ymin = 0.0,
	        ymax = 0.31,
	        xtick = {-0.5, -0.25, 0},
        	ytick  = {0, 0.05, 0.1, 0.15, 0.2, 0.25, 0.3},
	]
	
        \addplot [color=brown,
              solid,
              smooth,
              line width=2.0pt]
    table [x expr = \thisrowno{0}, y expr = \thisrowno{1}] {Sections/Shape-optimization-examples/Horn/Figures/280-4DV/shape_280_GIFT_4CV.txt};
  \addlegendentry{\textcolor{black}{Shape - present study}};
  
         \addplot [color= brown,
               only marks,
              mark=*,
              mark options={solid}]
    table [x expr = \thisrowno{0} , y expr = \thisrowno{1}] {Sections/Shape-optimization-examples/Horn/Figures/280-4DV/280-GIFT-CTP.txt};
  \addlegendentry{\textcolor{black}{Design variables - present study}};  
    
                \addplot [dashed,
              smooth,
              line width=2.0pt]
    table [x expr = \thisrowno{0} - 1, y expr = \thisrowno{1}] {Sections/Shape-optimization-examples/Horn/Figures/280-4DV/280-IGABEM-shape.txt};

    \addlegendentry{\textcolor{black}{IGABEM shape \cite{MOSTAFASHAABAN2020156}}};
    
        \addplot [color=black,
               only marks,
              mark=*,
              mark options={solid}]
    table [x expr = \thisrowno{0} - 1, y expr = \thisrowno{1}] {Sections/Shape-optimization-examples/Horn/Figures/280-4DV/280-IGABEM-cpt.txt};
    
\addlegendentry{\textcolor{black}{IGABEM design variables \cite{MOSTAFASHAABAN2020156}}};

 \addplot [color=red,
              solid,
              smooth,
              line width=2.0pt]
    table [x expr = \thisrowno{0} - 1, y expr = \thisrowno{1}] {Sections/Shape-optimization-examples/Horn/Figures/280-4DV/280-FEM-shape.txt};
    
    \addlegendentry{\textcolor{black}{FEM shape \cite{BARBIERI2013356} }};

\nextgroupplot[ title = {b) $f$=400 Hz}, 
             width=0.5\textwidth,
             height=0.3\textwidth,
	         xmin = 0.49 - 1,
	         xmax = 1.01 - 1,
        	ymin = 0.0,
	        ymax = 0.31,
	        xtick = {-0.5, -0.25, 0.0},
        	ytick  = {0, 0.05, 0.1, 0.15, 0.2, 0.25, 0.3},
	]
	
\addplot [color=brown,
              solid,
              smooth,
              line width=2.0pt]
    table [x expr = \thisrowno{0}, y expr = \thisrowno{1}] {Sections/Shape-optimization-examples/Horn/Figures/400-4DV/shape_400_GIFT_4CV.txt};

         \addplot [color= brown,
               only marks,
              line width=0.5 0pt,
              mark=*,
              mark options={solid}]
    table [x expr = \thisrowno{0}, y expr = \thisrowno{1}] {Sections/Shape-optimization-examples/Horn/Figures/400-4DV/400-GIFT-CTP.txt};

                \addplot [color=black,
              dashed,
              smooth,
              line width=2.0pt]
    table [x expr = \thisrowno{0} - 1, y expr = \thisrowno{1}] {Sections/Shape-optimization-examples/Horn/Figures/400-4DV/400-IGABEM-shape.txt};
    
         \addplot [color= black,
               only marks,
              line width=0.5 0pt,
              mark=*,
              mark options={solid}]
    table [x expr = \thisrowno{0} - 1, y expr = \thisrowno{1}] {Sections/Shape-optimization-examples/Horn/Figures/400-4DV/400-IGABEM-cpt.txt};

    \addplot [color=red,
              solid,
              smooth,
              line width=2.0pt]
    table [x expr = \thisrowno{0} - 1, y expr = \thisrowno{1}] {Sections/Shape-optimization-examples/Horn/Figures/400-4DV/400-FEM-shape.txt};

    \nextgroupplot[ title = {c) $f$=550 Hz}, 
             width=0.5 \textwidth,
             height=0.3\textwidth,
	         xmin = 0.49-1,
	         xmax = 1.01-1,
        	ymin = 0.0,
	        ymax = 0.31,
	        xtick = {-0.5, -0.25, 0},
        	ytick  = {0, 0.05, 0.1, 0.15, 0.2, 0.25, 0.3},
	]
	
	        \addplot [color=brown,
              solid,
              smooth,
              line width=2.0pt]
    table [x expr = \thisrowno{0}, y expr = \thisrowno{1}] {Sections/Shape-optimization-examples/Horn/Figures/550-4DV/shape_550_GIFT_4CV.txt};
    
         \addplot [color= brown,
               only marks,
              line width=0.5 0pt,
              mark=*,
              mark options={solid}]
    table [x expr = \thisrowno{0}, y expr = \thisrowno{1}] {Sections/Shape-optimization-examples/Horn/Figures/550-4DV/550-GIFT-CTP.txt};

                \addplot [color=black,
              dashed,
              smooth,
              line width=2.0pt]
    table [x expr = \thisrowno{0}-1, y expr = \thisrowno{1}] {Sections/Shape-optimization-examples/Horn/Figures/550-4DV/550-IGABEM-shape.txt};
    
         \addplot [color= black,
               only marks,
              line width=0.5 0pt,
              mark=*,
              mark options={solid}]
    table [x expr = \thisrowno{0}-1, y expr = \thisrowno{1}] {Sections/Shape-optimization-examples/Horn/Figures/550-4DV/550-IGABEM-cpt.txt};

    \addplot [color=red,
              solid,
              smooth,
              line width=2.0pt]
    table [x expr = \thisrowno{0}-1, y expr = \thisrowno{1}] {Sections/Shape-optimization-examples/Horn/Figures/550-4DV/550-FEM-shape.txt};

    \nextgroupplot[ title = {d) $f$=780 Hz}, 
             width=0.5 \textwidth,
             height=0.3\textwidth,
	         xmin = 0.49-1,
	         xmax = 1.01-1,
        	ymin = 0.0,
	        ymax = 0.31,
	        xtick = {-0.5, -0.25, 0},
        	ytick  = {0, 0.05, 0.1, 0.15, 0.2, 0.25, 0.3},
	]
	
	       \addplot [color=brown,
              solid,
              smooth,
              line width=2.0pt]
    table [x expr = \thisrowno{0}, y expr = \thisrowno{1}] {Sections/Shape-optimization-examples/Horn/Figures/780-4DV/shape_780_GIFT_4CV.txt}; 
    
         \addplot [color= brown,
               only marks,
              line width=0.5 0pt,
              mark=*,
              mark options={solid}]
    table [x expr = \thisrowno{0}, y expr = \thisrowno{1}] {Sections/Shape-optimization-examples/Horn/Figures/780-4DV/780-GIFT-CTP.txt};

                \addplot [color=black,
              dashed,
              smooth,
              line width=2.0pt]
    table [x expr = \thisrowno{0}-1, y expr = \thisrowno{1}] {Sections/Shape-optimization-examples/Horn/Figures/780-4DV/780-IGABEM-shape.txt};
    
         \addplot [color= black,
               only marks,
              line width=0.5 0pt,
              mark=*,
              mark options={solid}]
    table [x expr = \thisrowno{0}-1, y expr = \thisrowno{1}] {Sections/Shape-optimization-examples/Horn/Figures/780-4DV/780-IGABEM-cpt.txt};

     \addplot [color=red,
              solid,
              smooth,
              line width=2.0pt]
    table [x expr = \thisrowno{0}-1, y expr = \thisrowno{1}] {Sections/Shape-optimization-examples/Horn/Figures/780-4DV/780-FEM-shape.txt};

    \end{groupplot}
    \node at ($(7.5 cm ,-7 cm)$)  {\ref{grouplegendshape4}}; 
\end{tikzpicture}
\caption{Optimized horn shapes using 2 control points - 4 design variables.}
\label{fig:4DV-all-shape}
\end{figure}

\begin{figure}[ht!]
    \centering
    \begin{tikzpicture}[scale=1]
\begin{semilogyaxis}[cycle list name=exotic,
        xlabel = iterations,
        ylabel = $J(u)$,
	xmin = 1,
	xmax = 45,
	ymin = 1e-4,
	ymax = 0.5,
	legend style={at={(1.5,0.6)},anchor=south east}
    ]

     \addplot[blue, solid, line width=1.5pt, mark=*]  table [x expr = \thisrowno{0}, y expr = \thisrowno{1}] {Sections/Shape-optimization-examples/Horn/Figures/280-4DV/280-history.txt};
    
  \addlegendentry{$f$ = 280 Hz};
  
     \addplot[red, solid, line width=1.5pt, mark=*]  table [x expr = \thisrowno{0}, y expr = \thisrowno{1}] {Sections/Shape-optimization-examples/Horn/Figures/400-4DV/400-history.txt};
    
  \addlegendentry{$f$ = 400 Hz};
  
        \addplot[brown, solid, line width=1.5pt, mark=*]  table [x expr = \thisrowno{0}, y expr = \thisrowno{1}] {Sections/Shape-optimization-examples/Horn/Figures/550-4DV/550-history.txt};
    
  \addlegendentry{$f$ = 550 Hz};
  
          \addplot[cyan, solid, line width=1.5pt, mark=*]  table [x expr = \thisrowno{0}, y expr = \thisrowno{1}] {Sections/Shape-optimization-examples/Horn/Figures/780-4DV/780-history.txt};
    
  \addlegendentry{$f$ = 780 Hz}; 
\end{semilogyaxis}
\end{tikzpicture}
    \caption{Convergence of the objective function $J(u) = \mathcal{R}$ in terms of number of iterations for the horn optimization problem with 2 control points - 4 design variables.}
    \label{fig:obj_fx_number_vs_iter_horn_2ctp}
\end{figure}

\begin{figure}[!ht]
	\centering
	\begin{tikzpicture}[scale=1]
\begin{semilogyaxis}[cycle list name=exotic,
        xlabel= frequency (Hz),
        ylabel= $\mathcal{R}$,
	xmin = 1,
	xmax = 1050,
	ymin = 0.0001,
	ymax = 1.0,
	legend style={at={(1.6,0.5)},anchor=south east}
    ]

      \addplot [black, solid, smooth, line width=1.5pt]
    table [x expr = \thisrowno{0}, y=y] {Sections/Shape-optimization-examples/Horn/Figures/initial-shapes-spetrum/IGA.txt};
    
   \addlegendentry{Initial shape};

        \addplot [blue, solid, smooth, line width=1.5pt]
        table [x expr = \thisrowno{0}, y expr =  \thisrowno{1}] {Sections/Shape-optimization-examples/Horn/Figures/280-4DV/spectra_280_optimized_2CV.txt};
    
   \addlegendentry{$f$=280 Hz};
    
    \addplot [red, solid, smooth, line width=1.5pt]
    table [x expr = \thisrowno{0}, y expr =  \thisrowno{1}] {Sections/Shape-optimization-examples/Horn/Figures/400-4DV/spectra_400_optimized_2CV.txt};
    
    \addlegendentry{$f$=400 Hz};
    
    \addplot [brown, solid, smooth, line width=1.5pt]
    table [x expr = \thisrowno{0}, y=y] {Sections/Shape-optimization-examples/Horn/Figures/550-4DV/spectra_550_optimized_2CV.txt};
    
    \addlegendentry{$f$=550 Hz};
    
    \addplot [cyan, solid, smooth,line width=1.5pt]
    table [x expr = \thisrowno{0}, y=y] {Sections/Shape-optimization-examples/Horn/Figures/780-4DV/spectra_780_optimized_2CV.txt};
    
    \addlegendentry{$f$=780 Hz};

\end{semilogyaxis}
\end{tikzpicture}
\caption{Reflection spectra for the optimized horn shapes using 2 control points - 4 design variables.}
\label{fig:4DV-all-spectra}
\end{figure}


\subsubsection{Performance of the adaptive shape optimization algorithm}

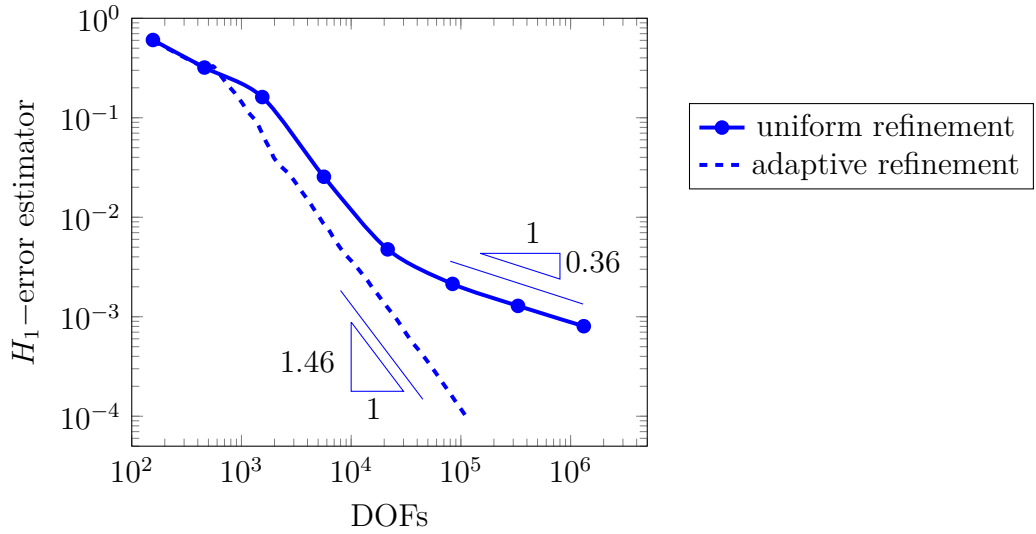
\begin{figure}[ht!]
    \centering
    \begin{tikzpicture}[scale=1]
\begin{loglogaxis}[cycle list name=exotic,
        xlabel = DOFs,
        ylabel = $H_{1}-$error estimator,
	xmin = 10^2,
	xmax = 5*10^6,
	ymin = 5e-5,
	ymax = 1.0,
	legend style={at={(1.75,0.6)},anchor=south east}
    ]

    

    

      \addplot [blue,mark = *, solid, smooth, line width=1.5pt]
    table [x expr = \thisrowno{0}, y=y] {Sections/Shape-optimization-examples/Horn/Figures/1000-2DV/1000-GIFT-uniform.txt};
    
  \addlegendentry{uniform refinement};

        \addplot [blue, dashed, smooth, line width=1.5pt]
        table [x expr = \thisrowno{0}, y expr =  \thisrowno{1}] {Sections/Shape-optimization-examples/Horn/Figures/1000-2DV/1000-GIFT-adaptive.txt};
    
   \addlegendentry{adaptive refinement};

    \addplot [color=blue,solid,forget plot]
  table[row sep=crcr]{%
8000	0.001835415 \\
45000	0.000147943 \\
};

    \addplot [color=blue,solid,forget plot]
  table[row sep=crcr]{%
10000	0.000883798\\
30000	0.000178129\\
10000	0.000178129\\
10000	0.000883798\\
};

    \addplot [color=blue,solid,forget plot]
  table[row sep=crcr]{%
80000	0.003614115 \\
1300000	0.00134137\\
};

    \addplot [color=blue,solid,forget plot]
  table[row sep=crcr]{%
150000	0.004335532\\
800000	0.002391098\\
800000	0.004335532\\
150000	0.004335532\\
};
\node[below, align=center, inner sep=0mm, text=black]
at (axis cs:1600000,4.5e-03,0) {$0.36$};
\node[left, align=right, inner sep=0mm, text=black]
at (axis cs:540000,7e-03,0) {$1$};

\node[below, align=center, inner sep=0mm, text=black]
at (axis cs:16000,1.5e-04,0) {$1$};
\node[left, align=right, inner sep=0mm, text=black]
at (axis cs:7200,3.5e-04,0) {$1.46$};

\end{loglogaxis}
\end{tikzpicture}
    \caption{Horn problem: convergence of the $H_{1}-$error estimator in terms of DOFs.}
    \label{fig:horn_comparison_unif_adaptive_error_est}
\end{figure}

\begin{figure}[ht!]
    \centering
    \begin{tikzpicture}[scale=1]
\begin{semilogxaxis}[cycle list name=exotic,
        xlabel= DOFs,
        ylabel= $J(u)  $,
	xmin = 10^2,
	xmax = 10^5,
	ymin = 0.07,
	ymax = 0.48,
	legend style={at={(1.95,0.5)},anchor=south east}
    ]

    

    
    

       \addplot [blue, solid , mark=*, line width=1.5pt]
    table [x expr = \thisrowno{0}, y expr =  \thisrowno{2}] {Sections/Shape-optimization-examples/Horn/Figures/1000-2DV/1000-GIFT-uniform.txt};
    
   \addlegendentry{uniform refinement};

        \addplot [blue, dashed, line width=1.5pt]
        table [x expr = \thisrowno{0}, y expr =  \thisrowno{2}] {Sections/Shape-optimization-examples/Horn/Figures/1000-2DV/1000-GIFT-adaptive.txt};
    
   \addlegendentry{adaptive refinement};   
   
       \addplot [red,dashed, line width = 1.pt]
    table [x expr = \thisrowno{0}, y=y] {Sections/Shape-optimization-examples/Horn/Figures/initial_convergence_study/convergence_1000_asymp.txt};
    
    \addlegendentry{$J(u) = 0.10281316$}
\end{semilogxaxis}
\end{tikzpicture}
    \caption{Horn problem: convergence of the objective function $J(u) = \mathcal{R}$ in terms of DOFs.}
    \label{fig:horn_comparison_unif_adaptive_cost_function}
\end{figure}
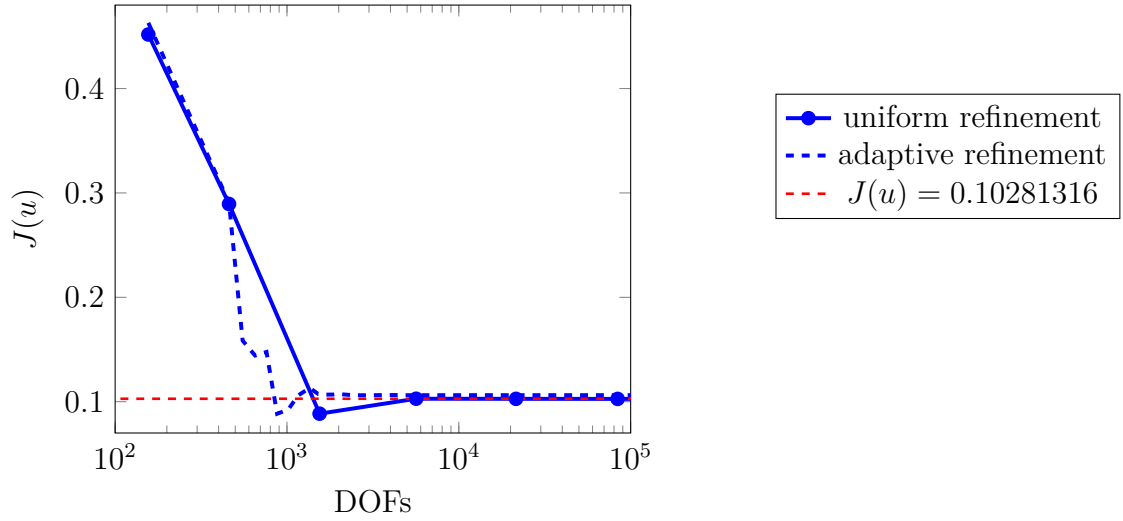

To exploit the capabilities of PHT-splines and the GIFT formulation, an adaptive optimization algorithm, Algorithm \ref{alg:adaptive_optimisation}, is studied in this section. First, Figures \ref{fig:horn_comparison_unif_adaptive_error_est} and \ref{fig:horn_comparison_unif_adaptive_cost_function} show the comparison between uniform and recovery-based adaptive refinement for the non-optimized geometry for $f=1000$ $\mathrm{Hz}$. As it can be seen in the figures, adaptive refinement overcomes uniform refinement in both error estimator and reaching the asymptotic value of the objective function.

\begin{figure}[ht!]
    \centering
    \begin{subfigure}[b]{0.48\textwidth}
        \includegraphics[width=\textwidth]{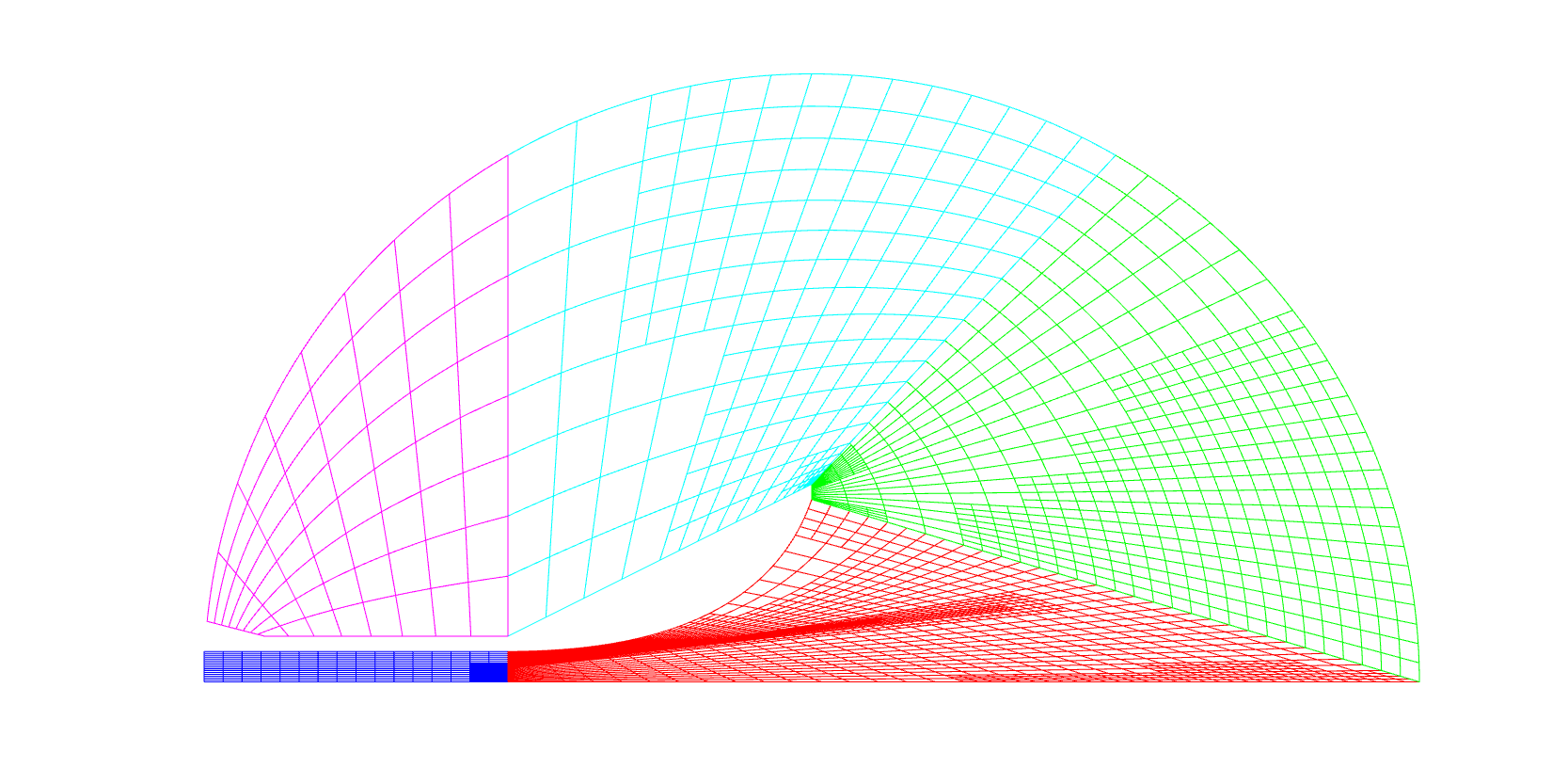}
        \caption{Adaptive mesh \vspace{0.3cm}}
    \end{subfigure}
    ~ 
    \begin{subfigure}[b]{0.48\textwidth}
        \includegraphics[width=\textwidth]{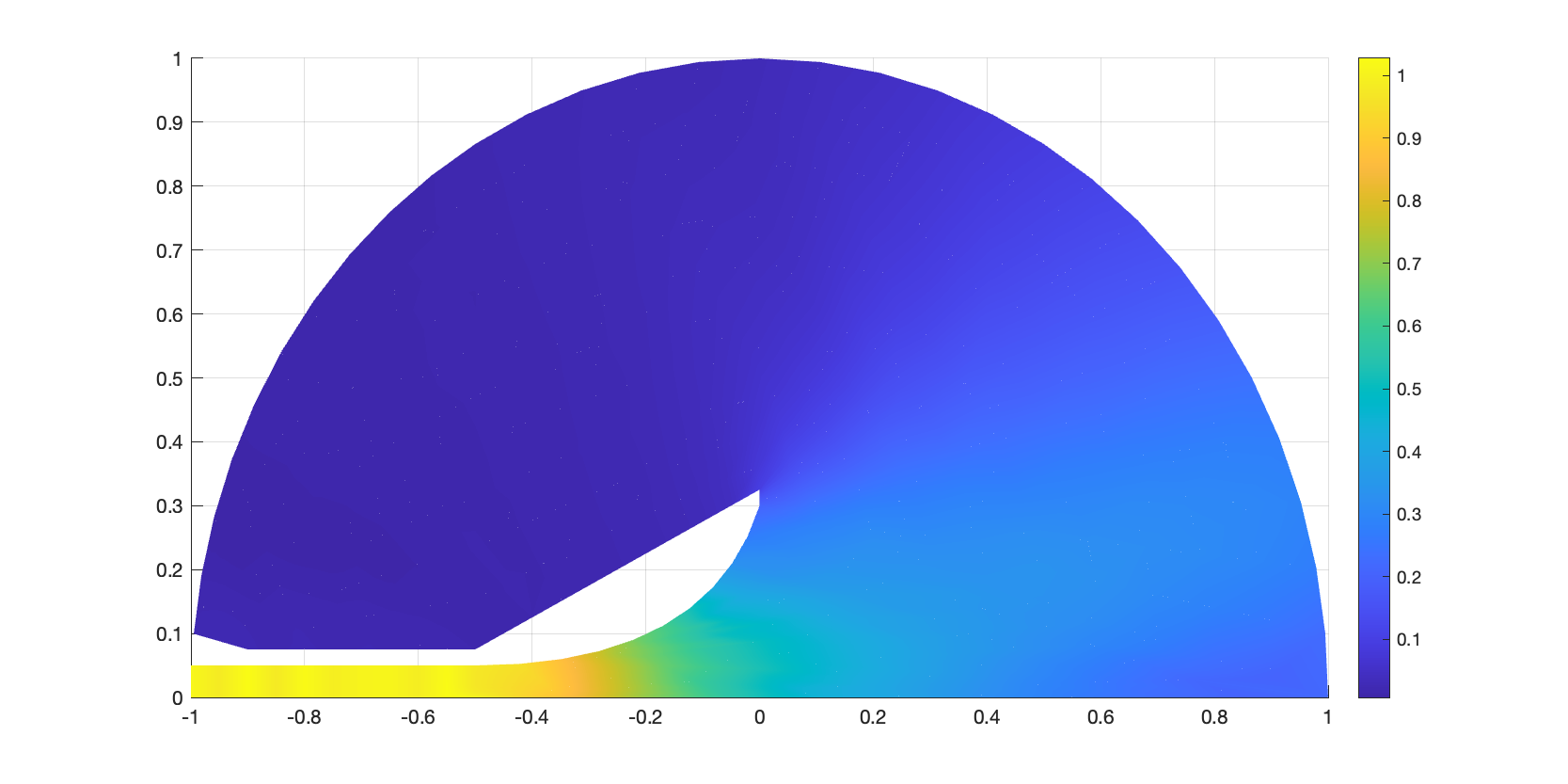}
       \caption{Optimized sound pressure (absolute value) for $f=1000$ Hz}
    \end{subfigure}    
 \caption{Horn example: adaptive mesh for the optimized geometry and the domain values. $f$=1000 Hz and 2 design variables are employed.}  \label{fig:horn_example_adaptive_mesh_adaptive_sol_optimized} 
\end{figure}

Table \ref{tab:horn_example_uniform_adaptive_comparison_table} shows the comparison between DOFs, computational time (s) and the optimal objective function for two different target tolerances, using $f=1000$ $\mathrm{Hz}$ and one control point (2 design variables). As it can be seen in the table, the adaptive optimization overcomes the uniform optimization in both, number of DOFs and computational time.
Figure \ref{fig:horn_example_adaptive_mesh_adaptive_sol_optimized} shows the adaptive optimized mesh and the optimized sound pressure obtained for $f=1000$ Hz and 1 control point. 

\begin{table}[ht!]
\centering
\begin{tabular}{lrrrr}
\multicolumn{1}{c}{} & \multicolumn{4}{c}{$\varepsilon_{\text{loop}} = 10^{-2}$}                                                                              \\ \cline{2-5} 
                     & \multicolumn{1}{l}{DOFs} & \multicolumn{1}{l}{ Time (s)} & \multicolumn{1}{l}{Objective function} & \multicolumn{1}{l}{Best solution}\\ \hline
Uniform              & 21516   & 171.225                 & $4.449\times 10^{-3}$ & [-0.0500, -0.0864]            \\  
Adaptive             & 2556    & 36.074                 & $4.953\times 10^{-3}$   & [-0.0500, -0.0902]          \\ \hline
Difference (\%)      & 88.12                       & 63.49                               & -11.33 & 3.81    \\ \hline
IGABEM \cite{MOSTAFASHAABAN2020156}      & -                      & -                               & $3.330\times 10^{-3}$  \\        
                     &         &                        &                       \\
\multicolumn{1}{c}{} & \multicolumn{4}{c}{$\varepsilon_{\text{loop}} = 10^{-3}$}                                                                              \\ \cline{2-5} 
                     & \multicolumn{1}{l}{DOFs} & \multicolumn{1}{l}{ Time (s)} & \multicolumn{1}{l}{ Objective function} & \multicolumn{1}{l}{Best solution} \\ \hline
Uniform              & 331788 & 5768.166             & $4.485\times10^{-3}$ & [-0.0500, -0.0882]             \\  
Adaptive             & 11348   & 186.901                & $4.485 \times 10^{-3}$ & [-0.0500, -0.0875] \\ \hline
Difference (\%)      & 96.58                       & 94.76                                & 0  & 0.69   
 \\ \hline
IGABEM \cite{MOSTAFASHAABAN2020156}      & -                      & -                               & $3.330\times 10^{-3}$  \\  
\end{tabular}
\caption{Horn example: uniform and recovery-based adaptive refinement results. $f$=1000 Hz and 2 design variables are employed.}
\label{tab:horn_example_uniform_adaptive_comparison_table}
\end{table}

Table \ref{tab:horn_example_2_uniform_adaptive_comparison_table} shows the uniform and adaptive results for 2 control points (4 design variables) and $f=780$ Hz. Similar conclusions can be drawn from this study.

\begin{table}[ht!]
\centering
\begin{tabular}{lrrrr}
\multicolumn{1}{c}{} & \multicolumn{3}{c}{$\varepsilon_{\text{loop}} = 10^{-3}$}                     \\ \cline{2-5} 
                     & DOFs    & Time (s) &  Objective function & Best solution\\ \hline
Uniform              & 331788 & 524.238              & $4.1562\times 10^{-4}$             &  [-0.050	-0.364	-0.108	-0.014] \\ \hline
Adaptive             & 12484   & 424.371                & $4.1999\times 10^{-4}$ & [-0.056	-0.341	-0.281	-0.001]         \\ \hline
Difference (\%)      & 96.24                       & 91.28                                & -0.968   
 \\ \hline
IGABEM \cite{MOSTAFASHAABAN2020156}      & -                      & -                               & $1.61\times 10^{-4}$  \\  
\end{tabular}
\caption{Horn example: uniform and recovery-based adaptive refinement results. $f$=780 Hz and 4 design variables are employed. $\varepsilon_{0} = 5\times 10^{-2}$}
\label{tab:horn_example_2_uniform_adaptive_comparison_table}
\end{table}




\cleardoublepage
\clearpage
\subsection{Sound barrier}
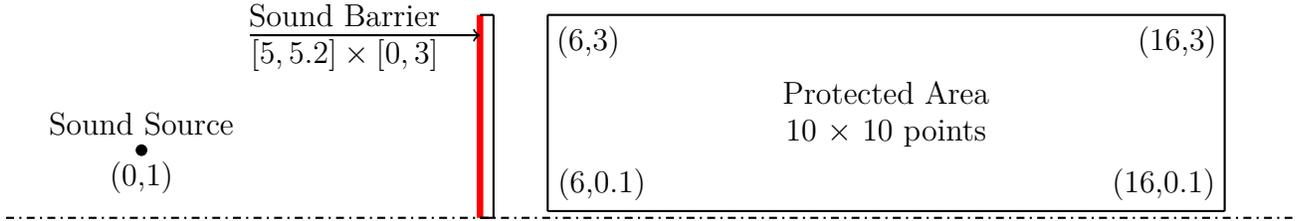
\begin{figure}[!ht]
\centering
\begin{tikzpicture}[scale = 0.9, classical/.style={thick}]
\draw[thick] (6,0.1)--(16,0.1) ;
\draw[thick] (6,0.1)--(6,3) ;
\draw[thick] (6,3)--(16,3) ;
\draw[thick] (16,0.1)--(16,3) ;

\draw[thick , red, line width=2.5pt] (5,0)--(5,3) ;
\draw[thick] (5,0)--(5.2,0) ;
\draw[thick] (5.2,0)--(5.2,3) ;
\draw[thick] (5,3)--(5.2,3) ;

 \node[mark size=2pt,color=black] at (0,1) {\pgfuseplotmark{*}};

\draw [thick,dash dot](-2,0) -- (17,0.0);

\node[] at (3,3) {Sound Barrier};
\node[] at (3,2.4) {$[5,5.2] \times [0,3]$};

\node[] at (0,1.4) {Sound Source};
\node[] at (0,0.6) {(0,1)};
\node[] at (11,1.85) {Protected Area};
\node[] at (11,1.25) {10 $\times$ 10 points};
\node[] at (6.8,0.5) {(6,0.1)};
\node[] at (15.1,0.5) {(16,0.1)};
\node[] at (6.6,2.6) {(6,3)};
\node[] at (15.3,2.6) {(16,3)};

\draw [thick] [->] plot coordinates { (1.6,2.7) (5,2.7)  };

\end{tikzpicture}
\caption{The vertical noise barrier problem.} \label{fig:wall_example_geo_measuements}
\end{figure}

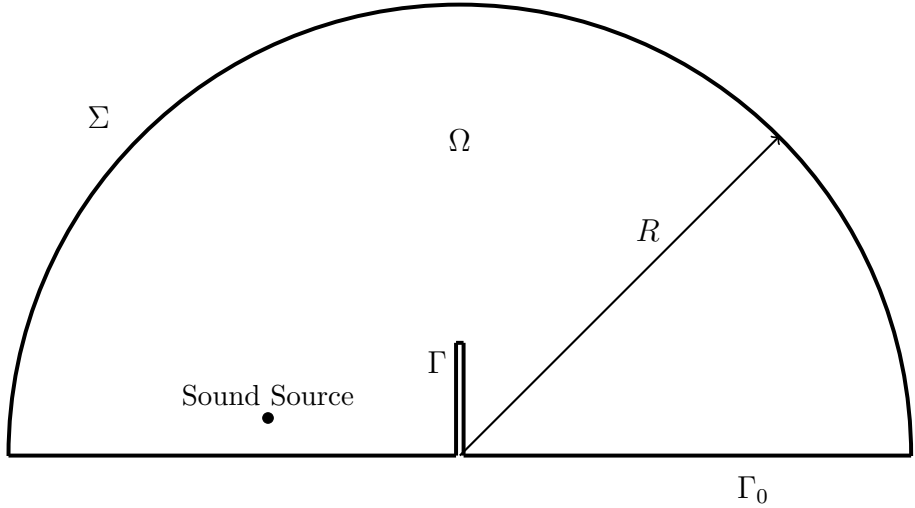
\begin{figure}[!ht]
	\centering
\centering
\begin{tikzpicture}[scale = 0.5, classical/.style={thick}]

\begin{scope}
    \clip  (-12.1,0) rectangle (12.1,12.05);
    \draw  [thick, line width=1.5pt] (0,0) circle(12.0);
     \draw [thick, line width=1.5pt](-0.1,0) -- (-0.1,3);  
     \draw [thick, line width=1.5pt](0.1,0) -- (0.1,3);  
     \draw [thick, line width=1.5pt](-0.1,3) -- (0.1,3);

 \end{scope}
     \draw [thick,line width=1.5pt](-12,0) -- (-0.1,0);
      \draw [thick,line width=1.5pt](12,0) -- (0.1,0);

\draw [thick,->](0,0) -- (8.5, 8.5);

 \node[mark size=2pt,color=black] at (-5.1,1) {\pgfuseplotmark{*}};

\node[] at (-5.1,1.6) {\small  {Sound Source}};

\node[] at (5,6) {$R$};
\node[] at (-9.6,9) {$\Sigma$};
\node[] at (7.8,-0.9) {$\Gamma_0$};
\node[] at (-0.6,2.5) {$\Gamma$};
\node[] at (0,8.4) {$\Omega$};
 
\end{tikzpicture}
	\caption{Computational domain for the vertical noise barrier problem with the truncation boundary.}
	\label{fig:barrier-bc}
\end{figure}

In the next example, we consider the shape optimization problem for a sound barrier, as shown in Figure \ref{fig:wall_example_geo_measuements}, where the design objective is to minimize the sound pressure in the protected area behind the barrier by changing the shape of the barrier front side (shown by the red line). The barrier occupies domain $[5, 5.2]\times[0, 3]$. A line source is located at $\textbf{x}_0 = (0, 1)$. The truncation domain is a circle of radius $R = 20$, centered at $(5.1, 0)$ with BGT-1 boundary condition prescribed on it. Figure \ref{fig:barrier-bc} shows the truncated domain with the barrier and the sound source. The front side of the barrier is parameterized by NURBS of degree $p=2$ and $N$ equally spaced control points. The end points, $(5,0)$ and $(5, 3)$ are fixed, and the shape is optimized by moving the inner control points along $x-$direction, i.e. by changing their coordinates $x_i$. 

\begin{figure}[ht!]
        \includegraphics[width=\textwidth]{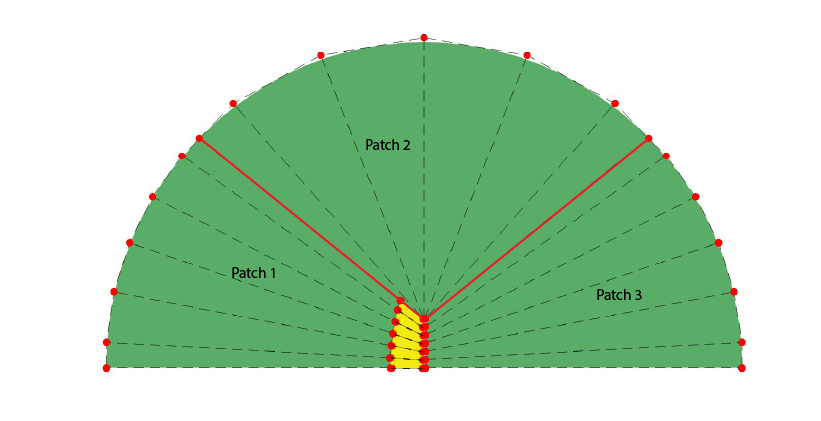}
\caption{Sound barrier optimization problem: NURBS parameterization.} \label{fig:barrier-nurbs}
\end{figure}


The full boundary value problem is formulated as follows:
\begin{equation}
    \begin{split}
        \Delta u + k^2 u &= \delta(\textbf{x} - \textbf{x}_0)
        \, \text{   in   }\Omega\\
        \dfrac{\partial u}{\partial \boldsymbol{n}} + \mathcal{B}u &= 0 \qquad \qquad \text{   on   }\Sigma\\
        \dfrac{\partial u}{\partial \boldsymbol{n}} &= 0 \qquad \qquad \text{   on   }\Gamma_0 \\
        \dfrac{\partial u}{\partial \boldsymbol{n}} &= 0 \qquad \qquad \text{   on   }\Gamma \\
    \end{split}
\end{equation}

where $\Gamma_0 = \{(x, y): y = 0\}$ is the symmetry line. Operator $\mathcal{B}$ represents BGT-1 condition on $\Sigma$. In this example, instead of solving the Helmholtz equation with the source function given by Dirac delta function, 
we decompose the solution into the Green's function (which satisfies Helmholtz equation with Dirac delta function and homogeneous Neumann boundary condition on $\Gamma_0$) and the unknown solution (which satisfies homogeneous Helmholtz equation in the domain and the modified boundary conditions). 

Let us assume that 
\begin{equation}
  u = u_G + \hat{u}  \label{eq:full_u}
\end{equation}
such that
\begin{equation}
    \begin{split}
        \Delta u_G + k^2 u_G &= \delta(\textbf{x} - \textbf{x}_0)
        \text{   in   }\Omega\\
        \dfrac{\partial u_G}{\partial \boldsymbol{n}} &= 0 \text{   on   }\Gamma_0 \\
    \end{split}
\end{equation}
Then $u_G = G(\textbf{x}, \textbf{x}_0)$ is Green's function in a half-plane, given by \cite{Green}
\begin{equation}
    G(\textbf{x}, \textbf{x}_0) = -\dfrac{i}{4}H_0^{(1)}(k|\textbf{x} - \textbf{x}_0|)-\dfrac{i}{4}H_0^{(1)}(k|\textbf{x} - \bar{\textbf{x}}_0|), \label{Green_function}
\end{equation}
where $\bar{\textbf{x}}_0 = (x_0, -y_0)$. Note that, $G(\textbf{x}, \textbf{x}_0)$ satisfies the Sommerfeld radiation condition at infinity. The remaining part of the solution satisfies:
\begin{equation}\label{eq:barrier}
    \begin{split}
        \Delta \hat{u} + k^2 \hat{u} &= 0
        \quad \quad \quad \, \, \text{   in   }\Omega\\
        \dfrac{\partial \hat{u}}{\partial \boldsymbol{n}} &= -\dfrac{\partial u_G}{\partial \boldsymbol{n}} \quad \text{   on   }\Gamma \\
        \dfrac{\partial \hat{u}}{\partial \boldsymbol{n}} &= 0 \quad \quad \quad \, \, \text{   on   }\Gamma_0 \\
        \dfrac{\partial \hat{u}}{\partial \boldsymbol{n}} + \mathcal{B} \hat{u} &= 0 \quad \quad \quad \, \, \text{   on   }\Sigma\\
    \end{split}
\end{equation}
The objective function is defined as
\begin{equation}
    J(u) =  \bar{u} u,
\end{equation}
where sound pressure $u$ (and its conjugate $\bar{u}$) is calculated at a grid of $10\times10$ points in domain $[6, 16]\times[0.1, 3]$.
The full optimization problem is formulated as follows:
\begin{equation}
\begin{split}
    \min\limits_{x_i} & J(u), \text{    subject to eq.}(\ref{eq:barrier}), (\ref{eq:full_u}), (\ref{Green_function})\\
    \text{Such that: } A(u) &\leq 0.6, \\
    x_i &\in [4.9, 5.1],
\end{split}
\end{equation}

where $A(u)$ is the area of the barrier. 

For GIFT implementation, we split the computational domain into three patches, as shown in Figure \ref{fig:barrier-nurbs}. Similarly to the horn example, we reduce the continuity of the geometry parameterization in the radial direction to localize changes in the stiffness matrix. Continuity of the geometry basis is reduced by inserting knot value at $0.1$ twice. Then all changes in the system matrix are localized in the elements adjacent to the left side of the barrier. This area is shown in yellow color in Figure \ref{fig:barrier-nurbs}.

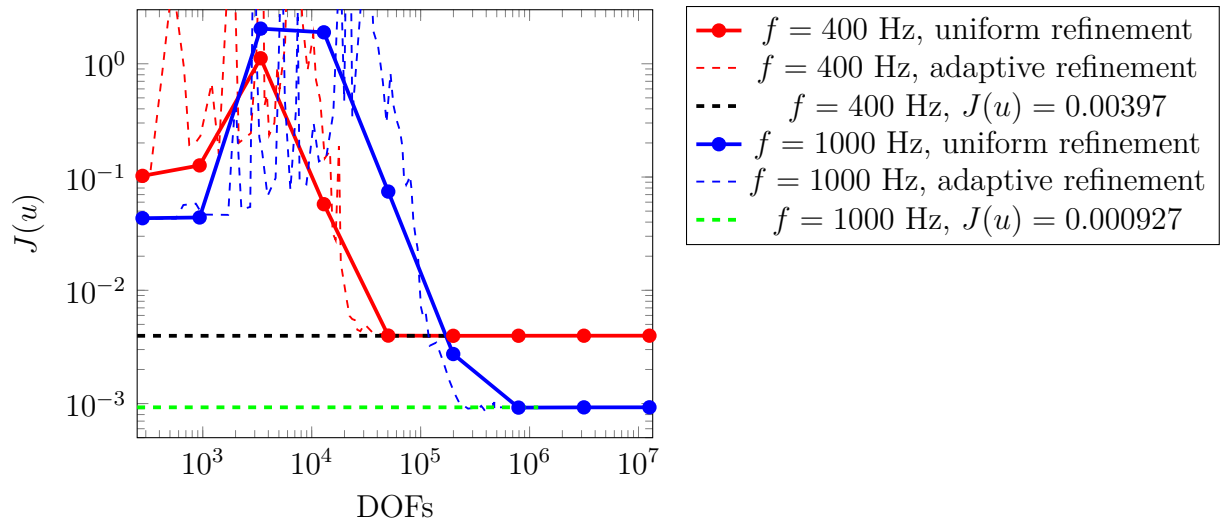
\begin{figure}[!ht]
	\centering
	\begin{tikzpicture}[scale=1]
\begin{loglogaxis}[cycle list name=exotic,
        xlabel = DOFs,
        ylabel = $J(u)$,
	xmin = 250,
	xmax = 13500000,
	ymin = 0.0005,
	ymax = 3,
	legend style={at={(2.1,0.45)},anchor=south east}
    ]
   \addplot [color = red,
             mark = *,
             solid,
             line width = 0.5mm]
    table [x expr = \thisrowno{0}, y expr = \thisrowno{1}] {Sections/Shape-optimization-examples/Sound_barrier/Sound_barrier_Figures/convergence_mesh_barrier_400.txt};
    
    \addlegendentry{$f = 400$ Hz, uniform refinement};

\addplot [color = red,
             dashed,
             line width = 0.25mm]
    table [x expr = \thisrowno{0}, y expr = \thisrowno{2}] {Sections/Shape-optimization-examples/Sound_barrier/Sound_barrier_Figures/convergence_mesh_barrier_400_adaptive_v2.txt};
    
    \addlegendentry{$f = 400$ Hz, adaptive refinement};
    
    \addplot [color = black,
    dashed,
    domain=250:200000,
    line width = 0.5mm] {0.003966089231123};
    
    \addlegendentry{$f = 400$ Hz, $J(u) = 0.00397$};

   \addplot [color = blue,
            mark = *,
             solid,
             line width = 0.5mm]
    table [x expr = \thisrowno{0}, y expr = \thisrowno{1}] {Sections/Shape-optimization-examples/Sound_barrier/Sound_barrier_Figures/convergence_mesh_barrier_1000.txt};
    
    \addlegendentry{$f = 1000$ Hz, uniform refinement};

   \addplot [color = blue,
             dashed,
             line width = 0.25mm]
    table [x expr = \thisrowno{0}, y expr = \thisrowno{2}] {Sections/Shape-optimization-examples/Sound_barrier/Sound_barrier_Figures/convergence_mesh_barrier_1000_adaptive_v2.txt};
    
    \addlegendentry{$f = 1000$ Hz, adaptive refinement};
    
    \addplot [color = green,
    dashed,
    domain=250:1200000,
    line width = 0.5mm] {0.00092696};
    
    \addlegendentry{$f = 1000$ Hz, $J(u) = 0.000927$};
    
\end{loglogaxis}
\end{tikzpicture}
\caption{Sound barrier example: convergence of the objective function $J(u)$ in terms of DOFs.} \label{fig:objective_convergence_barrier}
\end{figure}

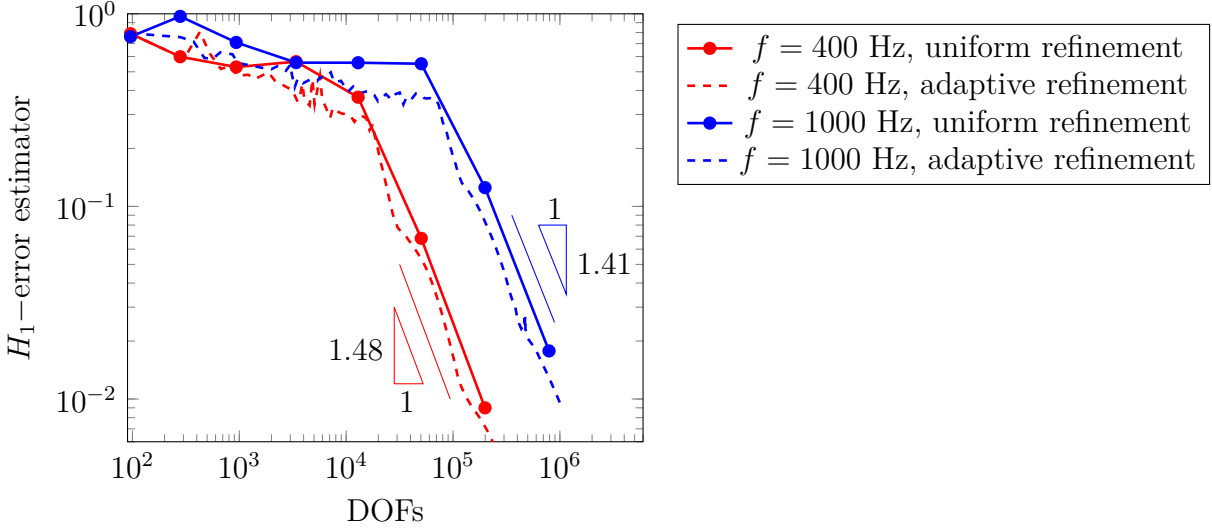
\begin{figure}[!ht]
	\centering
	\begin{tikzpicture}[scale=1]
\begin{loglogaxis}[cycle list name=exotic,
        xlabel = DOFs,
        ylabel = $H_{1}-$error estimator,
	xmin = 90,
	xmax = 6e6,
	ymin = 6e-3,
	ymax = 1,
	legend style={at={(2.1,0.6)},anchor=south east}
    ]
   \addplot [color = red,
             mark = *,
             solid,
             line width=1.0pt,
             mark size=2pt]
    table [x expr = \thisrowno{0}, y expr = \thisrowno{1}] {Sections/Shape-optimization-examples/Sound_barrier/Sound_barrier_Figures/convergence_mesh_barrier_400_v2.txt};
    
    \addlegendentry{$f = 400$ Hz, uniform refinement};
    
    \addplot [color = red,
             dashed,
             line width=1.0pt,
             mark size=1.5pt]
    table [x expr = \thisrowno{0}, y expr = \thisrowno{1}] {Sections/Shape-optimization-examples/Sound_barrier/Sound_barrier_Figures/convergence_mesh_barrier_400_adaptive_v2.txt};
    
    \addlegendentry{$f = 400$ Hz, adaptive refinement};

   \addplot [color = blue,
             mark = *,
             solid,
             line width=1.0pt,
             mark size=2pt]
    table [x expr = \thisrowno{0}, y expr = \thisrowno{1}] {Sections/Shape-optimization-examples/Sound_barrier/Sound_barrier_Figures/convergence_mesh_barrier_1000_v2.txt};
    
    \addlegendentry{$f = 1000$ Hz, uniform refinement};

   \addplot [color = blue,
             dashed,
             line width=1.0pt,
             mark size=1.5pt]
    table [x expr = \thisrowno{0}, y expr = \thisrowno{1}] {Sections/Shape-optimization-examples/Sound_barrier/Sound_barrier_Figures/convergence_mesh_barrier_1000_adaptive_v2.txt};
    
    \addlegendentry{$f = 1000$ Hz, adaptive refinement};

    \addplot [color=red,solid,forget plot]
  table[row sep=crcr]{%
31623	0.05 \\
94135	0.01 \\
};

    \addplot [color=red,solid,forget plot]
  table[row sep=crcr]{%
28184	0.03\\
52481	0.012\\
28184	0.012\\
28184	0.03\\
};

    \addplot [color=blue,solid,forget plot]
  table[row sep=crcr]{%
354813		0.09 \\
891251  	0.025 \\
};

    \addplot [color=blue,solid,forget plot]
  table[row sep=crcr]{%
630957	0.08\\
1148154	0.035\\
1148154	0.08\\
630957	0.08\\
};
\node[below, align=center, inner sep=0mm, text=black]
at (axis cs:0.9e6,0.11,0) {$1$};
\node[left, align=right, inner sep=0mm, text=black]
at (axis cs:4.8e6,0.05,0) {$1.41$};

\node[below, align=center, inner sep=0mm, text=black]
at (axis cs:37000,11e-03,0) {$1$};
\node[left, align=right, inner sep=0mm, text=black]
at (axis cs:22500,1.8e-02,0) {$1.48$};

\end{loglogaxis}
\end{tikzpicture}
\caption{Sound barrier example: convergence of the $H_{1}-$error estimator in terms of DOFs.} \label{fig:error_convergence_barrier}
\end{figure}

We start by conducting the convergence study for two values of frequency: $f = 400$ Hz and $f = 1000$ Hz. The results are shown in Figure \ref{fig:objective_convergence_barrier} together with the reference solutions from \cite{MOSTAFASHAABAN2020156} and \cite{CHEN2018507}, where we can see that the objective function converges to its asymptotic value for both, uniform and adaptive refinement of the solution mesh. Figure \ref{fig:error_convergence_barrier} shows the convergence of the $H_{1}-$error estimator with respect to the DOFs, for the same two frequency values. Overall, convergence studies show that adaptive refinement reaches the reference objective function slightly faster than the uniform refinement, while the convergence of the error estimator is the same in both cases. 

\subsubsection{Shape optimization results}
\begin{figure}[!ht]
	\centering
	\input{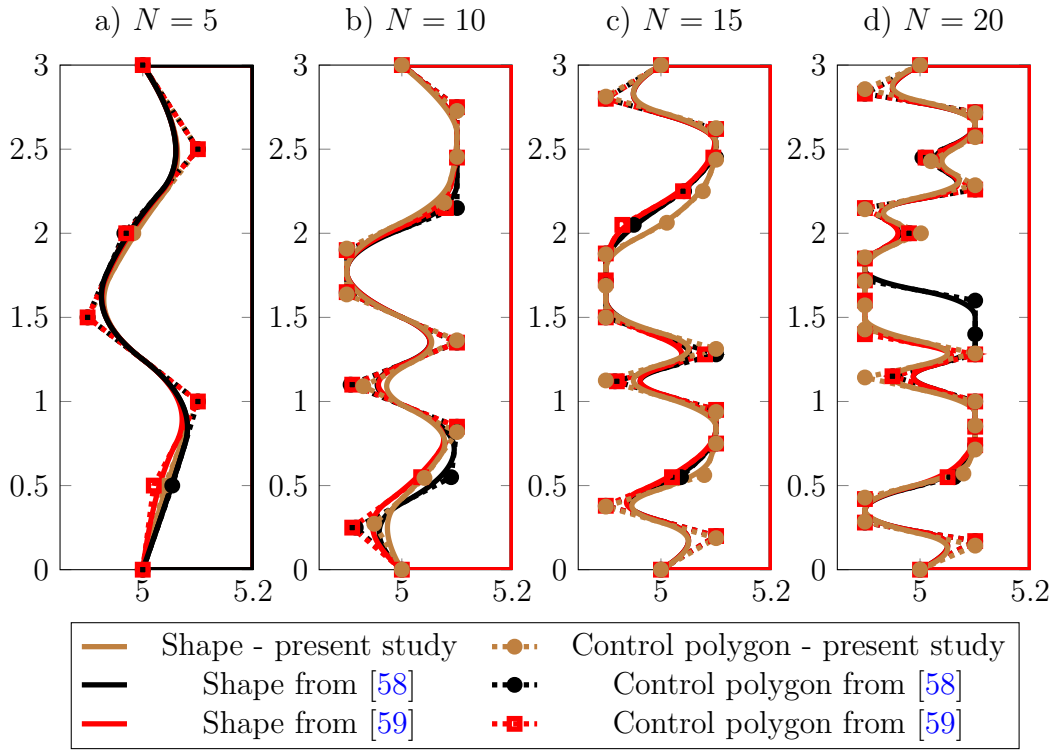}
\caption{Optimized barrier shapes for frequency 400 Hz and 5, 10, 15, and 20 design variables.} 
\label{fig:400-all-shape}
\end{figure}

The optimization results are compared with the reference solutions from the literature. This problem was studied in \cite{MOSTAFASHAABAN2020156} and \cite{CHEN2018507} with the isogeometric boundary element method (IGABEM). The results are compared in Figure \ref{fig:400-all-shape} for four different numbers of control points (design variables): $N = 5, 10, 15, 20$.

\begin{figure}[ht!]
\begin{subfigure}{0.48\textwidth}
	\centering
	\begin{tikzpicture}[scale=1]
\begin{axis}[cycle list name=exotic,
        xlabel = iterations,
        ylabel = $J(u)$,
	xmin = 1,
	xmax = 100,
	ymin = 0.0005,
	ymax = 0.004,
	legend style={at={(0.85,0.5)},anchor=south east}
    ]

\addplot[black, solid, smooth, line width=1.0pt, mark=*, mark size=1.0pt]  table [x expr = \thisrowno{0}, y expr = \thisrowno{1}] {Sections/Shape-optimization-examples/Sound_barrier/Sound_barrier_Figures/grad_history_fval_N_5_f_400.txt};
    \addlegendentry{$N$ = 5};
  
\addplot[blue, solid, smooth, line width=1.0pt, mark=*, mark size=1.0pt]  table [x expr = \thisrowno{0}, y expr = \thisrowno{1}] {Sections/Shape-optimization-examples/Sound_barrier/Sound_barrier_Figures/grad_history_fval_N_10_f_400.txt};
    \addlegendentry{$N$ = 10};
    
\addplot[red, solid, smooth, line width=1.0pt, mark=*, mark size=1.0pt]  table [x expr = \thisrowno{0}, y expr = \thisrowno{1}] {Sections/Shape-optimization-examples/Sound_barrier/Sound_barrier_Figures/grad_history_fval_N_15_f_400.txt};
    \addlegendentry{$N$ = 15};
  
\addplot[brown, solid, smooth, line width=1.0pt, mark=*, mark size=1.0pt]  table [x expr = \thisrowno{0}, y expr = \thisrowno{1}] {Sections/Shape-optimization-examples/Sound_barrier/Sound_barrier_Figures/grad_history_fval_N_20_f_400.txt};
    \addlegendentry{$N$ = 20}; 
    
\end{axis}
\end{tikzpicture}
\caption{Objective function $J(u)$ vs. number of iterations} \label{fig:PSO_obj}
\end{subfigure}
\begin{subfigure}{0.48\textwidth}
	\centering
	\begin{tikzpicture}[scale=1]
\begin{axis}[cycle list name=exotic,
        xlabel = iterations,
        ylabel = $A(u)$,
	xmin = 1,
	xmax = 100,
	ymin = 0.5,
	ymax = 0.6,
	legend style={at={(0.95,0.15)},anchor=south east}
    ]
\addplot[black, solid, smooth, line width=1.0pt, mark=*, mark size=1.0pt]  table [x expr = \thisrowno{0}, y expr = \thisrowno{2}] {Sections/Shape-optimization-examples/Sound_barrier/Sound_barrier_Figures/grad_history_fval_N_5_f_400.txt};
    \addlegendentry{$N$ = 5};   

\addplot[blue, solid, smooth, line width=1.0pt, mark=*, mark size=1.0pt]  table [x expr = \thisrowno{0}, y expr = \thisrowno{2}] {Sections/Shape-optimization-examples/Sound_barrier/Sound_barrier_Figures/grad_history_fval_N_10_f_400.txt};
    \addlegendentry{$N$ = 10};    
       
\addplot[red, solid, smooth, line width=1.0pt, mark=*, mark size=1.0pt]  table [x expr = \thisrowno{0}, y expr = \thisrowno{2}] {Sections/Shape-optimization-examples/Sound_barrier/Sound_barrier_Figures/grad_history_fval_N_15_f_400.txt};
    \addlegendentry{$N$ = 15};   

\addplot[brown, solid, smooth, line width=1.0pt, mark=*, mark size=1.0pt]  table [x expr = \thisrowno{0}, y expr = \thisrowno{2}] {Sections/Shape-optimization-examples/Sound_barrier/Sound_barrier_Figures/grad_history_fval_N_20_f_400.txt};
    \addlegendentry{$N$ = 20};    
\end{axis}
\end{tikzpicture}
\caption{Area $A(u)$ vs. number of iterations} \label{fig:PSO_area}
\end{subfigure}
\caption{Convergence process for the barrier problem.}
\end{figure}
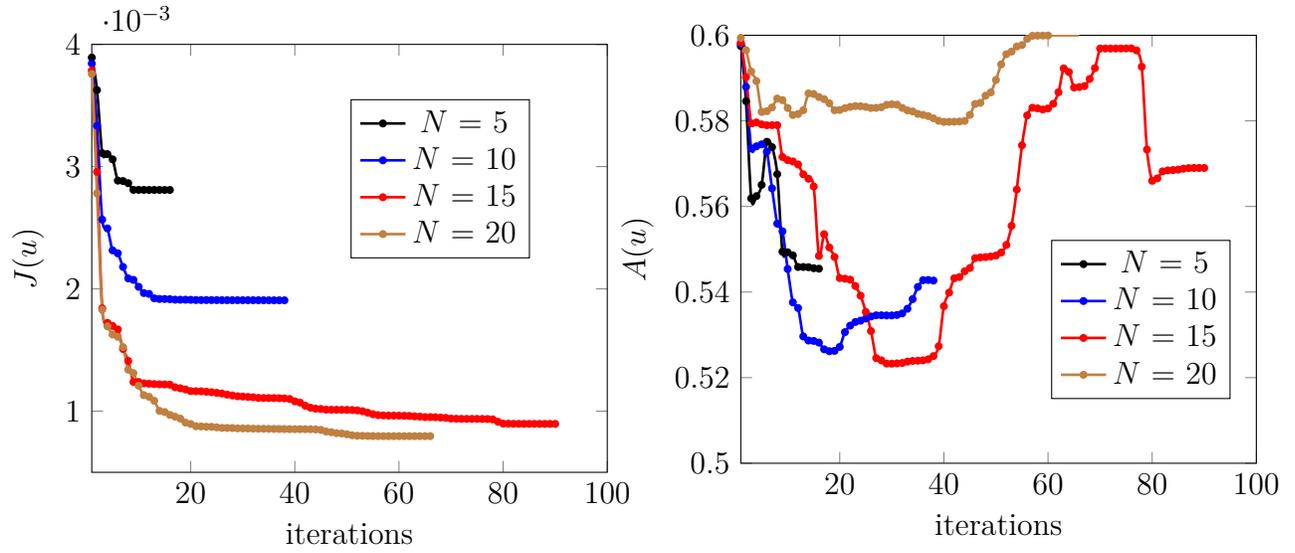

Convergence plots for the objective function and volume constraint are presented in Figures \ref{fig:PSO_obj} and \ref{fig:PSO_area}, respectively.

\begin{figure}[ht!]
    \centering
    \begin{subfigure}[b]{0.48\textwidth}
        \includegraphics[width=\textwidth]{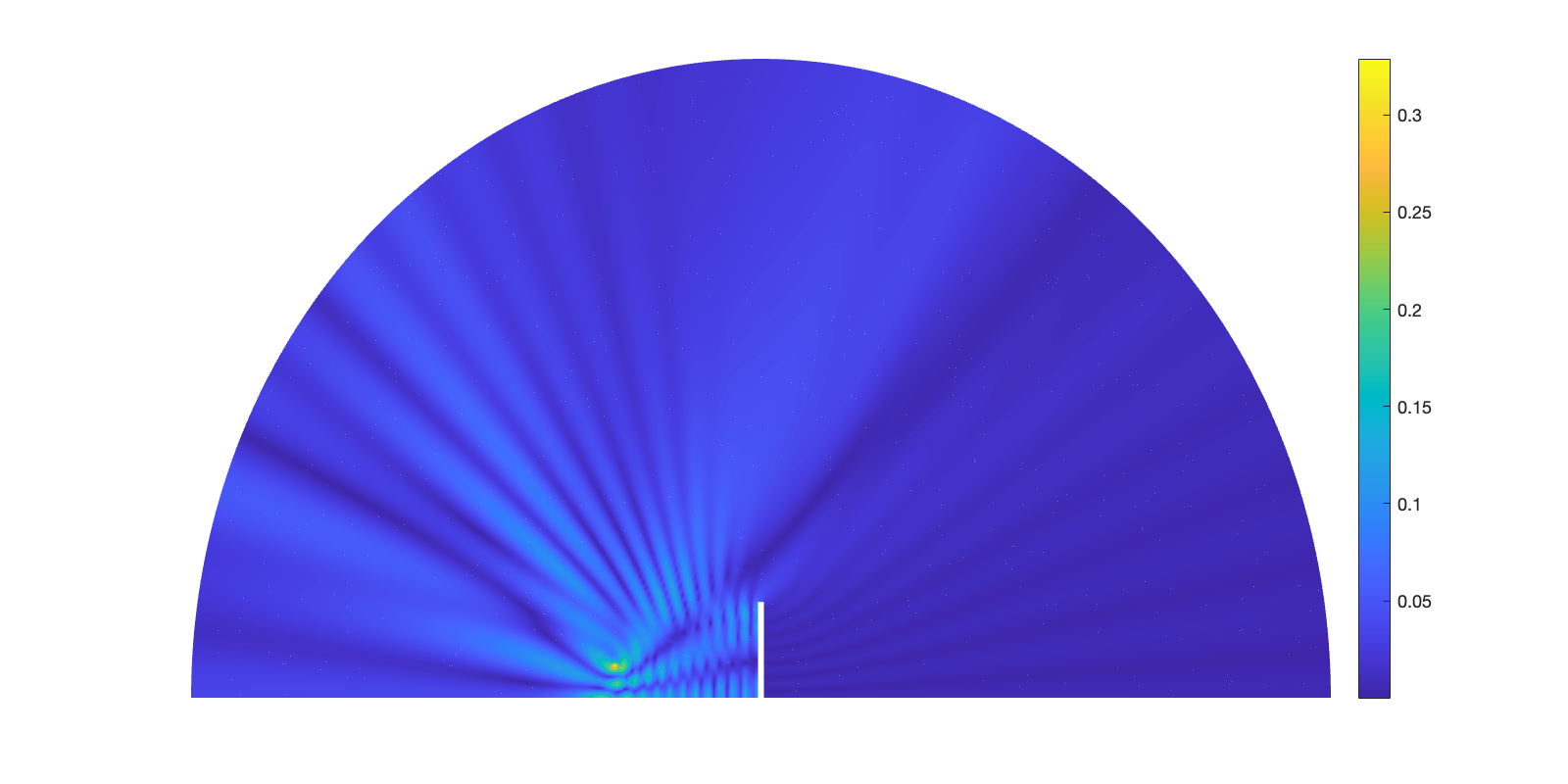}
        \caption{Initial sound pressure}
    \end{subfigure}
    ~ 
    \begin{subfigure}[b]{0.48\textwidth}
        \includegraphics[width=\textwidth]{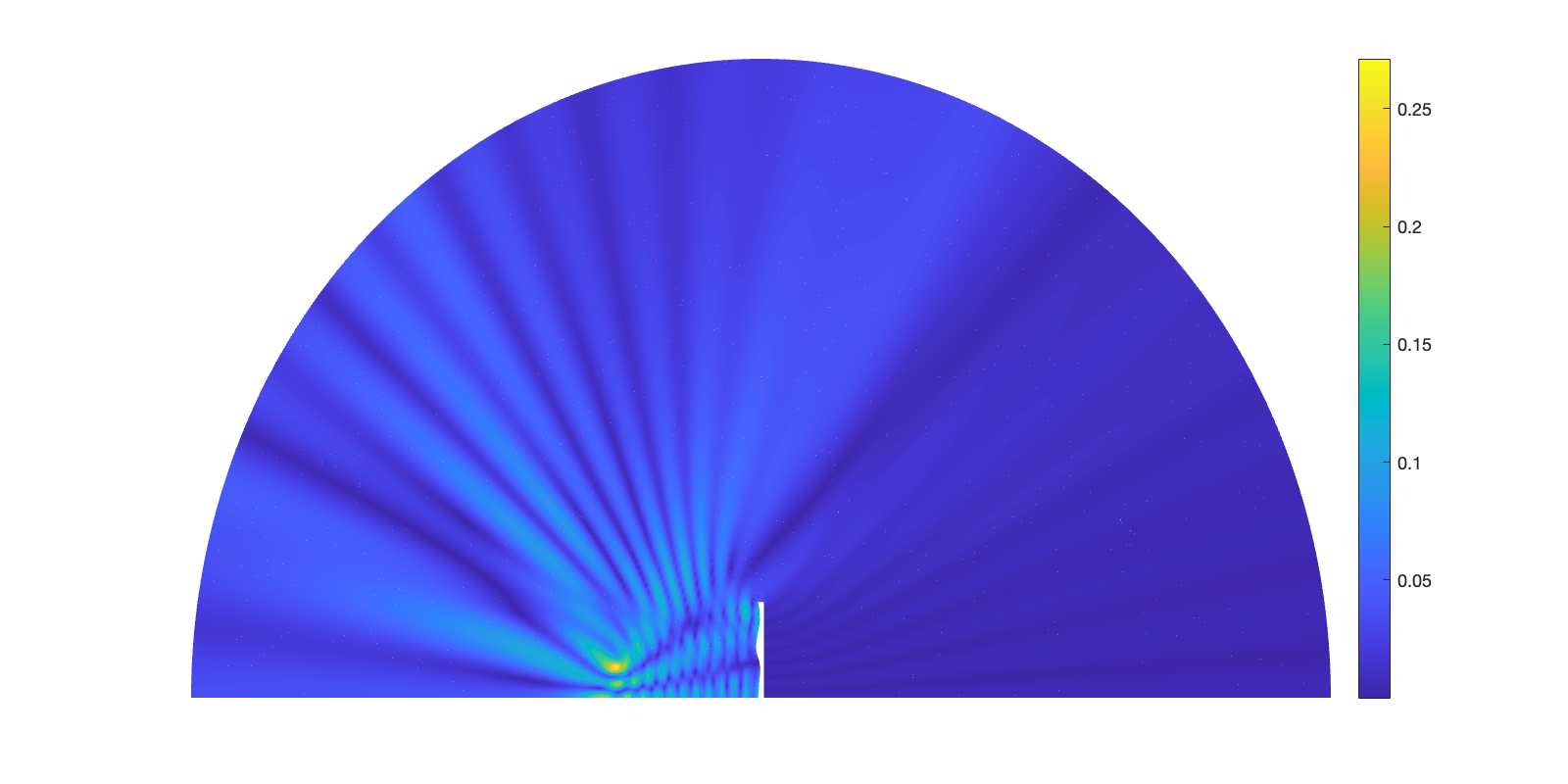}
       \caption{optimized sound pressure for $N=5$}
    \end{subfigure}
    ~ 
    \begin{subfigure}[b]{0.48\textwidth}
        \includegraphics[width=\textwidth]{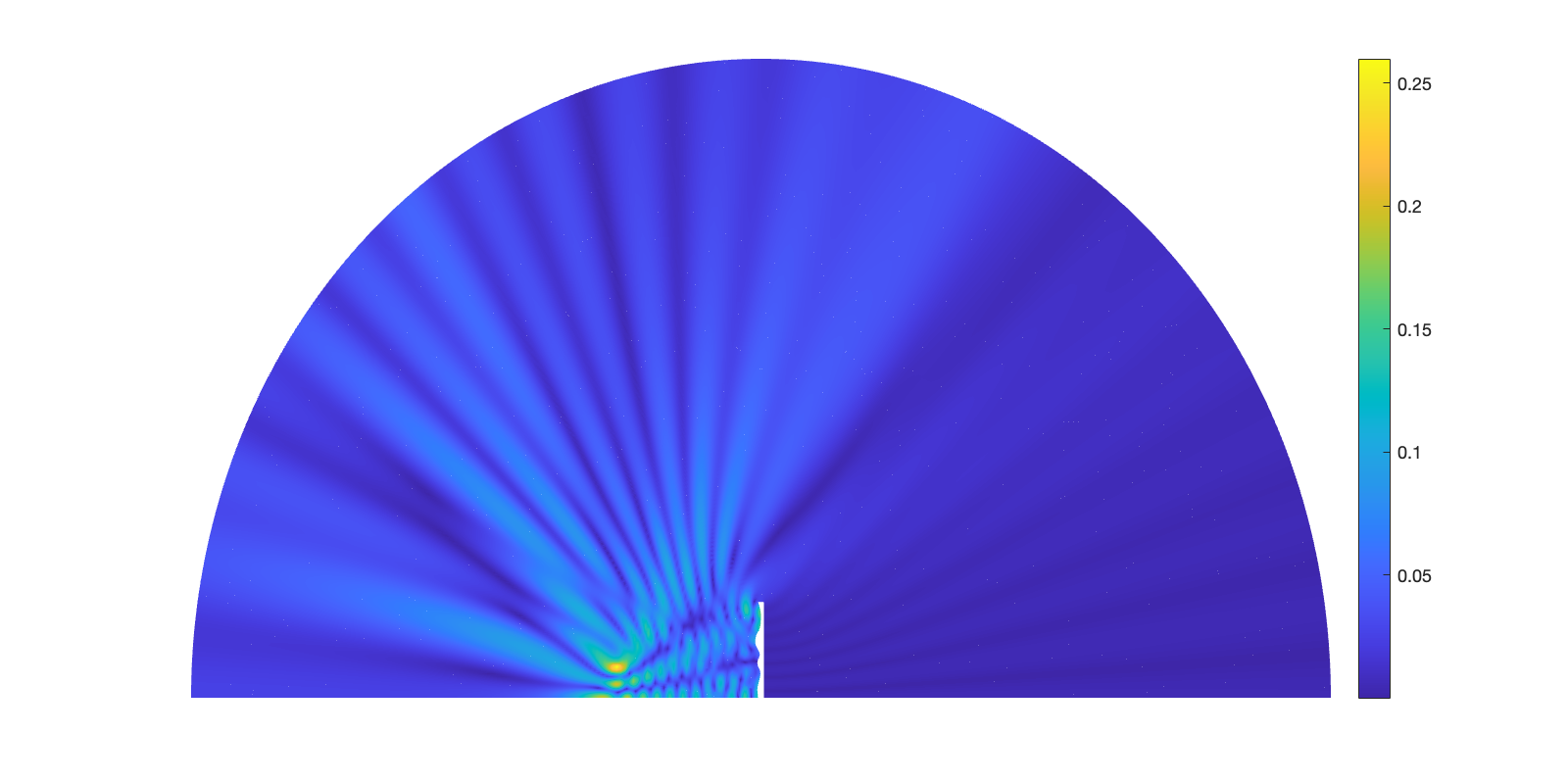}
       \caption{optimized sound pressure for $N=10$}
    \end{subfigure}
    ~ 
    \begin{subfigure}[b]{0.48\textwidth}
        \includegraphics[width=\textwidth]{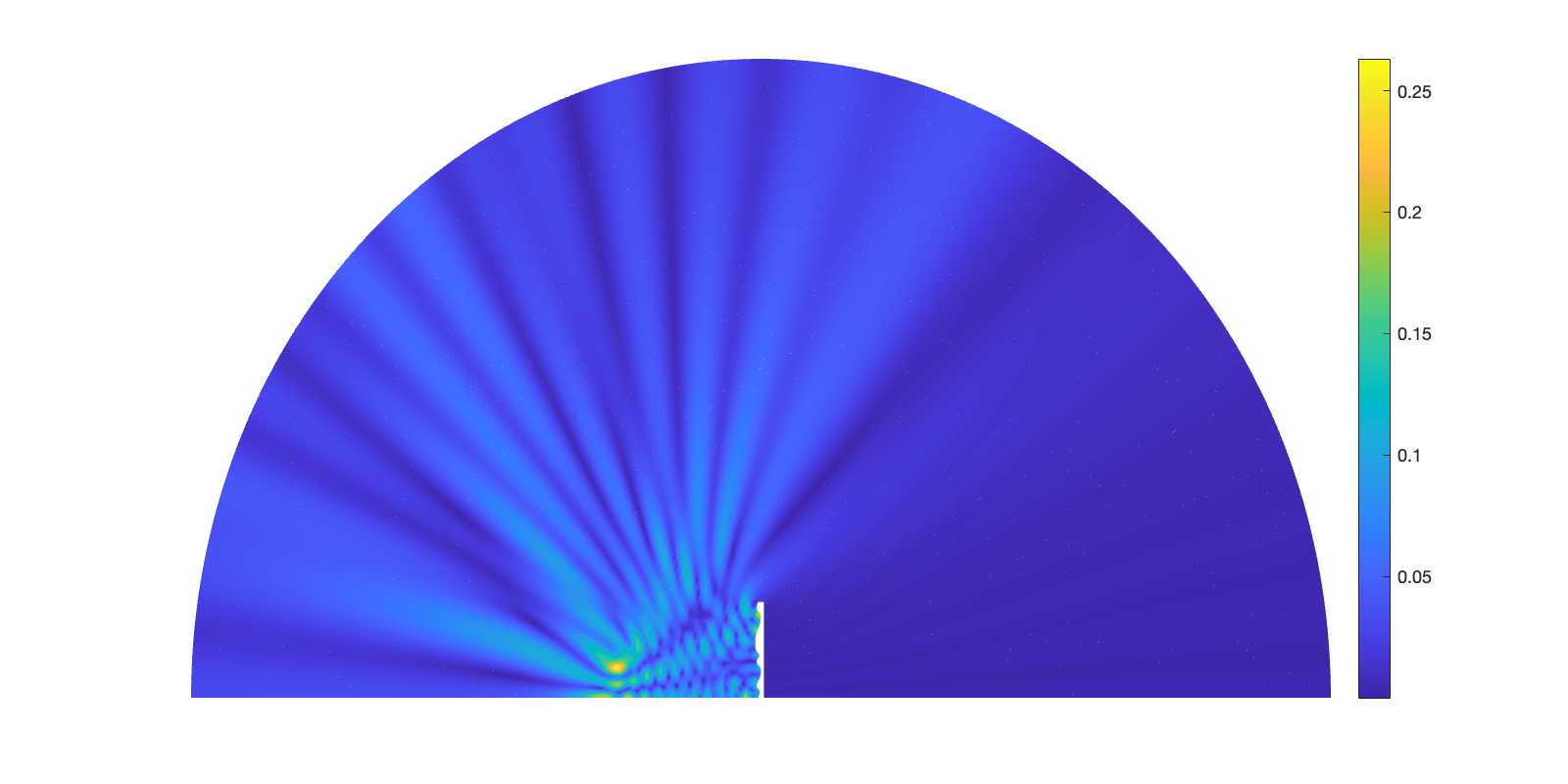}
       \caption{optimized sound pressure for $N=15$}
    \end{subfigure} 
        ~ 
    \begin{subfigure}[b]{0.48\textwidth}
        \includegraphics[width=\textwidth]{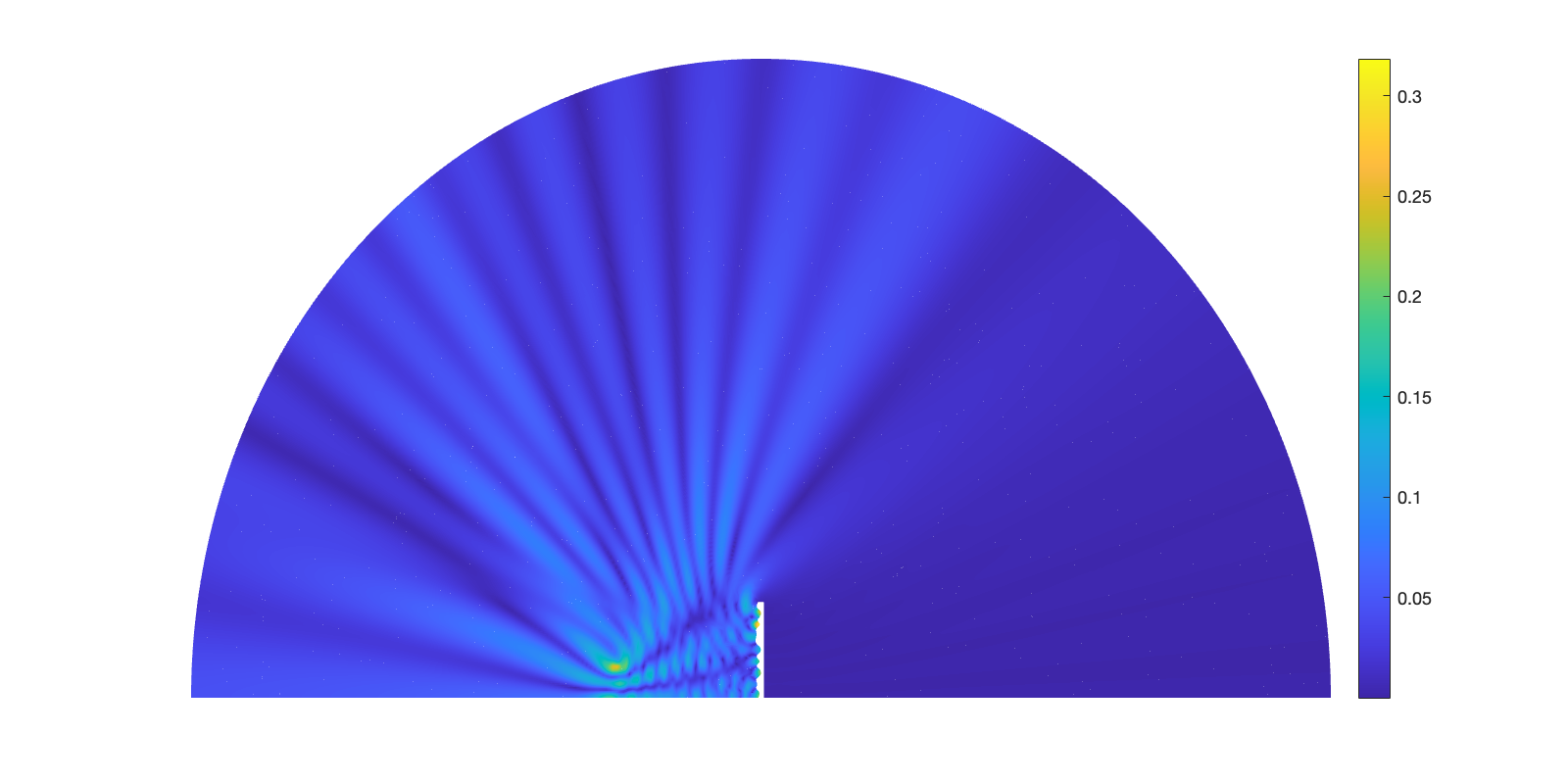}
       \caption{optimized sound pressure for $N=20$}
    \end{subfigure} 
    ~
    \caption{Sound barrier example: sound pressure distribution (absolute value) for the initial and optimized barriers using 5, 10, 15 and 20 design variables.}
    \label{fig:comparison_final_shapes_sound}
\end{figure}

The sound distribution for the optimized shapes is shown in Figure \ref{fig:comparison_final_shapes_sound}. 


\subsubsection{Performance of the adaptive shape optimization algorithm}

Uniform and adaptive optimization are tested by using Algorithm \ref{alg:adaptive_optimisation} for design variables $N=5$, 10, 15 and 20. The studies are performed in terms of DOFs, computational time (s), optimal objective function $J(u)$ and optimal solution. All the results are summarized in Table \ref{tab:barrier_example_comparison_uniform_adaptive_time_dof}. Results show agreement between the uniform and adaptive cases in terms of the objective function and the optimal solution. Regarding computational times, adaptive optimization overcomes uniform optimization for all the studied cases. It is interesting to notice that for cases $N = 15$ and 20, the differences in computational times are 18\% and 90\%, respectively. On the other hand, in terms of DOFs, uniform optimization seems to use less DOFs compared to adaptive optimization. This is explained by the results from the convergence study in Figure \ref{fig:error_convergence_barrier}, since there is no noticeable advantage to use adaptive refinement versus uniform refinement when analyzing the error-estimator versus DOFs. The same is observed in the adaptive meshes which the optimization algorithm generates.

\begin{figure}[ht!]
	\centering
	\begin{tikzpicture}[scale=1]
\begin{axis}[cycle list name=exotic,
    width=0.4\textwidth,
    height=0.8\textwidth,
	xmin = 4.8,
	xmax = 5.2,
	ymin = 0,
	ymax = 3.0,
	xtick = {5, 5.2},
	ytick  = {0, 0.5, 1, 1.5, 2, 2.5, 3},
     legend style={at={(2.9,0.5)},anchor=south east}
    ]

    \addplot [color=blue,
              solid,
              smooth,
             line width=2.0pt]
    table [x expr = \thisrowno{0}, y expr = \thisrowno{1}] {Sections/Shape-optimization-examples/Sound_barrier/Sound_barrier_Figures/opt_GRAD_GIFT_result_shape_uniform_n_10.txt};
    
    \addlegendentry{Shape - uniform refinement};
    
    \addplot [color=blue,
              dotted,
              line width=2.0pt,
              mark=*,
              mark options={solid}]
    table [x expr = \thisrowno{0}, y expr = \thisrowno{1}] {Sections/Shape-optimization-examples/Sound_barrier/Sound_barrier_Figures/cpt_GIFT_FD_N_10.txt};

    \addlegendentry{Control polygon - uniform refinement};
    
    \addplot [color=brown,
              solid,
              smooth,
              line width=2.0pt]
    table [x expr = \thisrowno{0}, y expr = \thisrowno{1}] {Sections/Shape-optimization-examples/Sound_barrier/Sound_barrier_Figures/opt_GRAD_GIFT_result_shape_adapt_n_10.txt};
    
    \addlegendentry{\textcolor{black} {Shape - adaptive refinement}};

    \addplot [color=brown,
              dotted,
              line width=2.0pt,
              mark=*,
              mark options={solid}]
    table [x expr = \thisrowno{0}, y expr = \thisrowno{1}] {Sections/Shape-optimization-examples/Sound_barrier/Sound_barrier_Figures/cpt_GIFT_FD_N_10_adapt.txt};
    
    \addlegendentry{\textcolor{black} {Control polygon - adaptive refinement}};

    \addplot [color=black,
              solid,
              smooth,
              line width=2.0pt]
    table [x expr = \thisrowno{0}, y expr = \thisrowno{1}] {Sections/Shape-optimization-examples/Sound_barrier/Sound_barrier_Figures/shape_Mostafa_N_10.txt};
    
    \addlegendentry{\textcolor{black} {Shape from \cite{MOSTAFASHAABAN2020156}}};
    
    \addplot [color = black,
              dotted,
              line width=2.0pt,
              mark=*,
              mark options={solid}]
    table [x expr = \thisrowno{0}, y expr = \thisrowno{1}] {Sections/Shape-optimization-examples/Sound_barrier/Sound_barrier_Figures/cpt_Mostafa_N_10.txt};
    
    \addlegendentry{\textcolor{black} {Control polygon from \cite{MOSTAFASHAABAN2020156}}};

    \addplot [color=red,
              solid,
              smooth,
              line width=2.0pt]
    table [x expr = \thisrowno{0}, y expr = \thisrowno{1}] {Sections/Shape-optimization-examples/Sound_barrier/Sound_barrier_Figures/shape_Chen_N_10.txt};
    
    \addlegendentry{\textcolor{black} {Shape from \cite{CHEN2018507}}};
    
    \addplot [color=red,
              dotted,
              line width=2.0pt,
              mark=square,
              mark options={solid}]
    table [x expr = \thisrowno{0}, y expr = \thisrowno{1}] {Sections/Shape-optimization-examples/Sound_barrier/Sound_barrier_Figures/cpt_Chen_N_10.txt};
    
    \addlegendentry{\textcolor{black} {Control polygon from \cite{CHEN2018507}}};

    
\end{axis}
\end{tikzpicture}
\caption{Optimized barrier shapes for frequency 400 Hz and 10 design variables. Comparison between uniform and adaptive refinement results.} 
\label{fig:400-10-vars-unif_vs_adapt}
\end{figure}
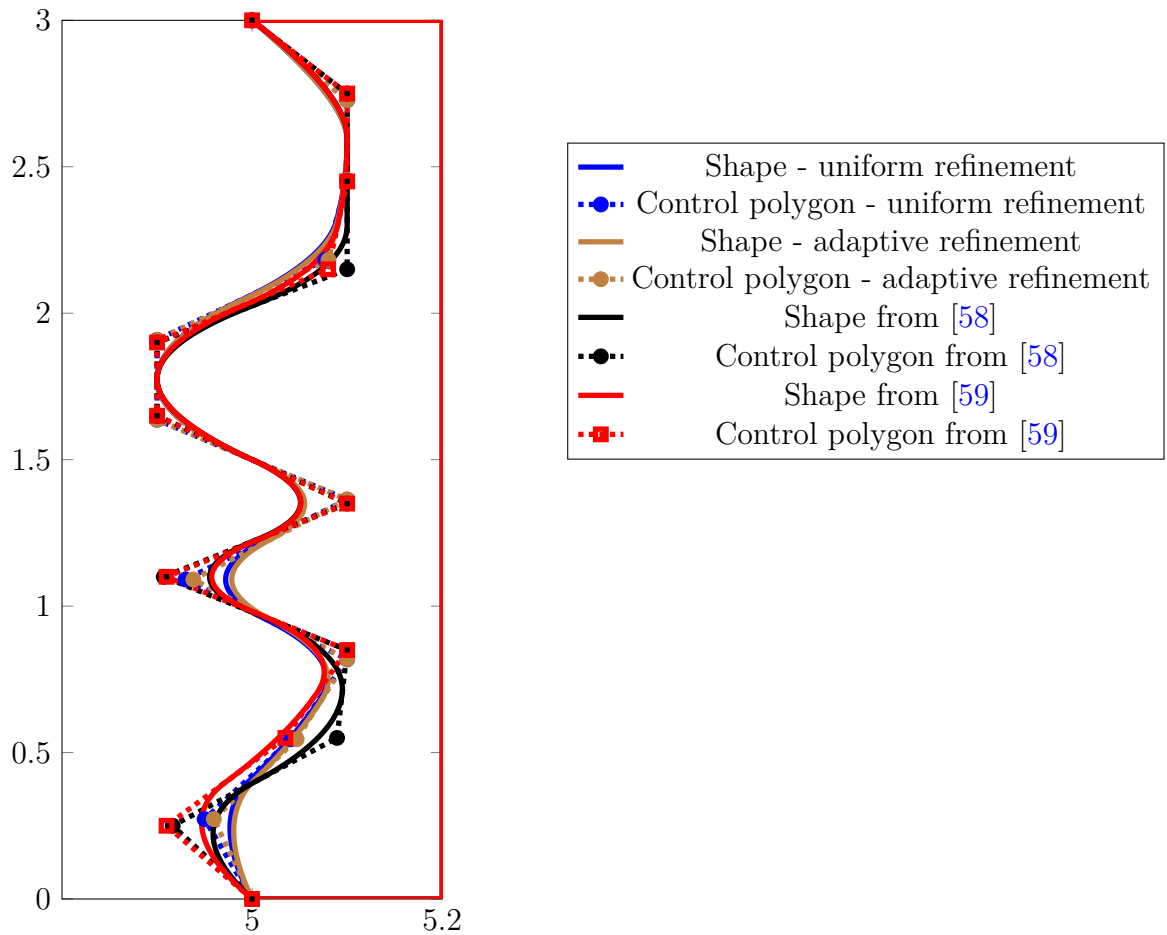

\begin{figure}[ht!]
    \centering
    \begin{subfigure}[b]{0.48\textwidth}
        \includegraphics[width=\textwidth]{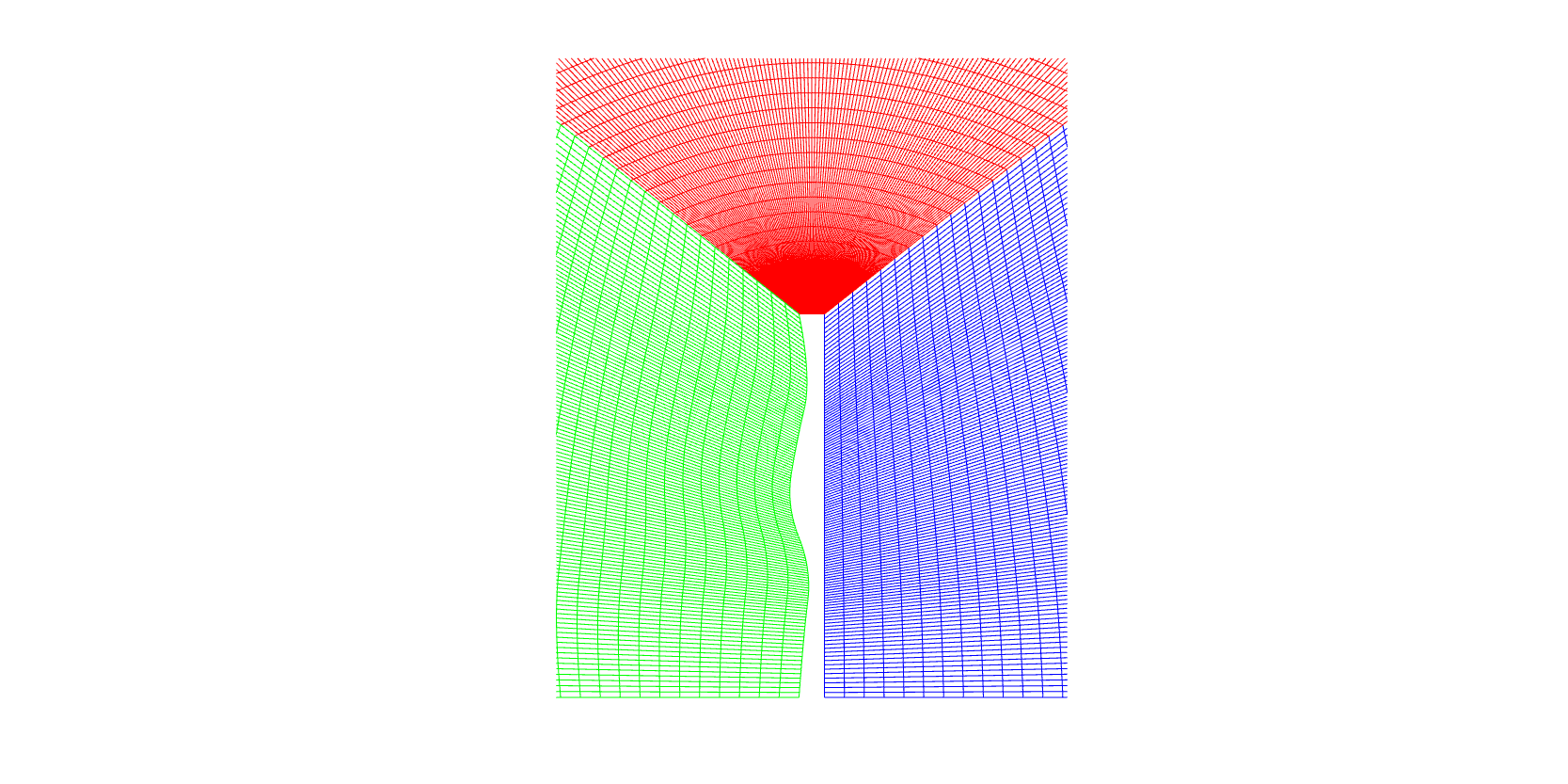}
        \caption{Uniform mesh: optimized shape for $N=5$}
    \end{subfigure}
    ~ 
    \begin{subfigure}[b]{0.48\textwidth}
        \includegraphics[width=\textwidth]{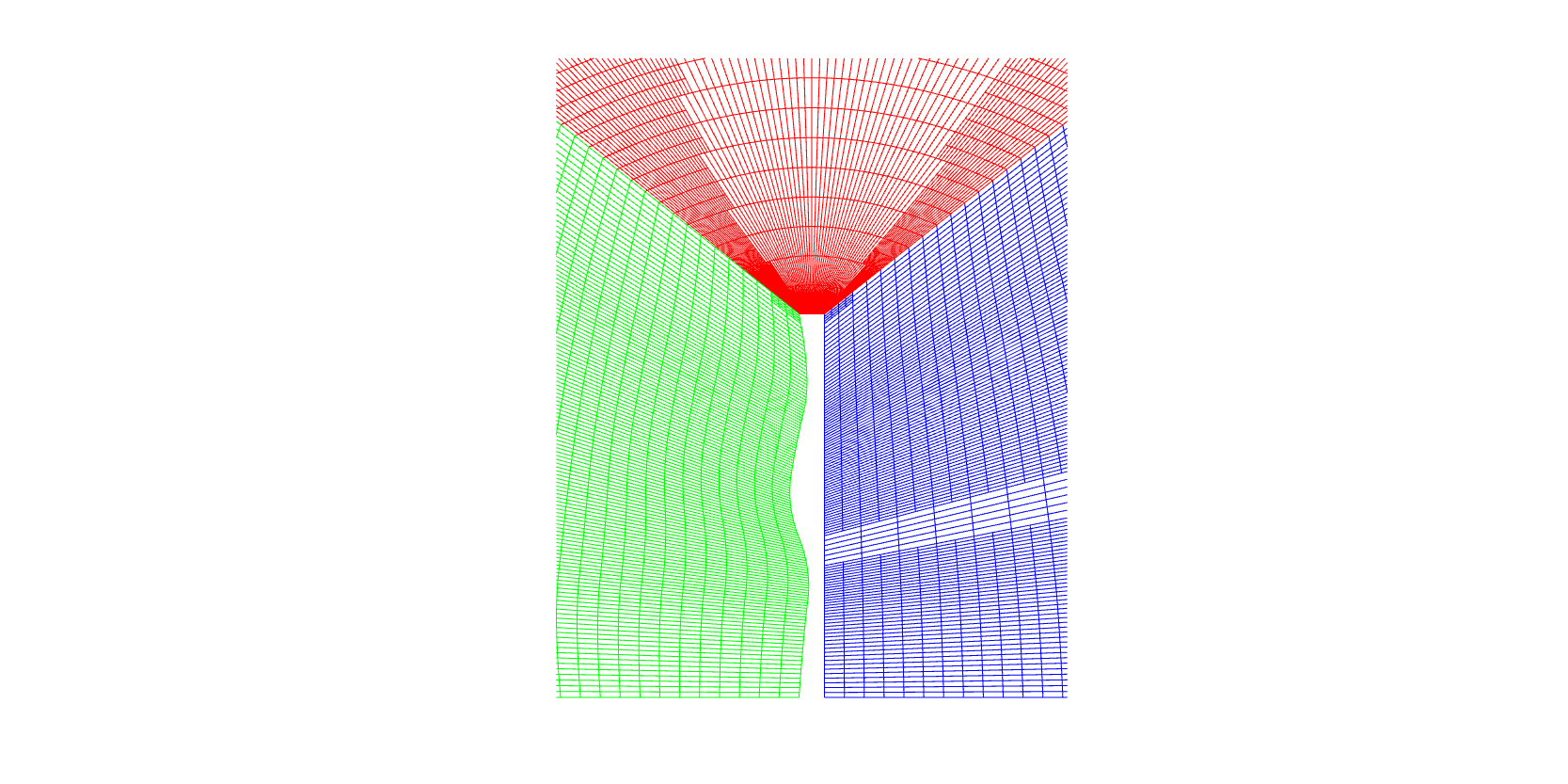}
      \caption{Adaptive mesh: optimized shape for $N=5$}
    \end{subfigure}
    ~ 
    \begin{subfigure}[b]{0.48\textwidth}
        \includegraphics[width=\textwidth]{Sections/Shape-optimization-examples/Sound_barrier/Sound_barrier_Figures/uniform_adapt_meshes/unif_5_bar_zoom.png}
        \caption{Uniform mesh: optimized shape for $N=10$}
    \end{subfigure}
    ~ 
    \begin{subfigure}[b]{0.48\textwidth}
        \includegraphics[width=\textwidth]{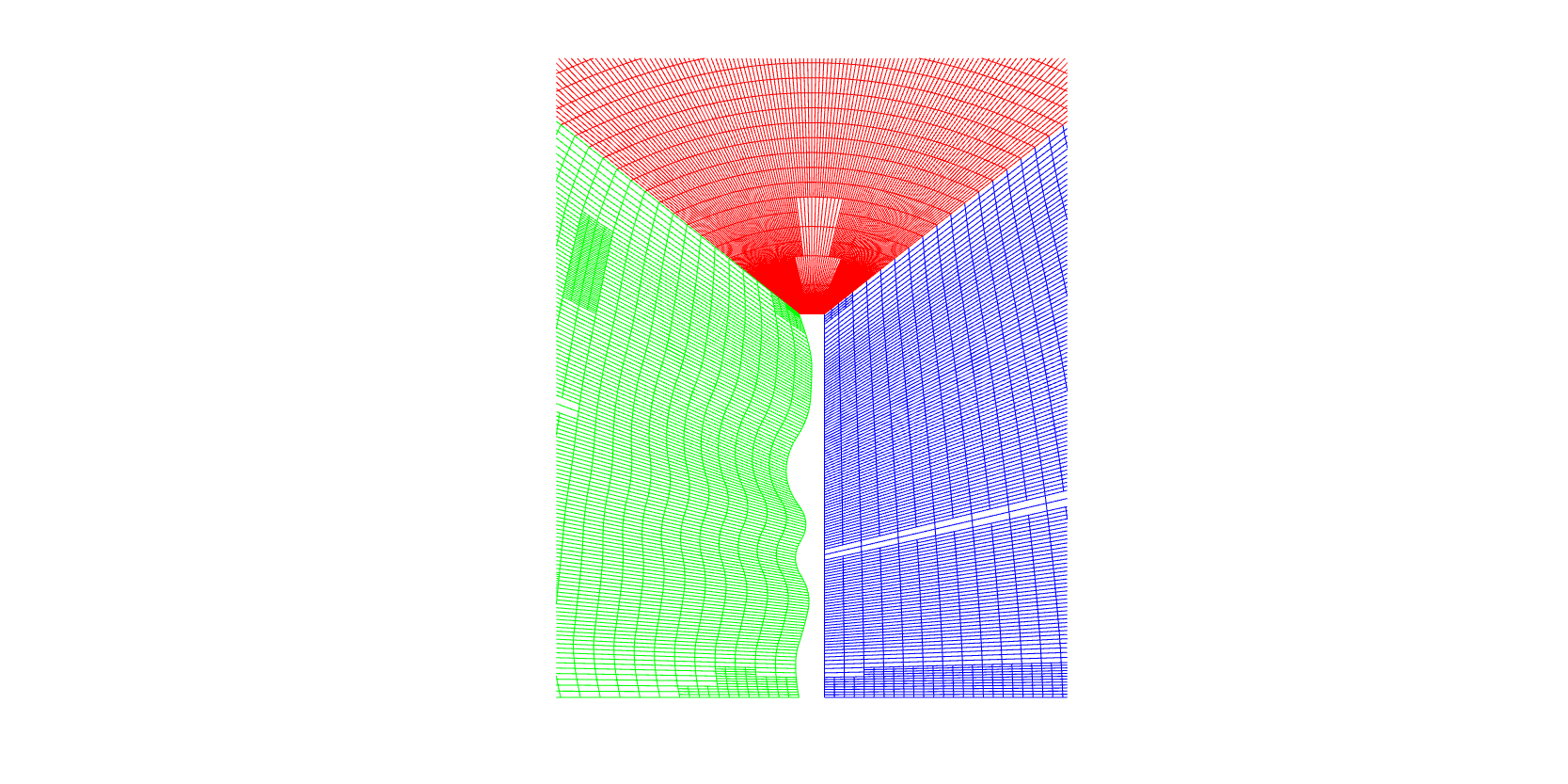}
      \caption{Adaptive mesh: optimized shape for $N=10$}
    \end{subfigure}
        ~ 
    \begin{subfigure}[b]{0.48\textwidth}
        \includegraphics[width=\textwidth]{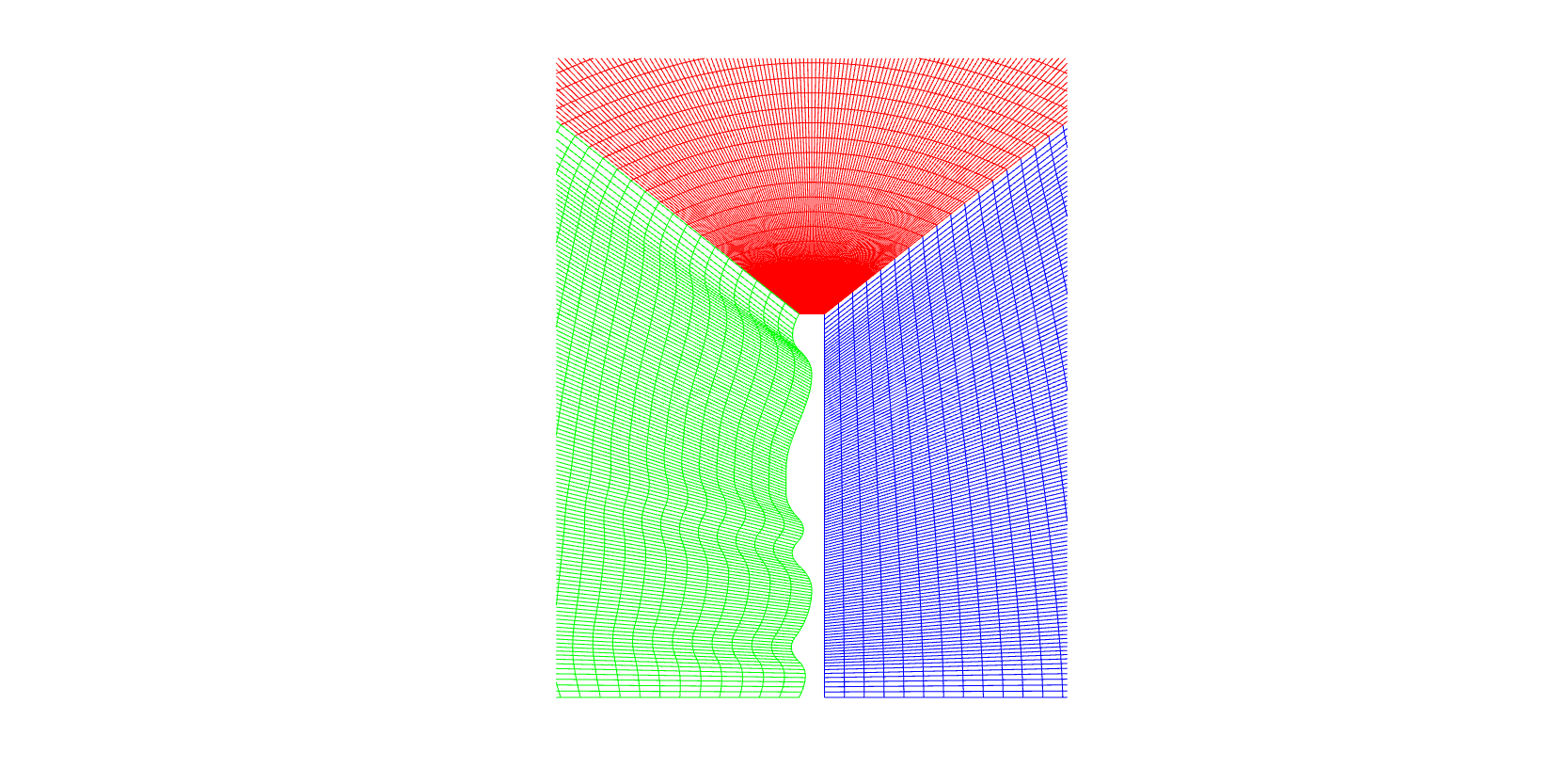}
        \caption{Uniform mesh: optimized shape for $N=15$}
    \end{subfigure}
    ~ 
    \begin{subfigure}[b]{0.48\textwidth}
        \includegraphics[width=\textwidth]{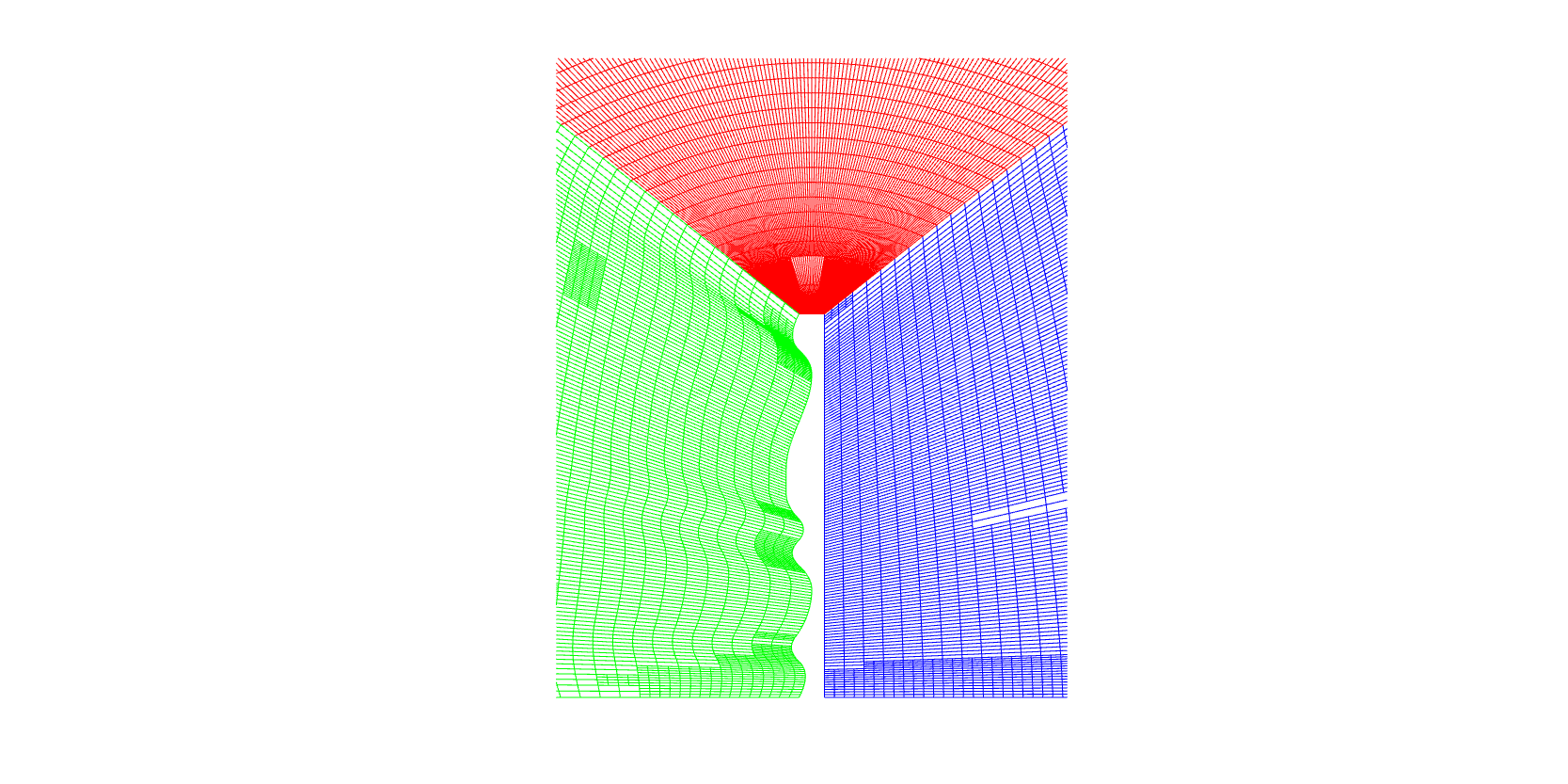}
      \caption{Adaptive mesh: optimized shape for $N=15$}
    \end{subfigure}
    ~
        \begin{subfigure}[b]{0.48\textwidth}
        \includegraphics[width=\textwidth]{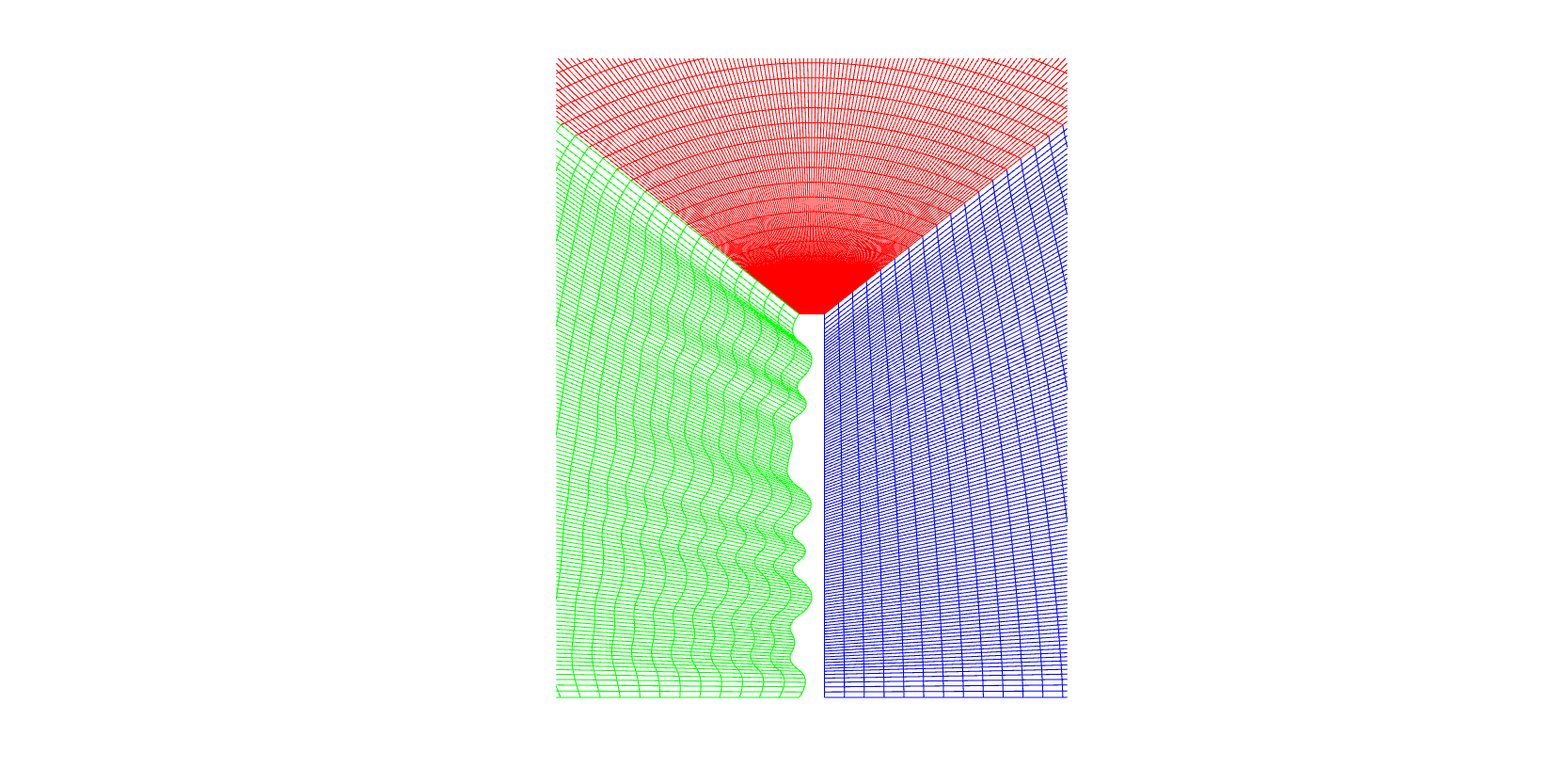}
        \caption{Uniform mesh: optimized shape for $N=20$}
    \end{subfigure}
    ~ 
    \begin{subfigure}[b]{0.48\textwidth}
        \includegraphics[width=\textwidth]{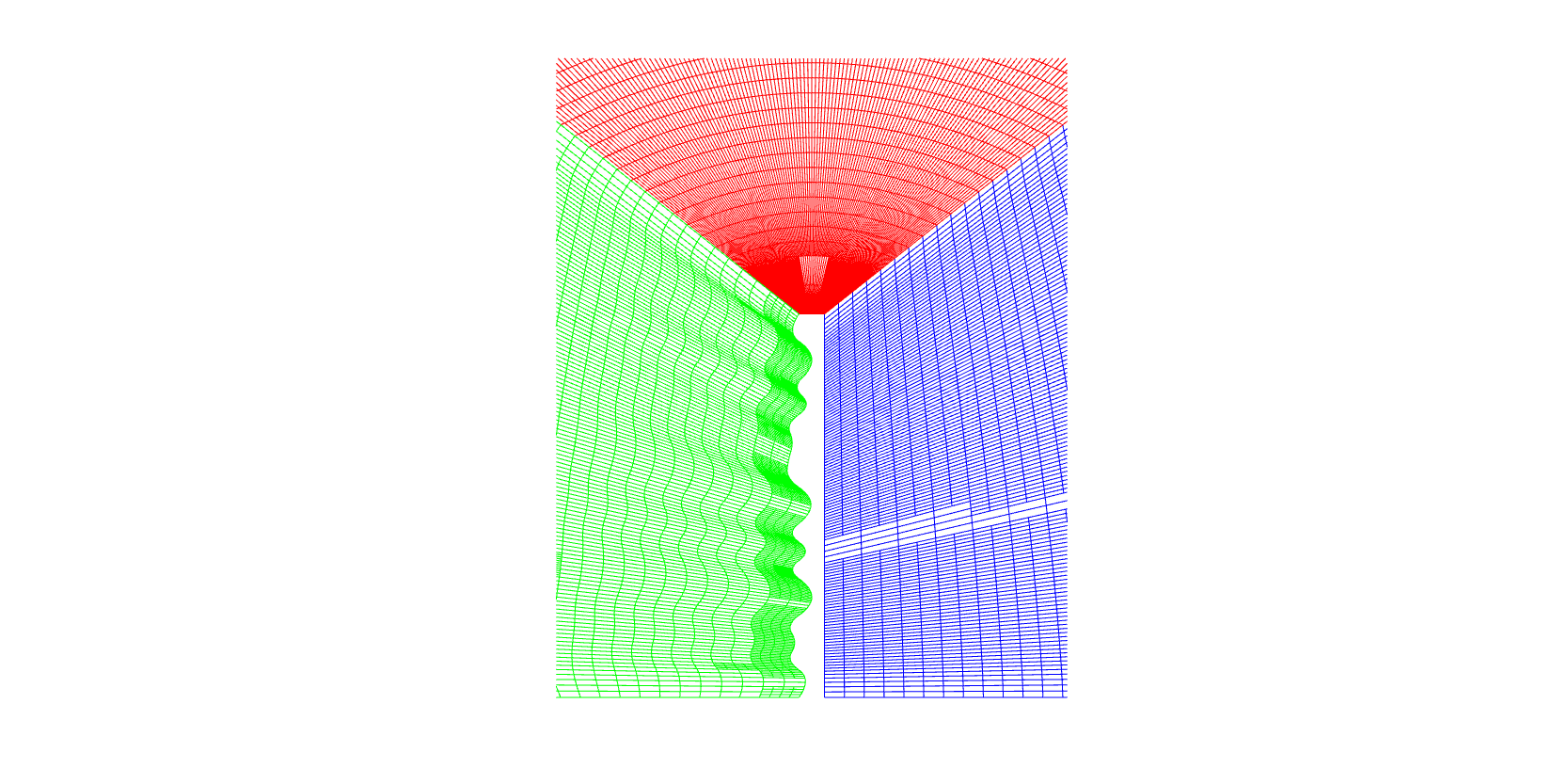}
      \caption{Adaptive mesh: optimized shape for $N=20$}
    \end{subfigure}
    ~
    \caption{Sound barrier example: optimized geometries, uniform and adaptive mesh, zoomed around the barrier area.}
    \label{fig:uniform_adapt_meshes_barrier_example}
\end{figure}

Figure \ref{fig:400-10-vars-unif_vs_adapt} shows the uniform and adaptive optimal shape for $N=10$, as well as the reference results from \cite{MOSTAFASHAABAN2020156} and \cite{CHEN2018507}. Again, uniform and adaptive optimization results as well as the reference results are in a good agreement. Uniform and adaptive final meshes are shown in Figure \ref{fig:uniform_adapt_meshes_barrier_example}.

\begin{landscape}
\begin{table}
\centering
\resizebox{\columnwidth}{!}{%
\begin{tabular}{lcrrr}
\cline{2-5}
                             & \multicolumn{4}{c}{5 Design Variables}                                                                                                                                                                                                               \\ \cline{2-5} 
                             & DOFs                        & \multicolumn{1}{c}{Time (s)} & \multicolumn{1}{c}{Objective function} & \multicolumn{1}{c}{Best Sol}                                                                                                                                 \\ \hline
\multicolumn{1}{r}{uniform}  & \multicolumn{1}{r}{199176} & 1801.761                 & 2.818$\times{10}^{-3}$                    & {[}0.000, -0.120, -0.200, 0.000, -0.063{]}                                                                                                                   \\
\multicolumn{1}{r}{adaptive} & \multicolumn{1}{r}{177976} & 1755.914                 & 2.814$\times{10}^{-3}$                     & {[}0.000, -0.120, -0.200, 0.000, -0.064{]}                                                                                                                   \\ 
\multicolumn{1}{r}{Difference (\%)} & \multicolumn{1}{r}{10.64} & 2.54                 & 0.14                    & -                                                                                                       \\ 
\multicolumn{1}{r}{IGABEM \cite{MOSTAFASHAABAN2020156}}  & \multicolumn{1}{r}{-} & -                 & 2.801$\times{10}^{-3}$                     & {-}                                                                                                                   \\\hline
                             & \multicolumn{1}{l}{}       & \multicolumn{1}{l}{}     & \multicolumn{1}{l}{}          & \multicolumn{1}{l}{}                                                                                                                                         \\ \cline{2-5} 
                             & \multicolumn{4}{c}{10 Design Variables}                                                                                                                                                                                                              \\ \cline{2-5} 
                             & DOFs                        & \multicolumn{1}{c}{Time (s)} & \multicolumn{1}{c}{Objective function} & \multicolumn{1}{c}{Best Sol}                                                                                                                                 \\ \hline
\multicolumn{1}{r}{uniform}  & \multicolumn{1}{r}{199176} & 4192.800                 & 1.886$\times{10}^{-3}$                     & {[}0.000, 0.000, -0.023, -0.200, -0.200, 0.000, -0.168, 0.000, -0.057, -0.149{]}                                                                             \\
\multicolumn{1}{r}{adaptive} & \multicolumn{1}{r}{211468} & 4189.357                 & 1.877$\times{10}^{-3}$                      & {[}0.000, 0.000, -0.023, -0.200, -0.200, 0.000, -0.170, 0.000, -0.059, -0.150{]}                                                                             \\ 
\multicolumn{1}{r}{Difference (\%)} & \multicolumn{1}{r}{-6.17} & 0.08                 & 0.48                    & -                                                                                                       \\ 
\multicolumn{1}{r}{IGABEM \cite{MOSTAFASHAABAN2020156}}  & \multicolumn{1}{r}{-} & -                & 1.904$\times{10}^{-3}$                      & {-}                                                                             \\ \hline
                             & \multicolumn{1}{l}{}       & \multicolumn{1}{l}{}     & \multicolumn{1}{l}{}          & \multicolumn{1}{l}{}                                                                                                                                         \\ \cline{2-5} 
                             & \multicolumn{4}{c}{15 Design Variables}                                                                                                                                                                                                              \\ \cline{2-5} 
                             & DOFs                        & \multicolumn{1}{c}{Time (s)} & \multicolumn{1}{c}{Objective function} & \multicolumn{1}{c}{Best Sol}                                                                                                                                 \\ \hline
\multicolumn{1}{r}{uniform}  & \multicolumn{1}{r}{199176} & 6098.439                 & 1.120$\times{10}^{-3}$                      & {[}-0.200, 0.000, 0.000, -0.060, -0.141, -0.200, -0.200, -0.200, -0.020, -0.200, 0.000, 0.000, -0.062, -0.200	, 0.000{]}                                     \\
\multicolumn{1}{r}{adaptive} & \multicolumn{1}{r}{232164} & 4987.424                 & 1.117$\times{10}^{-3}$                      & {[}-0.200, 0.000, 0.000, -0.059, -0.141, -0.200, -0.200, -0.200, -0.021, -0.200, 0.000, 0.000, -0.062, -0.200, 0.000{]}                                      \\ 
\multicolumn{1}{r}{Difference (\%)} & \multicolumn{1}{r}{-16.56} & 18.22                & 0.27                    & -                                                                                                       \\
\multicolumn{1}{r}{IGABEM \cite{MOSTAFASHAABAN2020156}} & \multicolumn{1}{r}{-} & -                & 1.145$\times{10}^{-3}$                      & {-}                                      \\ \hline
                             & \multicolumn{1}{l}{}       & \multicolumn{1}{l}{}     & \multicolumn{1}{l}{}          & \multicolumn{1}{l}{}                                                                                                                                         \\ \cline{2-5} 
                             & \multicolumn{4}{c}{20 Design Variables}                                                                                                                                                                                                              \\ \cline{2-5} 
                             & DOFs                        & \multicolumn{1}{c}{Time (s)} & \multicolumn{1}{c}{Objective function} & \multicolumn{1}{c}{Best Sol}                                                                                                                                 \\ \hline
\multicolumn{1}{r}{uniform}  & \multicolumn{1}{r}{199176}       & 89967.211                         &   3.804$\times{10}^{-4}$                             &          {[}-0.200, 0.000, 0.000, -0.152, 0.000, -0.200, -0.140, -0.185, -0.200, 0.000, -0.013, -0.200, 0.000, -0.200, 0.000, 0.000, -0.200, -0.113, -0.200, 0.000{]}                                                                                                                                                    \\
\multicolumn{1}{r}{adaptive} & \multicolumn{1}{r}{201564} & 9203.854                 & 3.664$\times{10}^{-4}$                    & {[}-0.200, 0.000, 0.000, -0.151, 0.000, -0.200, -0.141, -0.185, -0.200, 0.000, -0.014, -0.200, 0.000, -0.200, 0.000, 0.000, -0.200, -0.112, -0.200, 0.000{]} \\ 
\multicolumn{1}{r}{Difference (\%)} & \multicolumn{1}{r}{-1.20} & 89.77                & 3.75                    & -                                                                                                       \\
\multicolumn{1}{r}{IGABEM \cite{MOSTAFASHAABAN2020156}} & \multicolumn{1}{r}{-} & -                & 9.630$\times{10}^{-4}$                     & {-} \\ \hline
\end{tabular}%
}
\caption{Sound barrier example: uniform and recovery-based adaptive refinement results. Recovery-based error estimator tolerance is fixed to $\varepsilon_0 = \varepsilon_{\text{loop}} = 5\times{10}^{-2}$.}
\label{tab:barrier_example_comparison_uniform_adaptive_time_dof}
\end{table}
\end{landscape}

\section{Conclusions}\label{conclusions}
In this work we show application of Geometry Independent Field approximaTion (GIFT) for problems of acoustic shape optimization. Computational domain, including the optimized boundary, is modeled by NURBS, while the solution is approximated by adaptive PHT-splines. Accuracy of the solution at each step of the optimization process is controlled by the prescribed error tolerance. In three benchmark examples, the obtained results are shown to be in a good agreement with the reference solutions and it is shown that the adaptive algorithm can bring up to 96\% reduction in terms of the number of degrees of freedom and 94\% in terms of the computational time in comparison with uniform mesh refinement.       

\section{Acknowledgment}

This research was undertaken with the assistance of resources and services from the National Computational Infrastructure (NCI), which is supported by the Australian Government. Also The authors acknowledge support from the UNSW Resource Allocation Scheme managed by Research Technology Services at UNSW Sydney.
\bibliographystyle{unsrt}
\bibliography{bib_shape_opt.bib, bib_acoustics.bib, bib_IGA.bib, bib_shape_reconstruction.bib} 

\bibliography{main.bbl}

\end{document}